\documentclass[useAMS,usenatbib,usegraphicx]{mn2e}

\usepackage{aas_macros}

\title[Magnetic fields in massive star formation]{Magnetic fields during the early stages of massive star formation -- I. Accretion and disc evolution}
\author[D. Seifried et al.]
  {D.~Seifried,$^{1,2}$\thanks{dseifried@ita.uni-heidelberg.de} R.~Banerjee,$^{2,1}$ R.S.~Klessen,$^1$ D.~Duffin,$^3$ R.E.~Pudritz$^3$\\
  $^1$Institut f\"ur Theoretische Astrophysik, Universit\"at Heidelberg, Albert-Ueberle-Str. 2, 69120 Heidelberg, Germany \\
  $^2$Hamburger Sternwarte, Universit\"at Hamburg, Gojenbergsweg 112, 21029 Hamburg, Germany\\
  $^3$Department of Physics $\&$ Astronomy, McMaster University, Hamilton ON L8S 4M1, Canada}

\date{Accepted 2011 June 24.  Received 2011 June 22; in original form 2011 May 5}

\pagerange{\pageref{firstpage}--\pageref{lastpage}} \pubyear{2011}

\begin{document}

\label{firstpage}

\maketitle

\begin{abstract}
We present simulations of collapsing 100 M$_{\sun}$ mass cores in the context of massive star formation. The effect of variable initial rotational and magnetic energies on the formation of massive stars is studied in detail. We focus on accretion rates and on the question under which conditions massive Keplerian discs can form in the very early evolutionary stage of massive protostars. For this purpose, we perform 12 simulations with different initial conditions extending over a wide range in parameter space. The equations of magnetohydrodynamics (MHD) are solved under the assumption of ideal MHD. We find that the formation of Keplerian discs in the very early stages is suppressed for a mass-to-flux ratio normalised to the critical value $\mu$ below 10, in agreement with a series of low-mass star formation simulations. This is caused by very efficient magnetic braking resulting in a nearly instantaneous removal of angular momentum from the disc. For weak magnetic fields, corresponding to $\mu \ga 10$, large-scale, centrifugally supported discs build up with radii exceeding 100 AU. A stability analysis reveals that the discs are supported against gravitationally induced perturbations by the magnetic field and tend to form single stars rather than multiple objects. We find protostellar accretion rates of the order of a few 10$^{-4}$ M$_{\sun}$ yr$^{-1}$ which, considering the large range covered by the initial conditions, vary only by a factor of $\sim$ 3 between the different simulations. We attribute this fact to two competing effects of magnetic fields. On the one hand, magnetic braking enhances accretion by removing angular momentum from the disc thus lowering the centrifugal support against gravity. On the other hand, the combined effect of magnetic pressure and magnetic tension counteracts gravity by exerting an outward directed force on the gas in the disc thus reducing the accretion onto the protostars.
\end{abstract}


\section{Introduction}

The question of how massive stars form is still a highly debated field of research~\citep[e.g.][]{Zinnecker07}. It is believed that massive star formation takes place in high mass molecular cloud cores with masses ranging from roughly 100 M$_{\sun}$ up to a few 1000 M$_{\sun}$. Characteristic for such cores are sizes of few 0.1 pc and peak densities up to 10$^6$ cm$^{-3}$~\citep[e.g.][]{Beuther07}. Furthermore, from observations it is known that the interstellar medium as a whole is magnetised~\citep[see][for a recent overview]{Beck09}. Also the star forming cloud cores partly reveal a significant magnetisation. The importance of the magnetic field can be estimated by the mass-to-flux ratio $\mu$ normalised to the critical mass-to-flux ratio~\citep{Mouschovias76}:
\begin{equation}
 \mu = \frac{M_{core}}{\Phi_{\rmn{core}}}/\left(\frac{M}{\Phi}\right)_{\rmn{crit}} = \frac{M_{core}}{\int B_z dA}/\frac{0.13}{\sqrt{G}}\,.
 \label{eq:mu}
\end{equation}
Observed mass-to-flux ratios in high-mass star forming cores are typically only slightly supercritical with $\mu \la 5$ \citep{Falgarone08,Girart09,Beuther10} indicating a significant influence of magnetic fields on the star formation process. In magneto-hydrodynamical simulation, however, also higher values of $\mu$ up to $\sim$ 20, i.e. weaker magnetic fields, are found~\citep[e.g.][]{Padoan01,Tilley07}. Another common feature of star forming cores are their slow rotation velocities. Observed cores have rotational energies normalised to the gravitational energy which scatter around a mean of about 0.01~\citep{Goodman93,Pirogov03}.

In the field of low-mass star formation there is a great number of observations of discs and large-scale outflows which are the basic keystones of the
widely accepted disc accretion scenario for low-mass star formation~\citep[see, e.g. the reviews by][]{MacLow04,McKee07}. A similar formation scenario for massive stars is supported by a growing number of discs and bipolar outflows observed around high-mass protostellar objects~\citep[see][for recent reviews]{Beuther05,Cesaroni07}. Therefore, it is worthwhile to study the influence of magnetic fields and rotation on the formation of discs and outflows in the context of massive star formation with numerical simulations.

For low-mass star forming regions (M$_{\rmn{core}}$ $\sim$ 1 M$_{\sun}$), the influence of magnetic fields on the collapse of rotating cloud cores, the subsequent formation and evolution of discs and the launching of outflows has received extensive attention~\citep[e.g.][]{Allen03,Banerjee06b,Price07,Mellon08,Hennebelle08,Hennebelle08b,Hennebelle09,Duffin09,Machida10}. All authors find a more or less significant influence of magnetic fields on the evolution of discs surrounding the protostars. The perhaps most important result of these studies is that for a mass-to-flux ratio of $\mu \la 10$ the formation of Keplerian discs is largely suppressed.
This so-called magnetic braking catastrophe \citep{Allen03,Mellon08} turned the traditional angular momentum problem upside down: In highly magnetised cores magnetic braking seems to be so efficient that large-scale Keplerian discs, commonly observed around low-mass protostars, cannot form. Low-mass star formation simulations also reveal a strong impact of magnetic fields on the fragmentation properties of discs. In particular, strong magnetic fields tend to suppress disc fragmentation even in the presence of initial density perturbations~\citep[e.g.][]{Hosking04,Machida05,Hennebelle08b,Duffin09}. These results are in contrast with the observational fact that a large fraction of low-mass stars are binaries~\citep[e.g.][]{Duquennoy91}.

Numerical studies of the influence of magnetic fields on massive star formation, however, have received attention only recently~\citep{Banerjee07,Peters11,Hennebelle11}. \citet{Banerjee07} study the very early evolution of a protostar, its surrounding disc and the outflow performing a simulation with extremely high spatial resolution. The protostellar evolution over a timescale of some 10$^4$ yr is examined by \citet{Peters11} and \citet{Hennebelle11}, the latter authors focussing on the effect of magnetic fields and turbulence while \citet{Peters11} study the interplay of magnetic fields and radiation. Here, we systematically study the influence of rotation and magnetic fields on the formation and accretion history of massive stars and examine the question under which conditions massive Keplerian discs can form in an already very early stage of protostellar evolution. We perform a series of collapse simulations with different initial rotational and magnetic energies following the protostellar evolution over 4000 yr. Rotational and magnetic energies are selected in a way to cover a large range in parameter space in accordance with observations and numerical simulations. Thus, we are able to detect systematic dependencies on the initial conditions. Furthermore, the simulations serve as a useful guide to select representative parameter sets for subsequent and more detailed studies.

The paper is organised as follows. In Section \ref{sec:techniques} the numerical techniques and the initial conditions of the runs performed are described briefly. The results of the simulations are presented in Section \ref{sec:results}. First, the time evolution of the discs is presented for two representative simulations. Next, we analyse the velocity structure and the magnetic properties of the gas in the midplane. Afterwards, the accretion histories of the protostars as well as the effects of disc fragmentation are examined. In Section \ref{sec:dis} the results are discussed in a broader context and are compared to other numerical and observational studies before we summarise our results in Section \ref{sec:conclusion}.

\section{Numerical techniques and initial conditions} \label{sec:techniques}

\subsection{Numerical methods} \label{sec:methods}

We present 3-dimensional, magnetohydrodynamical (MHD) simulations of collapsing molecular cloud cores using the AMR code FLASH~\citep{Fryxell00}. We solve the equations of ideal MHD including self-gravity. The MHD-solver used preserves positive states and applies well for highly supersonic, astrophysical problems~\citep{Bouchut07,Waagan09,Bouchut10,Waagan11}. Using 13 levels of refinement results in a maximum spatial resolution of 4.7 AU. We apply a refinement criterion that guarantees that the Jeans length
\begin{equation}
 \lambda_{\rmn{J}} = \sqrt{\frac{\pi c_\rmn{s}^2}{G \rho}}
\end{equation}
is always resolved with at least 8 grid cells. Here $c_{\rmn{s}}$ denotes the sound speed and $G$ the gravitational constant.
A resolution of 4.7 AU guarantees that we can resolve the Jeans length at $\sim 10^{-13}$ g cm$^{-3}$ corresponding to the density limit where molecular gas starts to get optically thick (see also the Appendix). In addition, a second refinement criterion implemented in FLASH is applied making use of the second derivative of the density field. This criterion guarantees that protostellar discs and outflows are resolved on the highest level of refinement allowing us to draw detailed conclusions about their properties.

We follow the collapse of the molecular cloud core with increasingly higher resolution until the maximum refinement level is reached. As soon as the density exceeds the critical value of
\begin{equation}
 \rho_{\rmn{crit}} = 1.78 \cdot 10^{-12} \rmn{g\,cm^{-3}}
 \label{eq:crit}
\end{equation}
we create a sink particle~\citep[for details see][]{Federrath10}. Subsequently, all gas with $\rho > \rho_{\rmn{crit}}$ within a radius of $r_{\rmn{sink}} = 12.6$ AU around the sink particle is accreted onto the sink provided that it is gravitationally bound.
The density cut-off at $\rho_{\rmn{crit}}$ guarantees that the Jeans length is resolved everywhere by about 8 grid cells. The magnetic field is not changed during the accretion process to avoid the violation of the divergence-free condition. Keeping the magnetic field unchanged is motivated also physically by the onset of ambipolar diffusion at densities of $\sim$ 10$^{-12}$ g cm$^{-3}$~\citep[e.g.][]{Nakano02}. Ambipolar diffusion would allow the gas to slip against the field lines, hence leaving them in the environment of the protostar. We briefly mention that due to numerical diffusion the increase of the magnetic field strength in the centre ceases at some point during the simulations therefore avoiding an unlimited growth of the Alfv\'enic velocity
\begin{equation}
 v_{\rmn{A}} = \frac{B}{\sqrt{4 \pi \rho}}\,.
\label{eq:alf}
\end{equation}

In order to follow the thermodynamical evolution of the gas properly, we apply a cooling routine which includes the effects of molecular cooling and dust cooling as well as a treatment for optically thick gas~\citep[see][for details]{Banerjee06a}. In this routine the cooling rates are computed after each hydrodynamical timestep and the state variables are updated accordingly for all grid cells. Using a subcycling scheme ensures that during one subcycle timestep the thermal energy of each cell does not change by more than 20\%. We mention that the cooling is not applied to the ambient medium enclosing the molecular cloud core to prevent it from collapsing as well.

As soon as an outflow is launched from the protostellar disc, we introduce a minimum gas density threshold within a radius of 67 AU around the simulation centre. Within this radius the gas density is kept artificially above a minimum value of $1 \cdot 10^{-15}$ g cm$^{-3}$. Whenever a cell density falls below this threshold during one timestep, mass is added to this particular cell until the limit of $1 \cdot 10^{-15}$ g cm$^{-3}$ is reached. Hence, we avoid the hydrodynamical timestep falling to prohibitively small values which otherwise would result from high Alfv\'enic velocities caused by low densities (see Eq.~\ref{eq:alf}). The effects of this artificial density threshold will be discussed in detail in Section~\ref{sec:numerics}.

\subsection{Initial conditions}

With the numerical methods described above we are able to follow the collapse of a magnetised molecular cloud core down to a resolution of 4.7 AU. The core has a mass of 100 M$_{\sun}$ and a diameter of 0.25 pc surrounded by gas with a density 100 times lower than the density at the core edge. The cubic simulation box has a length of 0.75 pc with the core located in its centre. Initially the core is resolved by a grid with a spacing of 302 AU fulfilling the applied refinement criteria. To assure pressure equilibrium at the edge of the core, the ambient gas has a temperature of 2000 K which is 100 times higher than the initially constant core temperature of 20 K. Calculating the Jeans mass using a temperature of 20 K and the average density of the core shows that the core contains about 56 Jeans masses and is therefore highly gravitationally unstable. The core itself has a density profile
\begin{equation}
 \rho(r) \propto r^{-1.5}.
\end{equation}
Hence, the density in the centre\footnote{To avoid unphysically high densities in the interior of the core, we cut off the $r^{-1.5}$-profile at a radius of 0.0125 pc. Within this radius the density distribution follows a parabola $\rho(r) \propto (1-(r/r_0)^2)$.} and at the edge of the core are initially $2.3 \cdot 10^{-17}$ g cm$^{-3}$ and $4.2 \cdot 10^{-19}$ g cm$^{-3}$. For a mean molecular weight of $\mu_{\rmn{mol}} = 2.3$ adopted in this work, this corresponds to a particle density of $n = 6.0 \cdot 10^6$ and $1.1 \cdot 10^5$ cm$^{-3}$, respectively. The initial core mass and size are comparable to that of observed massive molecular cloud cores~\citep[e.g.][and references therein]{Beuther07}, where masses of 100 - 1000 M$_{\sun}$ and sizes of a few 0.1 pc are found. The adopted density profile is also in good agreement with observations finding exponents around -1.5~\citep[e.g.][]{Beuther02b,Pirogov03,Pirogov09}. In addition, the initial setup described above is comparable to recent studies on massive star formation in mass and/or density profile~\citep{Krumholz07,Krumholz09,Girichidis11,Hennebelle11,Peters10a,Peters10b,Peters11} which allows for quantitative comparison.

Beside the mass distribution, which is not changed in this work, there are two more crucial parameters affecting the evolution of the cloud core, namely the magnetic field strength and the rotational energy. Initially, all simulations have a magnetic field pointing into the z-direction where $B_z$ declines outwards with the cylindrical radius R as
\begin{equation}
 B_z \propto R^{-0.75}.
\end{equation}
This corresponds to a constant $\beta_{\rmn{plasma}} = P/(B^2/8\pi)$ in the equatorial plane where $P$ is the thermal pressure. To guarantee $\nabla \bmath{B} = 0$, in the initial configuration $B_z$ is constant along the z-direction. We note that the chosen field configuration is compromise between observations and a numerically feasible setup. However, by the time the first sink particles form, the field has reached a self-consistent, hourglass-like configuration as observed in massive star forming regions~\citep[e.g.][]{Girart09,Alves11}. Therefore, we expect the chosen initial configuration of the magnetic field not to change the results significantly compared to an initially hourglass-shaped configuration. Initially, the cores are also rotating rigidly with the rotation axis parallel to the magnetic field. Neither the density field nor the velocity field have any perturbations at the beginning of the simulation so that a uniform and inward directed collapse can be expected.

\begin{figure}
 \includegraphics[width=84mm]{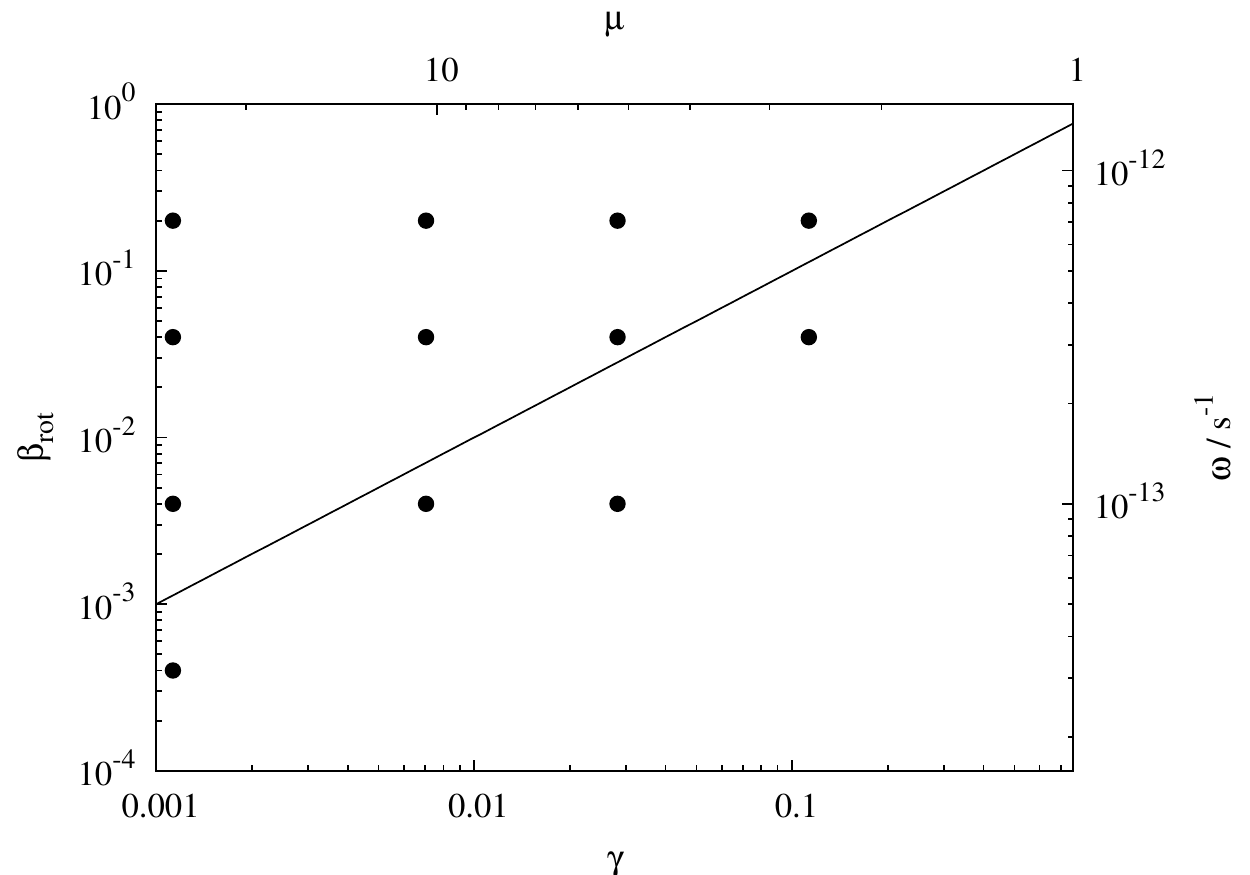}
 \caption{Initial values of the 12 simulations performed. The initial rotational ($\beta_{\rmn{rot}}$) and magnetic energy ($\gamma$) are normalised to the gravitational energy. As can be seen, the starting points bracket the curve where rotational and magnetic energy are equal (black line). The second axes show the corresponding normalised mass-to-flux ratio $\mu$ and the rotation frequency $\omega$.}
 \label{fig:models}
\end{figure}

We performed 12 simulations changing the initial magnetic field strength and the initial rotation frequency $\omega$. For comparative purposes we also performed a simulation with zero magnetic field and zero rotation. The initial values of all simulations presented here are shown in Fig.~\ref{fig:models}. As the magnetic field is changing with radius, Fig.~\ref{fig:models} shows the initial ratio of magnetic energy to gravitational energy $\gamma$ and the corresponding normalised mass-to-flux ratio (Eq.~\ref{eq:mu}) averaged over the entire core. The initial magnetic field strength at the centre of the cores ranges from $100 \mu$G up to 1 mG. For comparison with $\gamma$, Fig.~\ref{fig:models} also shows the ratio of rotational energy to gravitational energy $\beta_{\rmn{rot}}$ beside the angular frequency. Both $\gamma$ and $\beta_{\rmn{rot}}$ are calculated numerically and are always well below 1 as the gravitational energy is significantly larger than the magnetic and rotational energies. In addition, it can be seen from Fig.~\ref{fig:models} that magnetic and rotational energies extend over more than two orders of magnitude and bracket the line where $\gamma$ equals $\beta_{\rmn{rot}}$. The initial values and the names of all runs are listed in Table~\ref{tab:models}. In the nomenclature of each run the first number give the mass-to-flux ratio and the second number $\beta_{\rmn{rot}}$ multiplied by a factor of 100.
\begin{table}
 \caption{Performed simulations with normalised initial mass-to-flux-ratio $\mu$, ratio of magnetic to gravitational energy $\gamma$, magnetic field strength in the centre B, ratio of rotational to gravitational energy $\beta_{\rmn{rot}}$, and the corresponding angular frequency $\omega$.}
 \label{tab:models}
 \begin{tabular}{@{}lccccc}
  \hline
  Run & $\mu$  & $\gamma$ & B ($\mu$G) & $\beta_{\rmn{rot}}$ & $\omega$ (10$^{-13}s^{-1}$) \\
  \hline
  26-20 & 26 & $1.13 \cdot 10^{-3}$ & 132 & $2 \cdot 10^{-1}$ & 7.07 \\
  26-4  & 26 &  $1.13 \cdot 10^{-3}$ & 132 & $4 \cdot 10^{-2}$ & 3.16 \\
  26-0.4 & 26 &  $1.13 \cdot 10^{-3}$ & 132 & $4 \cdot 10^{-3}$ & 1.00 \\
  26-0.04 & 26 &  $1.13 \cdot 10^{-3}$ & 132 & $4 \cdot 10^{-4}$ & 0.316 \\
  \hline
  10-20 & 10.4 & $7.06 \cdot 10^{-3}$ & 330 & $2 \cdot 10^{-1}$ & 7.07 \\
  10-4 & 10.4 &  $7.06 \cdot 10^{-3}$ & 330 & $4 \cdot 10^{-2}$ & 3.16 \\
  10-0.4 & 10.4 &  $7.06 \cdot 10^{-3}$ & 330 & $4 \cdot 10^{-3}$ & 1.00 \\
  \hline
  5.2-20 & 5.2 & $2.82 \cdot 10^{-2}$ & 659 & $2 \cdot 10^{-1}$ & 7.07 \\
  5.2-4 & 5.2 &  $2.82 \cdot 10^{-2}$ & 659 & $4 \cdot 10^{-2}$ & 3.16 \\
  5.2-0.4 & 5.2 &  $2.82 \cdot 10^{-2}$ & 659 & $4 \cdot 10^{-3}$ & 1.00  \\
  \hline
  2.6-20 & 2.6 & $1.13 \cdot 10^{-1}$ & 1\,318  & $2 \cdot 10^{-1}$ & 7.07 \\
  2.6-4 & 2.6 &  $1.13 \cdot 10^{-1}$ & 1\,318 & $4 \cdot 10^{-2}$ & 3.16 \\
  \hline
  inf-0 & $\infty$ & 0 & 0 & 0 & 0 \\
  \hline
 \end{tabular}
\end{table}

The considered values of rotational energies coincide reasonably well with values from observations, ranging from $10^{-4}$ up to 1.4 with a mean around 0.01~\citep{Goodman93,Pirogov03}.
Furthermore, observations of high-mass star forming regions typically reveal mass-to-flux ratios $\mu \la$ 5 and magnetic field strengths between a few 100 $\mu$G and a few mG~\citep{Lai01,Curran07,Falgarone08,Girart09,Beuther10}. Here, we explore this range with a number of simulations ($\mu \le 5.2$) but also consider initial configurations with significantly weaker magnetic fields ($\mu > 10$). We recognize that turbulent motions are significant for massive clumps~\citep[e.g.][]{Caselli95}. However, since the focus of this work is to investigate the formation of discs in magnetised environments, we will delay the inclusion of turbulence to a subsequent paper.

\section{Results} \label{sec:results}

\begin{table}
 \caption{Formation time t$_0$ of the first sink particle, total mass (the mass in brackets is the mass of the most massive sink if more than one is formed) of all sink particles after 4000 yr (3000 yr for the runs 2.6-20 and 2.6-4), the corresponding time averaged total accretion rate and the number of sinks created.}
 \label{tab:msinks}
 \begin{tabular}{@{}lcccc}
  \hline
  Run & t$_0$  & M$_{\rmn{sink}}$  & $\dot{\rmn{M}}_{\rmn{acc}}$   & N$_{\rmn{sinks}}$ \\
      &      (kyr)       &     (M$_{\sun}$)     &   (10$^{-4}$ M$_{\sun}$/yr) & \\
  \hline
  26-20 & 15.7   & 1.26 (1.03) & 3.14 & 13 \\
  26-4 & 15.2    & 2.08 (1.79) & 5.19 & 10 \\
  26-0.4 & 15.0  & 2.93 & 7.32 & 1 \\
  26-0.04 & 15.0 & 3.39 & 8.47 & 1 \\
  \hline
  10-20 & 15.8   & 1.28 & 3.19 & 1 \\
  10-4 & 15.3    & 2.23 & 5.57 & 1 \\
  10-0.4 & 15.2  & 2.98 & 7.46 & 1 \\
  \hline
  5.2-20 & 16.7  & 1.78 & 4.45 & 1 \\
  5.2-4 & 16.2   & 2.28 & 5.71 & 1 \\
  5.2-0.4 & 16.1 & 2.55 & 6.37 & 1 \\
  \hline
  2.6-20 & 21.3  & 1.30 & 4.33 & 1 \\
  2.6-4 & 20.5   & 1.48 & 4.93 & 1 \\
  \hline
  inf-0 & 15.1 & 3.55 & 8.86 & 1 \\
  \hline
 \end{tabular}
\end{table}
In this section, we present the results of our collapse simulations where we in particular focus on the evolution of the discs and the accretion onto the protostars. The evolution and properties of the outflows launched from the protostellar discs will be discussed in a subsequent paper. In the following, we mainly limit our consideration to the phase after the first sink particle has been formed and only shortly describe the initial collapse phase. After the formation of the first sink particle the simulations run for further 4000 yr (except the runs 2.6-20 and 2.6-4 which are run for 3000 yr only) to study the time evolution of the outflow, the formation of a protostellar disc and of potential secondary sink particles as well as the accretion history of the sink particles. The computational cost of all simulations presented here sum up to about 1\,000\,000 CPU-h.

The typical timescale for the collapse of the considered cloud core is its free-fall time
\begin{equation}
 \tau_{\rmn{ff}} = \sqrt{\frac{3 \pi}{32 G \rho}} = 13.9\, \rmn{kyr}
\end{equation}
where we have used the central density of $\rho = 2.3 \cdot 10^{-17}$ g cm$^{-3}$. In Table~\ref{tab:msinks} we list the formation time t$_0$ of the first sink particle for all runs performed. As can be seen, t$_0$ is longer than $\tau_{\rmn{ff}}$ by a factor of 1.1 to 1.4. 
As run inf-0 has a prolonged collapse time compared to $\tau_{\rmn{ff}}$ as well, we performed a careful numerical test showing that t$_0$ can be reduced by a few 100 yr using a higher initial resolution. However, the main reason for the delay of the collapse is probably not a numerical issue but rather the build-up of a strong pressure gradient in the centre of the core counteracting gravity. In contrast, the unexpected fact that t$_0$ for run inf-0 is somewhat longer than for the runs 26-0.4 and 26-0.04 is most like a numerical issue. Turning on weak magnetic field in a test run with no rotation results in a slightly shortened collapse time compared to inf-0. This suggest that the somewhat longer t$_0$ in run inf-0 is an intrinsic effect of the numerical scheme we will not follow up further here. However, in each simulation subset (equal $\mu$ or $\beta_{\rmn{rot}}$) there are physically well motivated trends in t$_0$ recognisable. As can be inferred from Table~\ref{tab:msinks}, the collapse gets slowed down with increasing magnetic field strength and increasing amount of rotational energy, both counteracting gravity. Interestingly, the variation of the rotational energy by two orders of magnitude at a fixed magnetic field strength changes t$_0$ by no more than roughly 0.8 kyr, i.e. by $\sim$ 5\%. In contrast, increasing the magnetic energy by a factor of 100 for fixed rotational energies prolongates the collapse time by roughly 5 kyr. However, as all cores are supercritical ($\mu > 1$) and have rotational energies well below the gravitational energy (see Figure~\ref{fig:models}), one cannot expect the collapse time t$_0$ to be significantly longer than the free-fall time. In the following, all times refer to the time elapsed since the formation of the first sink particle at t$_0$.

\subsection{Global disc properties} \label{sec:time}

As a first step, we analyse the time evolution of two representative simulations after the formation of the first sink particle. For this purpose, we consider run 26-4 and run 5.2-4 with equal initial rotational energies but different magnetic field strengths.
To begin with, we compare the properties of the gas in a well defined region of interest. This region is described by a cylinder around the simulation centre with a height of 47 AU above and below the midplane. We mention that, although this height does not reflect the real disc scale height, we use it for three reasons. Firstly, this choice provides a numerically simple calculation of the desired quantities. Furthermore, it guarantees that for different simulations and times identical areas are compared. Finally, it is used as we cannot determine a well defined disc scale height as even in a single run the disc height varies in time and radial position. For example, the frequently used approximation of the scale height~\citep[e.g.][]{Cesaroni07}
\begin{equation}
 H = \frac{c_{\rmn{s}}}{\omega}
\end{equation}
varies between a few AU and $\sim$ 100 AU. This is in accordance with other estimates of the disc height made the authors. Hence we argue that our choice of 47 AU, located in the middle of this range, is reasonable.

In the following we will denote this region simply as the disc. The quantities in the disc are averaged vertically and azimuthally before consideration.
\begin{figure*}
 \includegraphics[width=168mm]{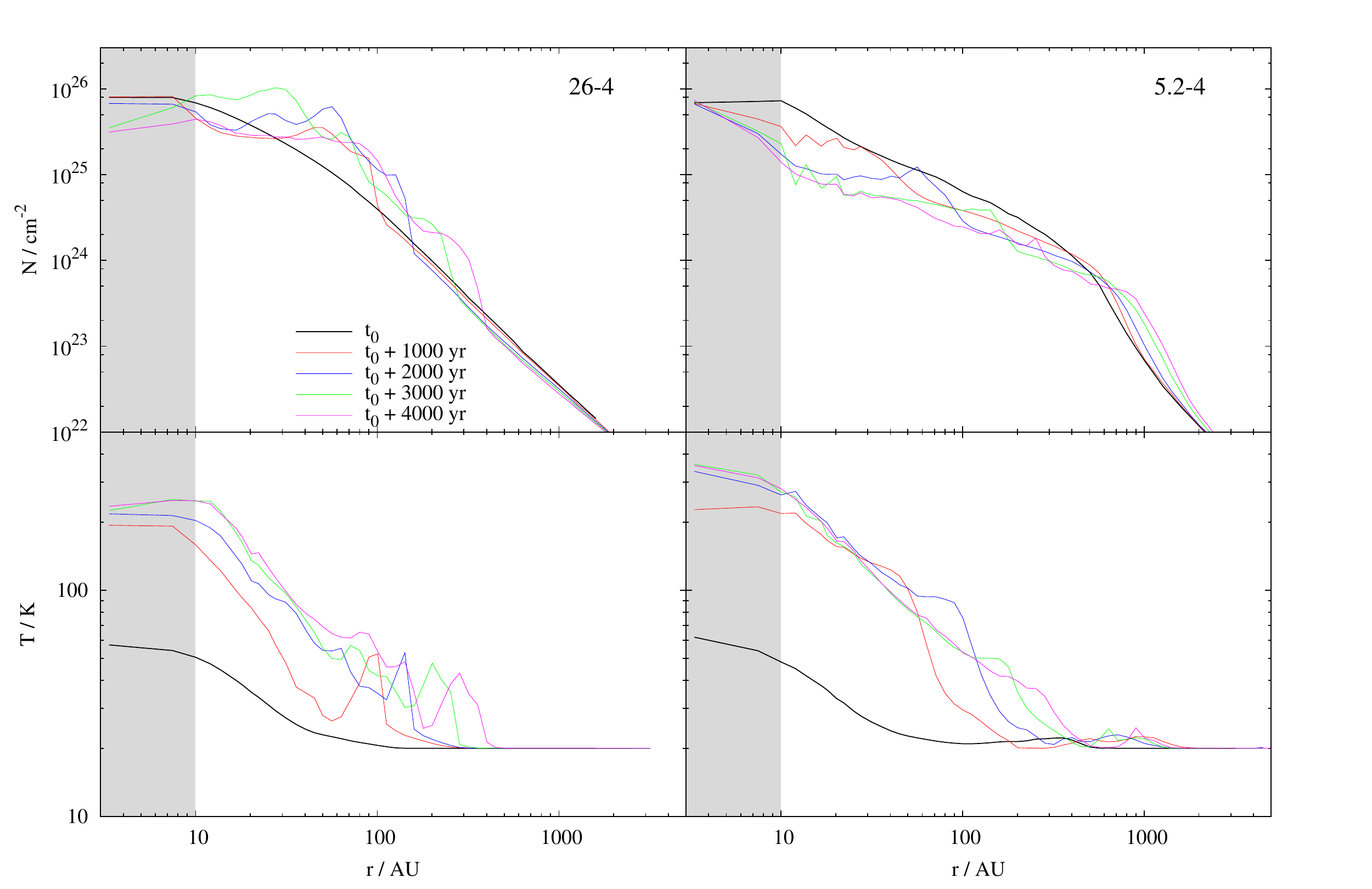}
 \caption{Radial profile of the column density (upper panel) and mass weighted temperature (bottom panel) for run 26-4 (left panel) and run 5.2-4 (right panel) at five different times after the formation of first sink particle at time t$_0$. The profiles are calculated by averaging azimuthally and vertically over a disc with a height of 47 AU above and below the midplane. The region below 10 AU is subject to resolution effects. Therefore, here and in the following plots we shaded this area to guide the readers eye.}
 \label{fig:disk_evol}
\end{figure*}
From the results shown in Fig.~\ref{fig:disk_evol}, it can be inferred that in run 26-4 an accretion shock occurs where both, column density and temperature experience a sudden increase. This is also true for run 5.2-4, although here the jumps in density and temperature is somewhat smoother. In both cases, however, the shock front moves outwards as time evolves. In run 26-4 the density profile inside the accretion shock is nearly flat with values around a few 10$^{25}$ cm$^{-2}$ corresponding to densities of $\ga$ 10$^{-13}$ g cm$^{-3}$. For run 5.2-4 the density profile is declining outwards and seems to decrease slightly over time. In both cases the mass-weighted temperature in the midplane increases by about one order of magnitude up to a few 100 K.

The temperature increase in the inner region is caused by the gas getting optically thick at density around 10$^{-13}$ g cm$^{-3}$~\citep[see][for details of the applied cooling function]{Banerjee06a}. Therefore, the gas in the disc looses its ability to cool efficiently and cannot fast enough radiate away the thermal energy transfered to it by compression work. This results in temperatures of up to a few 100 K in the inner disc region consistent with observational results~\citep[see e.g. the review of][and references therein]{Cesaroni07}. Further out in the disc the gas experiences a strong temperature increase due to shock heating. At the accretion shock kinetic energy is converted quite fast into thermal energy. Assuming that gas gets decelerated by $\sim$ 1 km s$^{-1}$ (see Fig.~\ref{fig:disk_evol}) and the corresponding kinetic energy is immediately transformed into thermal energy, this would result in a temperature increase of roughly 90 K. This is in accordance with the observed increase of about 30 K regarding to the fact that a good fraction of the energy is radiated away and that the conversion into heat does not happen all at once.

\begin{figure*}
 \includegraphics[width=168mm]{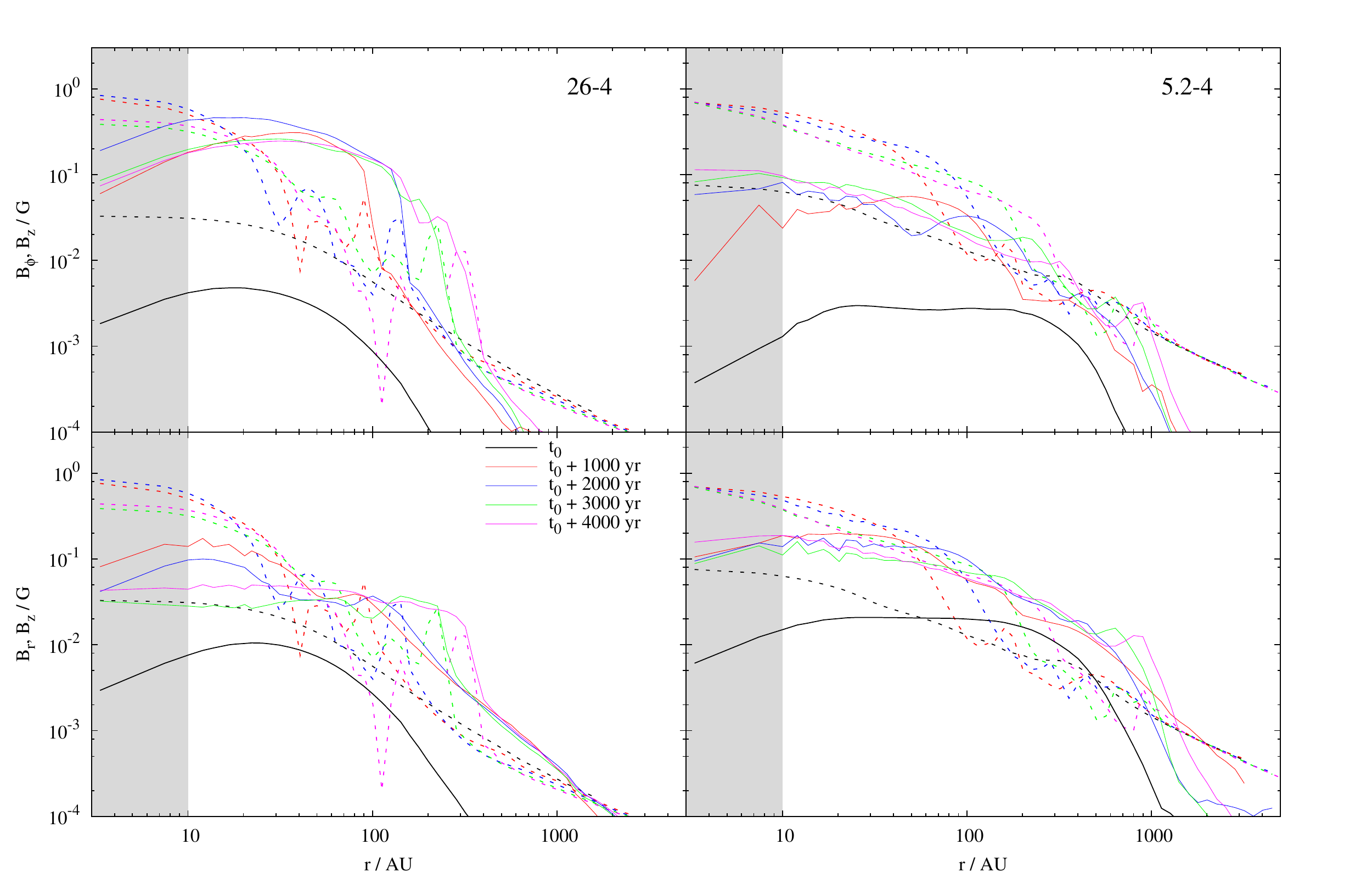}
 \caption{Radial profile of the toroidal component (top) and the radial component (bottom) of the magnetic field for run 26-4 (left) and run 5.2-4 (right) at the same times as in Fig.~\ref{fig:disk_evol}. For comparative purposes $B_z$ (dashed lines) is shown in both panels as well. In both runs $B_{\rmn{r}}$ is comparable in magnitude to $B_z$ whereas $B_{\phi}$ is larger than $B_z$ for run 26-4 and smaller than or comparable to $B_z$ for run 5.2-4.}
 \label{fig:mag_evol}
\end{figure*}
\begin{figure*}
 \includegraphics[width=168mm]{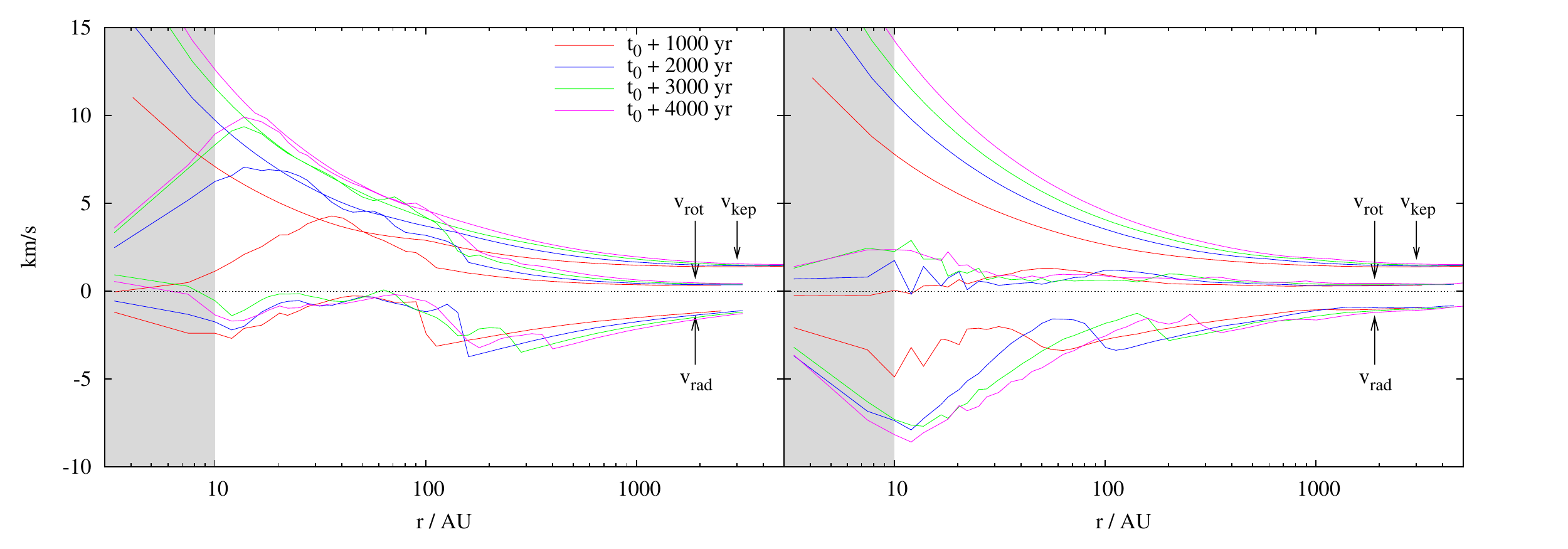}
 \caption{Radial profile of the Keplerian velocity, rotation velocity and radial velocity (negative values) for run 26-4 (left) and run 5.2-4 (right). The velocities are averaged in the same way as in Fig.~\ref{fig:disk_evol} and shown for same snapshots except that of t = t$_0$. For run 26-4 a rotationally supported structure evolves while in run 5.2-4 the rotation is clearly sub-Keplerian with radial infall close to free-fall.}
 \label{fig:vel_evol}
\end{figure*}

In Fig.~\ref{fig:mag_evol} we show the magnetic properties of the discs. For the calculation of $B_{\rmn{r}}$ and $B_{\phi}$ the absolute values are used for averaging as both components have opposite signs above and below the disc and hence would cancel themselves out. For comparative purposes we plot $B_z$ (initially the only component) and either $B_{\phi}$ or $B_{\rmn{r}}$ in each panel. For the comparison we only consider the radial range above 10 AU as the region further inside is subject to resolution effects. As can be seen, the different components in the midplane reach values up to a about 1 G. $B_z$ is somewhat larger in run 5.2-4 than in run 26-4 due to the five times higher, initial field strength in run 5.2-4. In both runs, however, $B_{\rmn{r}}$ (bottom panel of Fig.~\ref{fig:mag_evol}) is of the order of $B_z$. $B_{\rmn{r}}$ is created by the inwards drag of the magnetic field during the collapse and later during the accretion process. In contrast to $B_{\rmn{r}}$, the toroidal components of both runs differ significantly (see top panel of Fig.~\ref{fig:mag_evol}). In run 26-4 $B_\phi$ is the dominant component in the region within the accretion shock being larger than $B_z$ and $B_{\rmn{r}}$ by up to one order of magnitude. This is due to the fast rotation of the disc winding up the poloidal field (see Fig.~\ref{fig:vel_evol} and text below). In contrast, in the strongly magnetised run 5.2-4, $B_\phi$ is mostly smaller than $B_z$ which is attributed to the lower rotation velocity in this case (see Fig.~\ref{fig:vel_evol}). We also mention that all components show signs of an accretion shock moving outwards with time in accordance with the density and temperature field shown in Fig.~\ref{fig:disk_evol}. This is caused by the tight coupling of magnetic fields and matter due to ideal MHD. Furthermore, a comparison with the thermal pressure shows that in both runs the magnetic pressure in the region within the accretion shock is larger than or at least equal to the thermal pressure. This implies that the magnetic field must play a significant role in the evolution of the circumstellar disc and in the accretion history of the protostars.

In Fig.~\ref{fig:vel_evol} the time evolution of the rotation ($v_{\rmn{rot}}$), radial ($v_{\rmn{rad}}$) and Keplerian velocity ($v_{\rmn{kep}}$) is shown for both runs.
For run 26-4 (left panel) a rotationally supported disc builds up with rotation velocities close to the Keplerian velocity. The maximum radius, where $v_{\rmn{rot}}$ equals $v_{\rmn{kep}}$ increases over time consistent with the build-up of a Keplerian disc. At the same time the radial velocity in the inner region nearly drops to zero. We emphasise that considering the behaviour within the central 10 AU is not conclusive as we reach the resolution limit at such small radii. The overall situation changes dramatically when considering run 5.2-4 (right panel). Here, no rotationally supported disc builds up with $v_{\rmn{rot}}$ staying significantly below $v_{\rmn{kep}}$ all the time. In fact, the absolute value of $v_{\rmn{rad}}$ is nearly as high as $v_{\rmn{kep}}$ denoting gas which is almost in free-fall. Nevertheless, a flat disc-like density structure builds up.

\begin{figure*}
 \includegraphics[width=56mm]{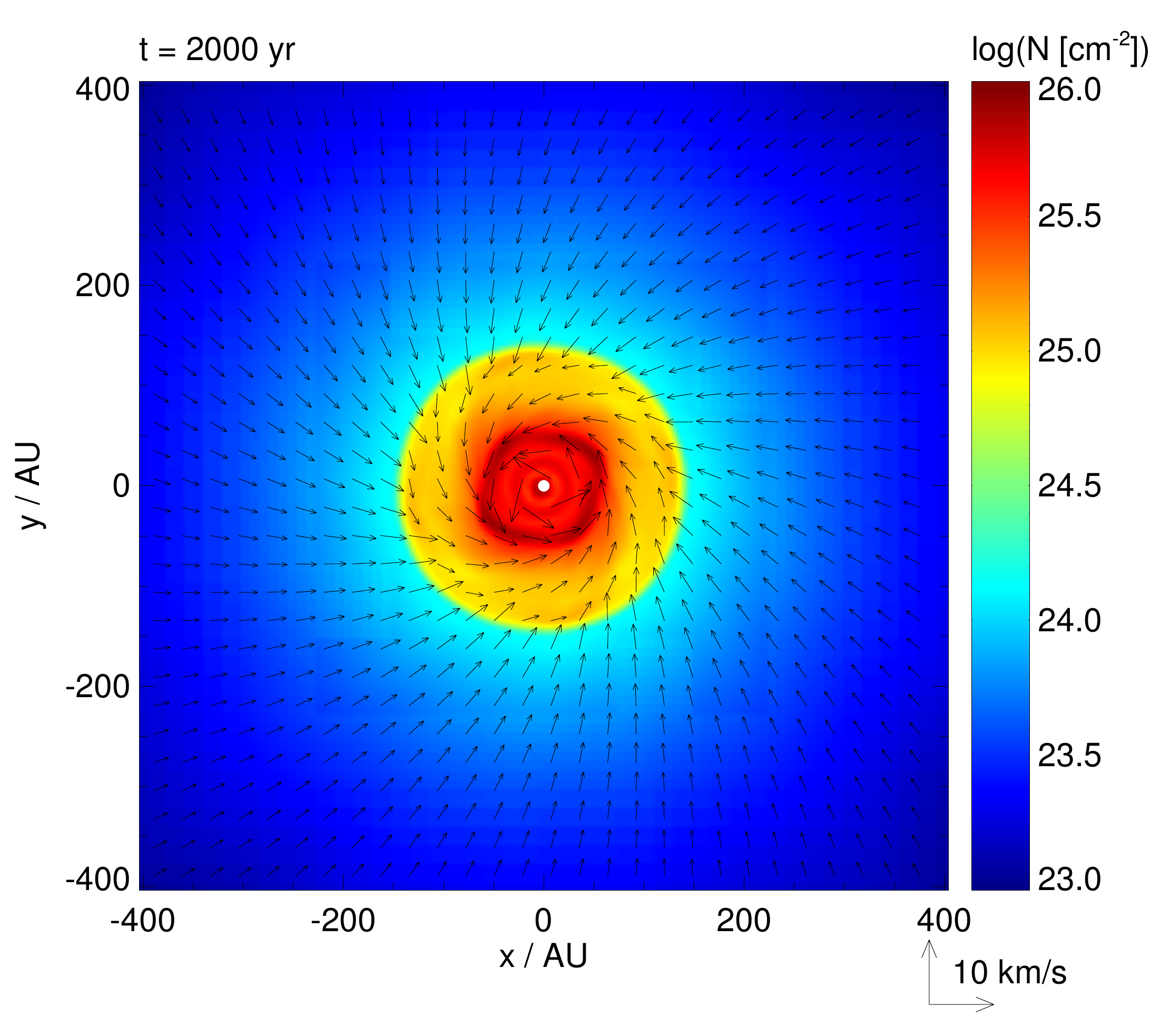}
 \includegraphics[width=56mm]{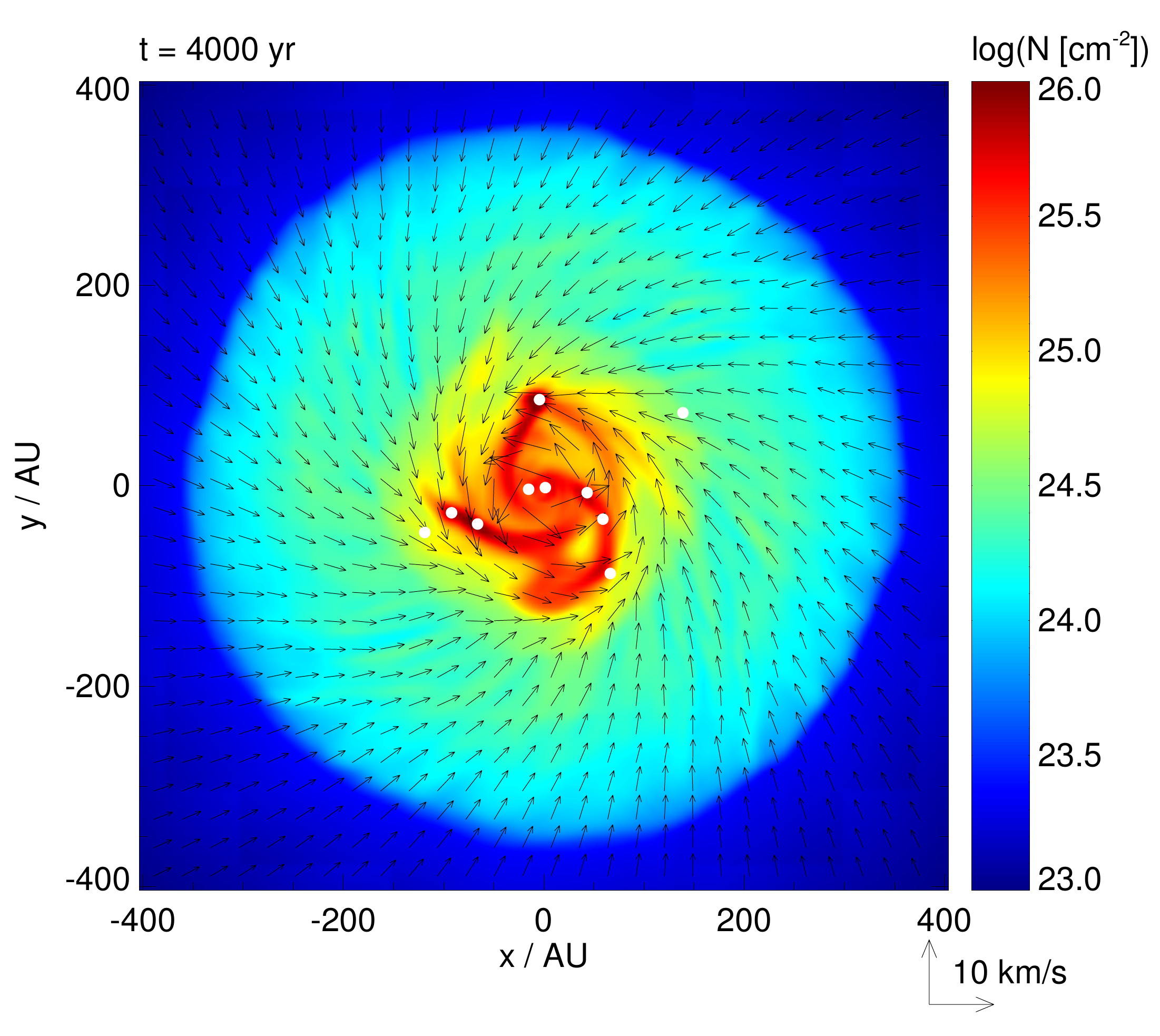}
 \includegraphics[width=56mm]{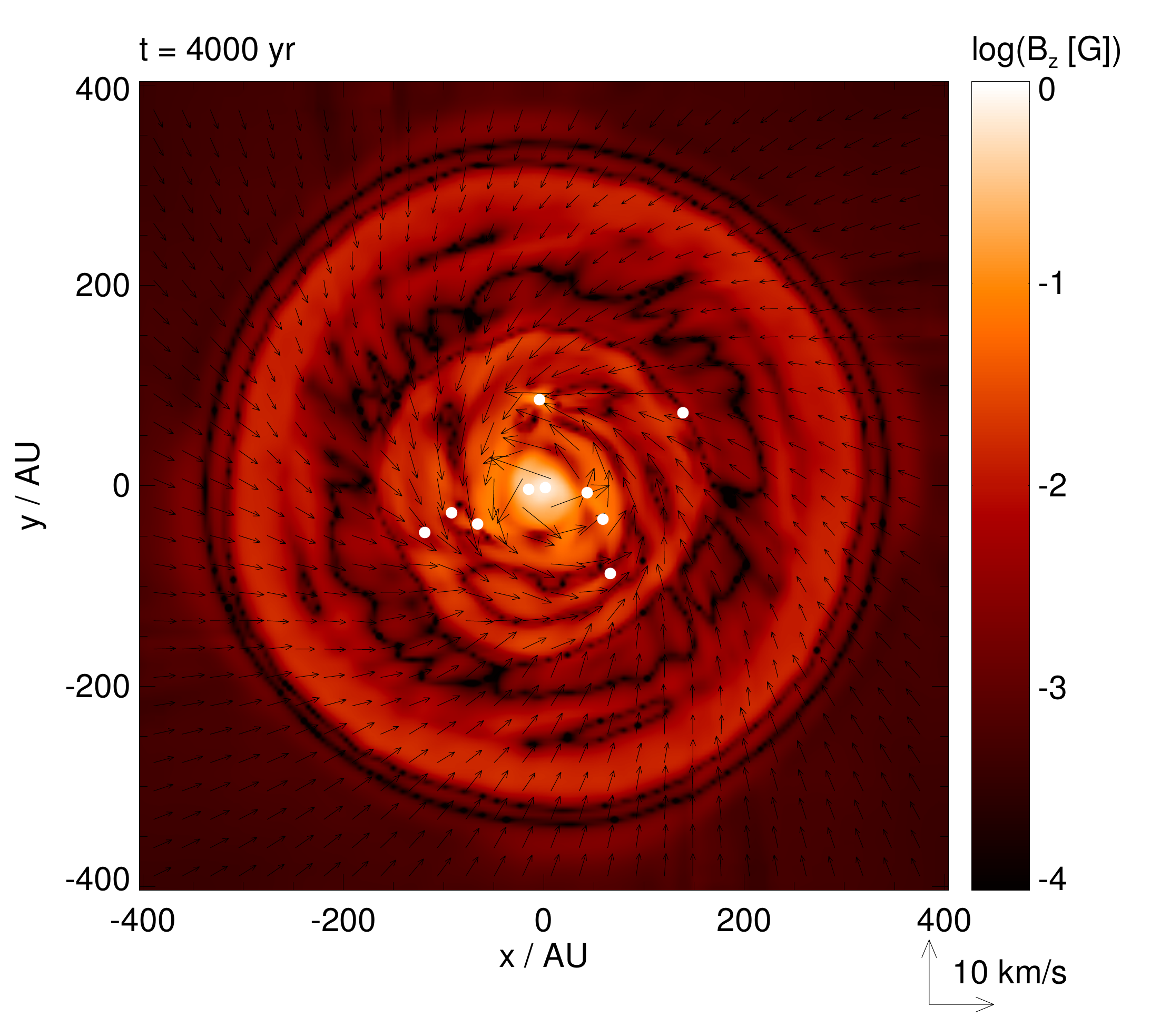} \\
 \includegraphics[width=56mm]{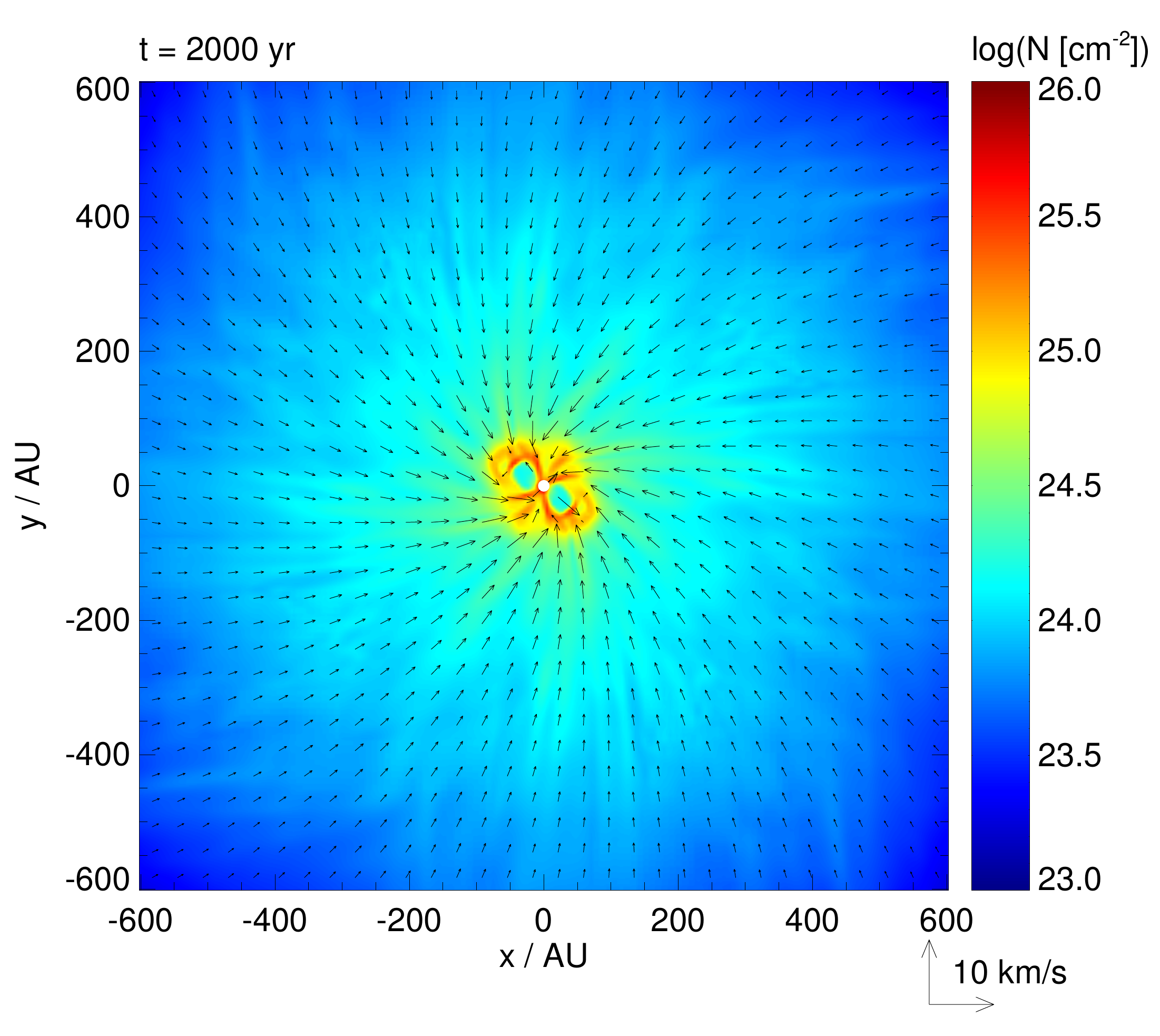}
 \includegraphics[width=56mm]{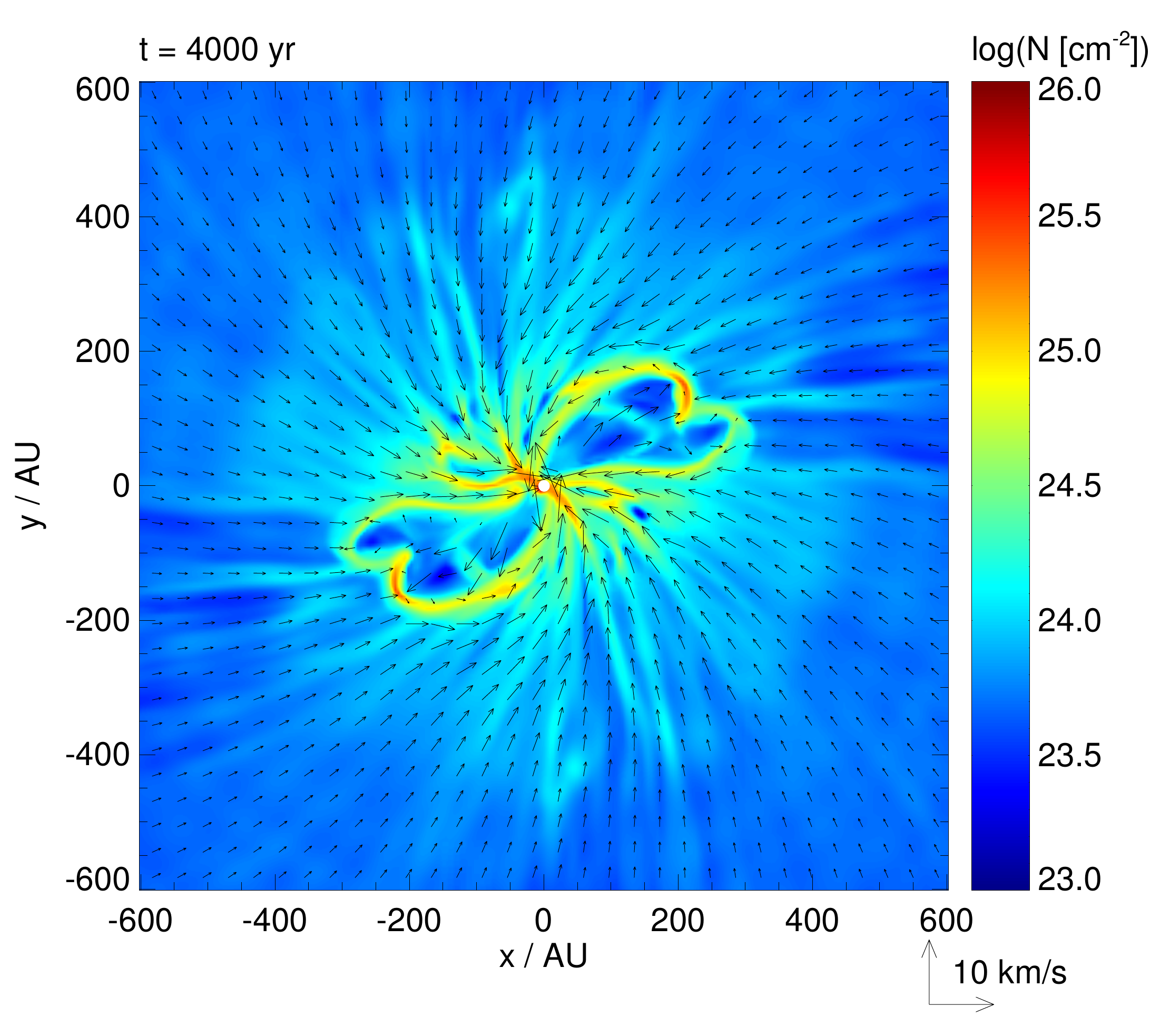}
 \includegraphics[width=56mm]{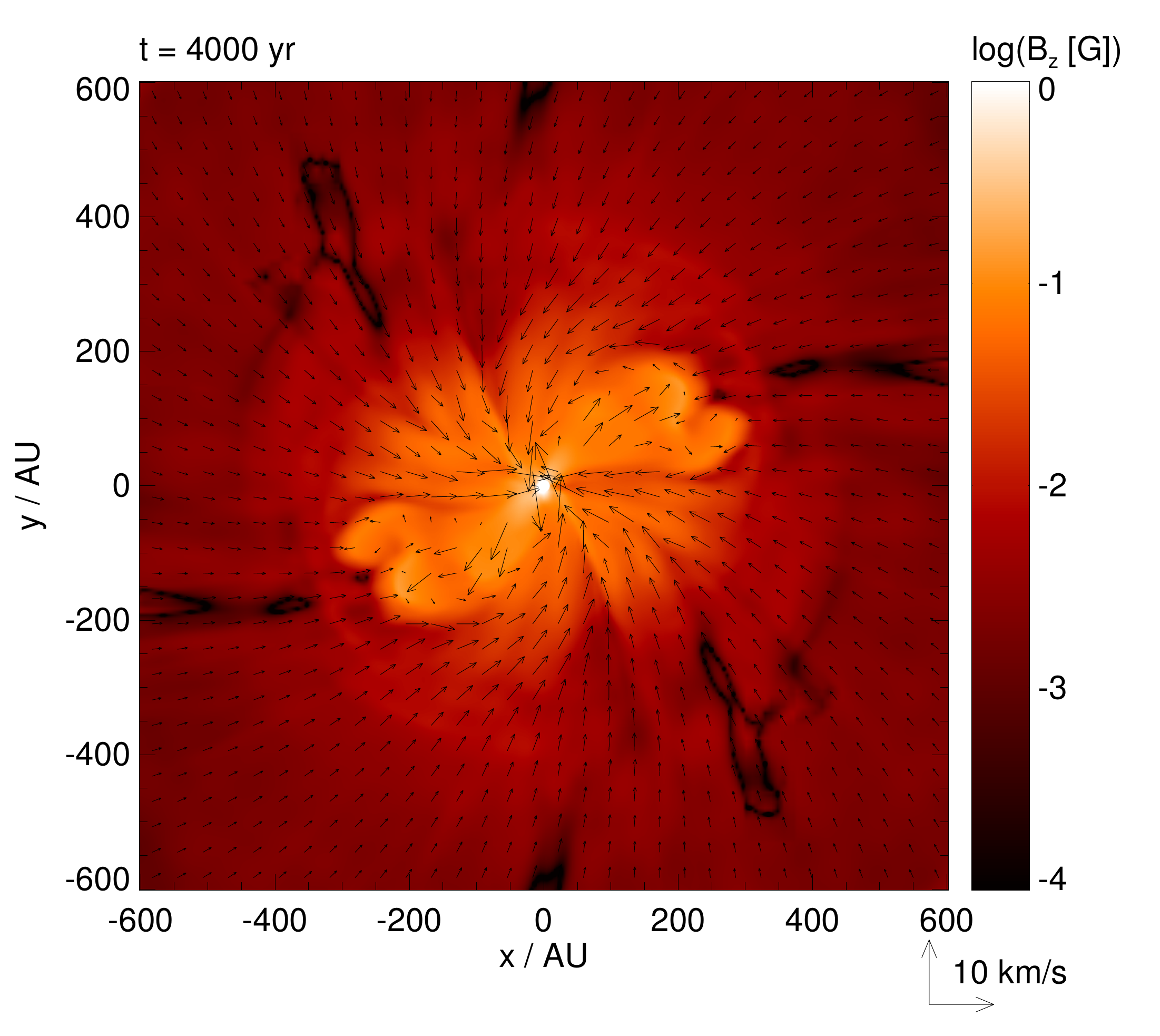} 
 \caption{Column density of the disc after 2000 yr (left) and 4000 yr (middle) for run 26-4 (top) and run 5.2-4 (bottom). White dots mark the projected positions of the sink particles, black vectors the velocity field in the midplane. In run 26-4 the disc is well defined with its inner region being subject to fragmentation. In contrast, in run 5.2-4 a disc with sub-Keplerian motions forms. The bubble like structures in this run are caused by the high magnetic pressure in the midplane driving material outwards (see also Section~\ref{sec:numerics}). Right: z-component of the magnetic field in the midplane for run 26-4 (top) and run 5.2-4 (bottom) at the end of the simulation, i.e. after 4000 yr. The tight correlation between field strength and matter is due to the condition of ideal MHD. Note the different spatial scales in the upper and lower panel.}
 \label{fig:disk}
\end{figure*}

\begin{figure}
 \includegraphics[width=84mm]{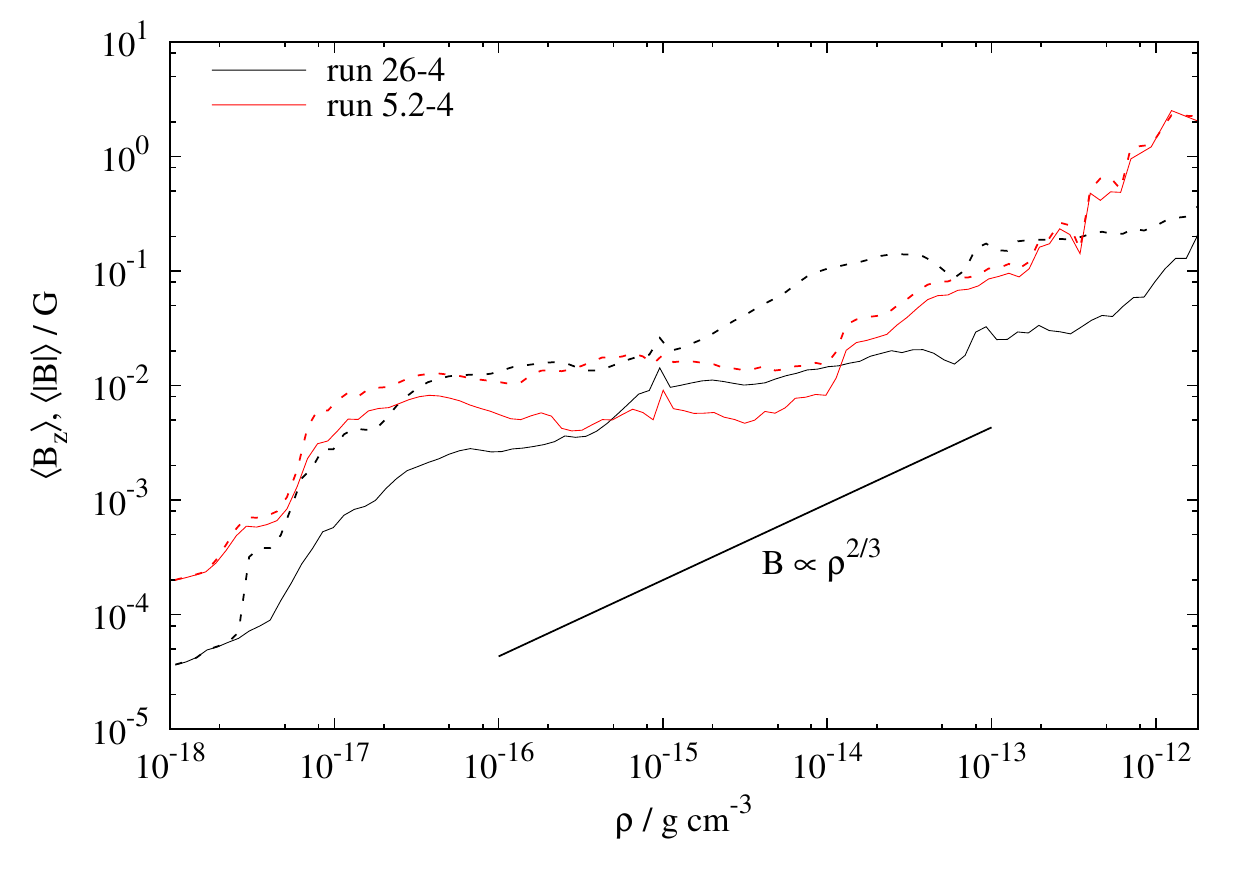}
 \caption{Average of $B_z$ (solid lines) and of the total magnetic field strength (dashed lines) over density for run 26-4 (black) and run 5.2-4 (red) after 4000 yr. The average is calculated in density bins of equal size in log-space. The magnetic field is strongly coupled to the gas and scales roughly as B $\sim \rho^{2/3}$ over more than 6 orders of magnitude in density.}
 \label{fig:mag_vs_dens}
\end{figure}

The dramatically different evolution of the gas in the midplane of both runs is also indicated in Fig.~\ref{fig:disk}. Here we show the integrated column density, the magnetic field strength and the velocity field along a slice in the midplane.
For run 26-4 (upper panel of Fig.~\ref{fig:disk}) a well defined Keplerian disc with more or less sharp boundaries develops. The velocity field within the disc shows a significant rotational component as already seen in Fig.~\ref{fig:vel_evol}. Despite the strong fragmentation occurring in the inner 200 AU, the overall disc-like structure is maintained. In contrast, in run 5.2-4 (lower panel of Fig.~\ref{fig:disk}) there is no evidence for the development of a Keplerian disc and the column density increases more or less smoothly towards the centre with nearly radial infall. The right panel of Fig.~\ref{fig:disk} shows the z-component of the magnetic field in the midplane. As can be seen, a field strength of up to 1 G is reached with slightly higher values for run 5.2-4. The close coupling of magnetic fields and matter due to ideal MHD is especially pronounced in the top panel of Fig.~\ref{fig:disk} where the strong increase in the column density at r $\sim$ 350 AU coincides quite well with a jump in $B_z$. 
In Fig.~\ref{fig:mag_vs_dens} the coupling of gas and magnetic field over the whole density range is demonstrated even more clearly. Apart from local variations, the average of $B_z$ and $\left | \rmn{B} \right |$, calculated in density bins of equal size in log-space, scales roughly as $\rho^{2/3}$ or slightly weaker over more than 6 order of magnitude in density. This is in accordance with the scaling of a magnetic field in case of a spherical collapse under the condition of ideal MHD.

The bubble like features seen in run 5.2-4 can be explained as follows: during the accretion process the magnetic field is dragged inwards with the gas. As we consider ideal MHD, the field lines cannot diffuse outwards leading to a constant increase in the field strength in the centre. Beyond a certain point in time, the magnetic pressure is strong enough to effectively counteract gravity and starts to push material outwards. Indeed, analysing the $B_z$ component of the magnetic field in the bottom right panel of Fig.~\ref{fig:disk} shows that the outwards moving gas is associated with a strong $B_z$.
This causes $B_z$ to decrease slightly in the very centre whereas at larger radii the magnetic field strength increases over time (see right panel of Fig.~\ref{fig:mag_evol}).

The two simulations presented here show significant differences although the initial magnetic field strength is varied by a factor of 5 only. These systematic differences also show up in a more or less pronounced way when comparing all simulations with each other. We will discuss these differences in detail in the next section where the complete set of simulations with varying initial conditions is considered.

\subsection{Velocity structure} \label{sec:vel}

\begin{figure*}
 \includegraphics[width=168mm]{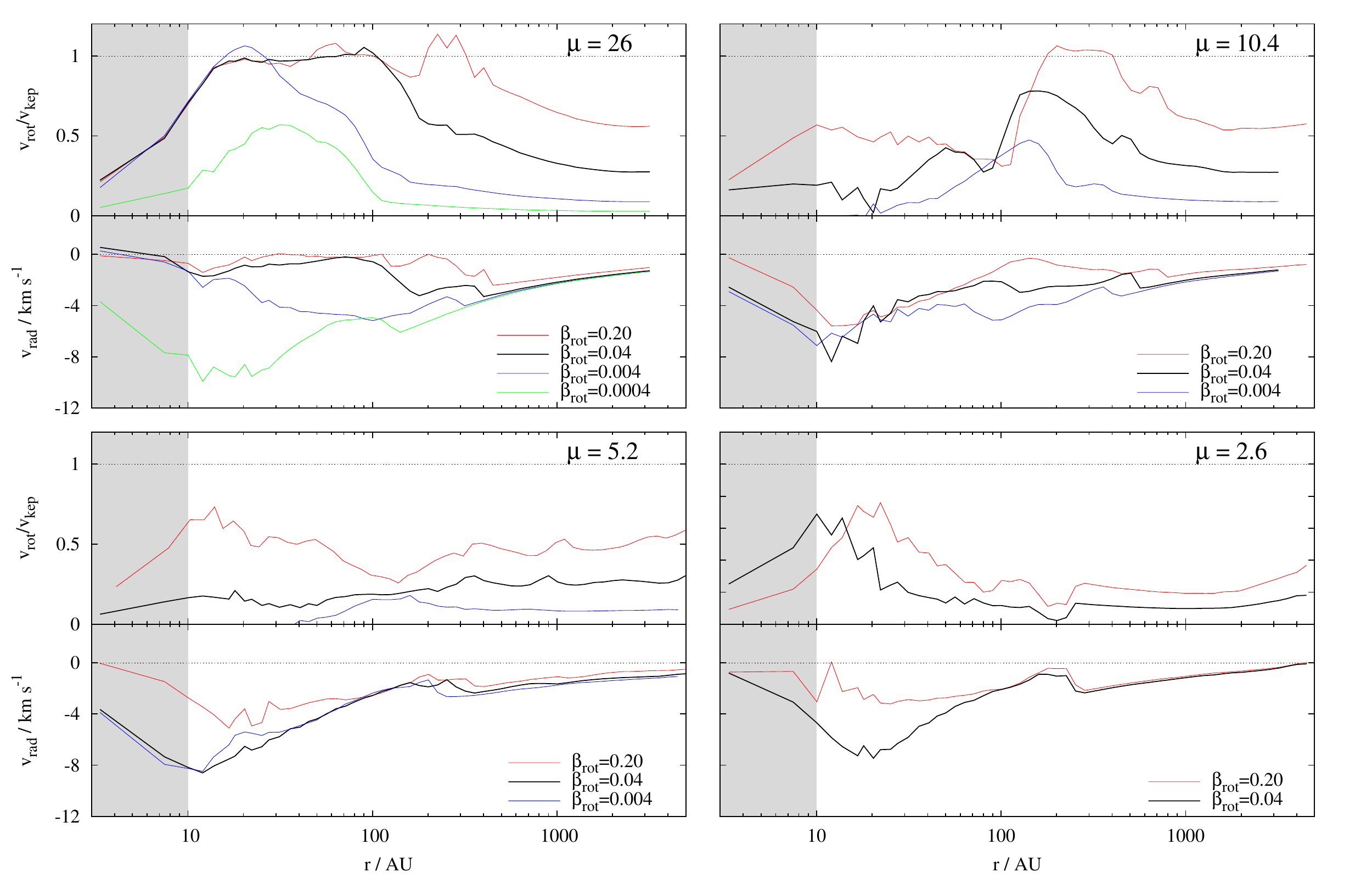}
 \caption{Radial profile of the velocity field for the simulations with $\mu$ = 26 (top left), 10.4 (top right), 5.2 (bottom left) and 2.6 (bottom right) after 4000 yr (3000 yr for $\mu$ = 2.6). For better comparison between the individual simulations the rotation velocity $v_{\rmn{rot}}$ is normalised to the Keplerian velocity $v_{\rmn{kep}}$. The plotted quantities are again averaged azimuthally and vertically in a disc with a height of 47 AU. For higher $\beta_{\rmn{rot}}$ the ratio $v_{\rmn{rot}}$/$v_{\rmn{kep}}$ is closer to unity necessary for a Keplerian disc to form. Only in case of weak magnetic fields (top row) the rotation velocity reaches the Keplerian velocity while it is significantly below $v_{\rmn{kep}}$ for strong magnetic fields. As expected, the absolute value of $v_{\rmn{rad}}$ increases with decreasing $v_{\rmn{rot}}$/$v_{\rmn{kep}}$ due to a lower centrifugal support against gravity.}
 \label{fig:vel}
\end{figure*}

Next, we focus on the effect of the initial conditions on the velocity field in the midplane around the central sink particle. We omit the time evolution as the behaviour of each individual run is qualitatively similar to one of the two runs shown before and concentrate on the situation after 4000 yr. As already shown in Section \ref{sec:time}, even small changes in the initial configuration of the core cause characteristical differences. The effects of rotational and magnetic energy on the velocity structure at the end of the simulations can be seen in Fig.~\ref{fig:vel}.
Beside the radial velocity, the ratio $v_{\rmn{rot}}$/$v_{\rmn{kep}}$ is shown as well. Decreasing the initial amount of rotational energy for runs with fixed magnetic field strength (see individual panels of Fig.~\ref{fig:vel}) reduces the centrifugal support against gravity resulting in lower values of $v_{\rmn{rot}}$/$v_{\rmn{kep}}$ and consistently in higher infall velocities. Additionally, the differences for runs with varying $\beta_{\rmn{rot}}$ but fixed $\mu$ seem to decrease when the initial field strength is increased. Furthermore, comparing runs with fixed $\beta_{\rmn{rot}}$ but varying field strength shows that the centrifugal support, i.e. $v_{\rmn{rot}}$/$v_{\rmn{kep}}$ decreases with increasing field strength. This effect is pronounced strongest for high initial rotational energies.

\begin{figure}
 \includegraphics[width=84mm]{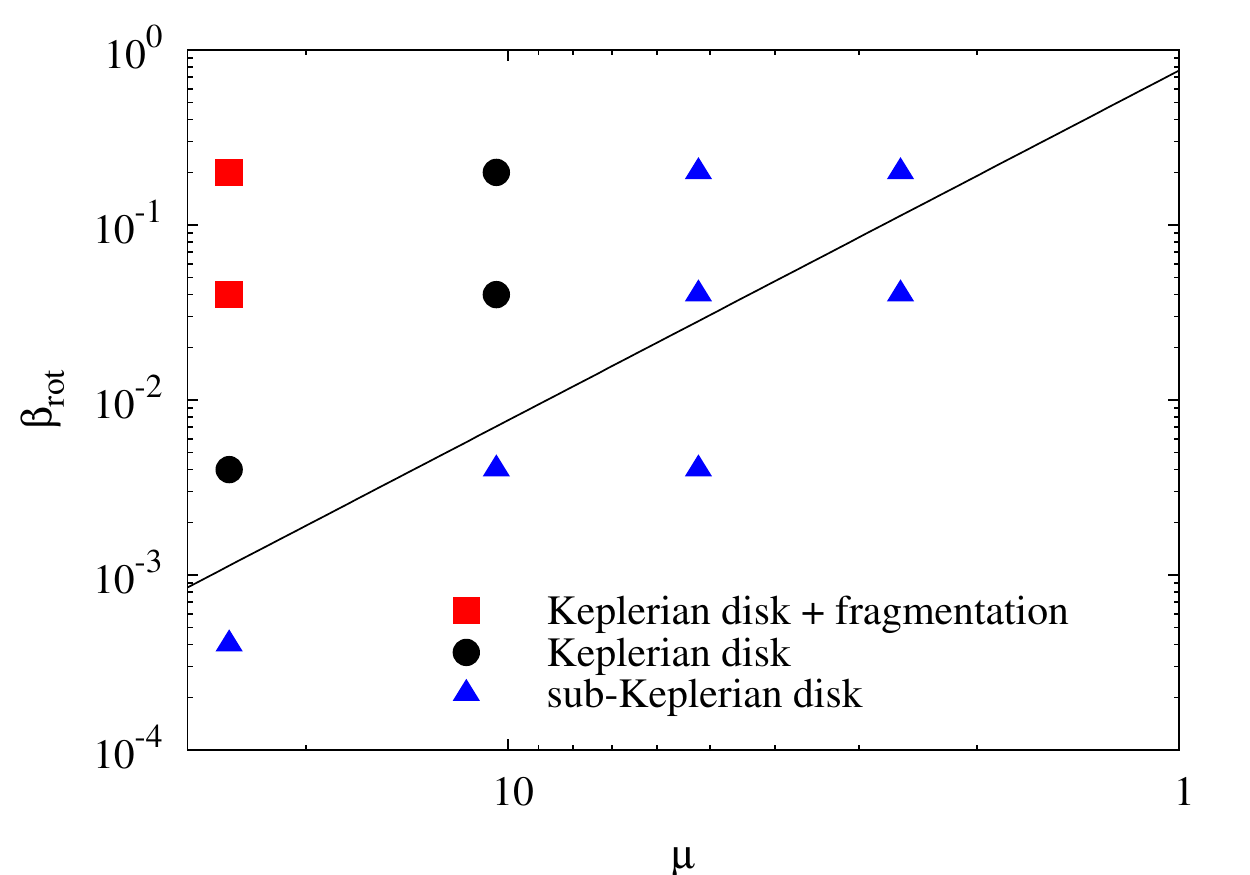}
 \caption{Phase diagram of magnetic field and rotational energy showing the results of the simulations concerning the question of disc formation. For $\mu < 10$ no centrifugally supported discs form (blue triangles) as well as for the slowly rotating cores in the runs 26-0.04 and 10-0.4. For weak magnetic fields Keplerian discs (black circles) form. Two of these discs are subject to fragmentation (red squares). The black solid line shows the curve where rotational and magnetic energy are equal.}
 \label{fig:diskoverview}
\end{figure}

In principle, the simulations can be divided into two sets depending on the initial mass-to-flux-ratio $\mu$. In general, for runs with $\mu > 10$ centrifugally supported discs develop whereas for $\mu < 10$ the formation of Keplerian discs is largely suppressed and only sub-Keplerian discs form at this early stage (see Section~\ref{sec:diskevol} for a detailed comparison with other numerical work). An overview over the observed dependency of disc formation and fragmentation on the initial conditions is given in Fig.~\ref{fig:diskoverview}.
We note that in the runs 26-0.04 and 10-0.4 the structure formed in the midplane strongly resembles that of centrifugally supported discs but with rotational velocities well below the Keplerian velocity. Thus, we do not denote them as Keplerian discs.

In the following, we qualitatively describe the track of a fluid particle moving along the midplane towards the centre. For all simulations considered in this work, the fluid particle first gets accelerated inwards until a radius of some 100 AU (depending on the specific simulation considered) is reached. At this radius the gas experiences a deceleration, meaning its infall motion slows down. This region can be identified as a so-called magnetic barrier~\citep{Mellon08} and is found in all simulations. This kind of barrier is different from a centrifugal barrier as here the rotation velocity is well below the Keplerian velocity which would be necessary to balance gravity. The reason for deceleration at such a magnetic barrier is twofold. Firstly, an outward directed thermal pressure gradient accounts for a part of the deceleration. This can be inferred from the significant increase in density and temperature at the corresponding radius (compare Fig.~\ref{fig:disk_evol}). Furthermore, the magnetic field itself slows down the infall motion via the combination of an outward directed magnetic pressure gradient and the effect of magnetic tension which can be inferred from Fig.~\ref{fig:mag_evol}. All components of the magnetic field experience a sudden increase in the region of deceleration, thus resulting in outward directed magnetic forces. Both magnetic pressure and magnetic tension increase in strength for smaller $\mu$ with the latter starting to dominate for low $\mu$. For cases with $\mu < 1$ not considered here, we expect the collapse of the core perpendicular to the field lines to be prevented completely by the Lorentz force~\citep{Mouschovias76}.

For radii within the magnetic barrier the velocity profiles start to differ significantly from each other. For weak magnetic fields (top panel of Fig.~\ref{fig:vel}) the gas infall speed stays roughly constant. Due to angular momentum conservation this results in an increased rotation velocity which in consequence leads to another slow down of the infall motion. For the highest rotational energies a centrifugal barrier is encountered bringing the infall completely to halt. As seen in the top panel of Fig.~\ref{fig:vel}, the occurrence and extension of the centrifugal barrier depend strongly on the initial setup. On the other hand, an increase in centrifugal support and hence a slowdown of infall gives magnetic braking more time to operate~\citep{Mouschovias80}. This is seen in the sudden drop of $v_{\rmn{rot}}$/$v_{\rmn{kep}}$ for runs with $\mu = 10.4$ causing a loss of centrifugal support and thus again speed-up of the infall motion. This is also the case for runs with low $\mu$ (bottom panel of Fig.~\ref{fig:diskoverview}) where magnetic braking starts to act efficiently directly after passing the magnetic barrier so that no centrifugal barrier occurs any more. The innermost drop of $v_{\rmn{rad}}$ (r $<$ 10 AU) without a corresponding increase in $v_{\rmn{rot}}$/$v_{\rmn{kep}}$ is certainly caused by the limiting effect of numerical resolution as this region is only marginally resolved by about 3 grid cells.

\subsection{Torques} \label{sec:tor}

\begin{figure*}
 \includegraphics[width=72mm]{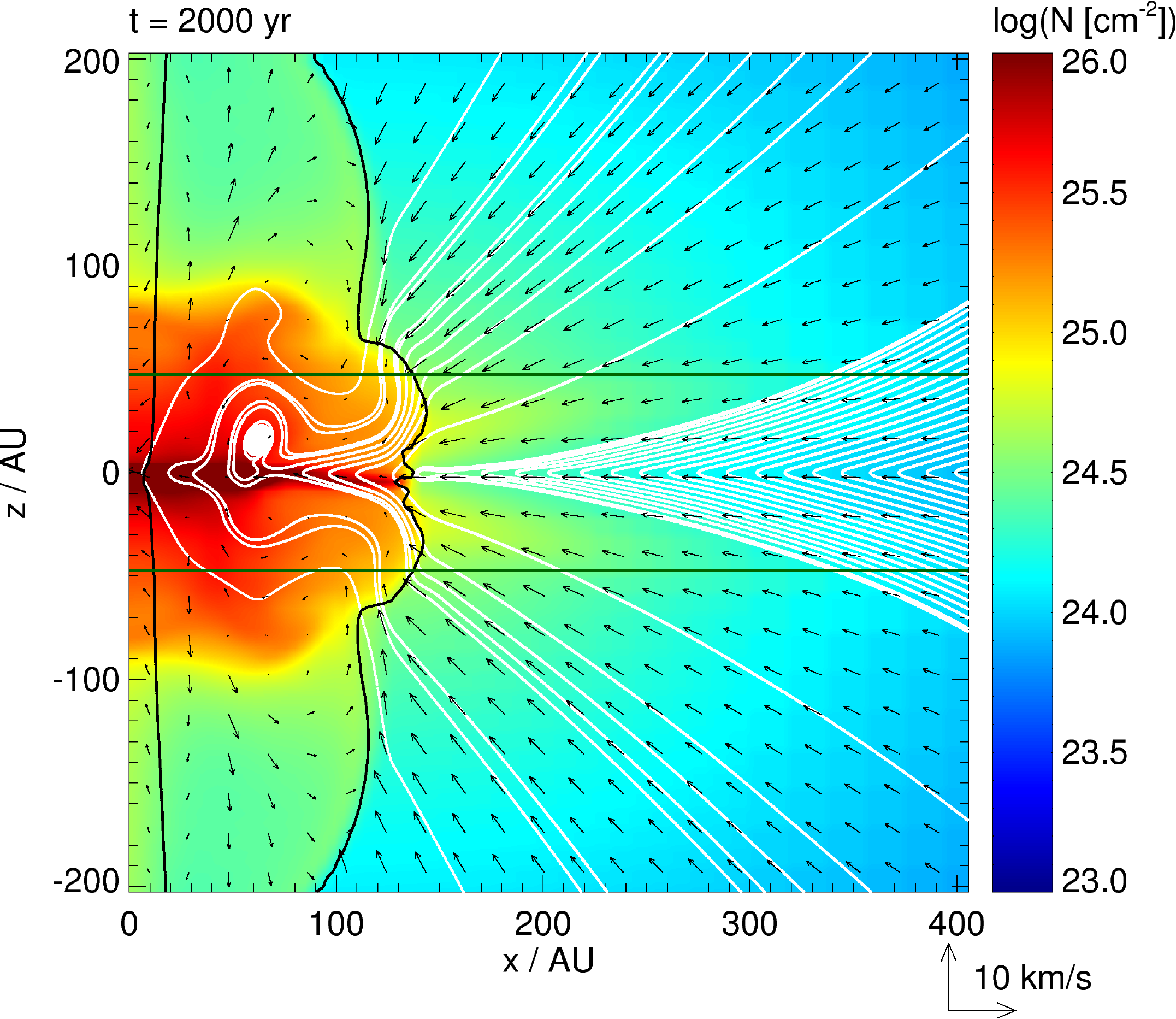}
 \includegraphics[width=72mm]{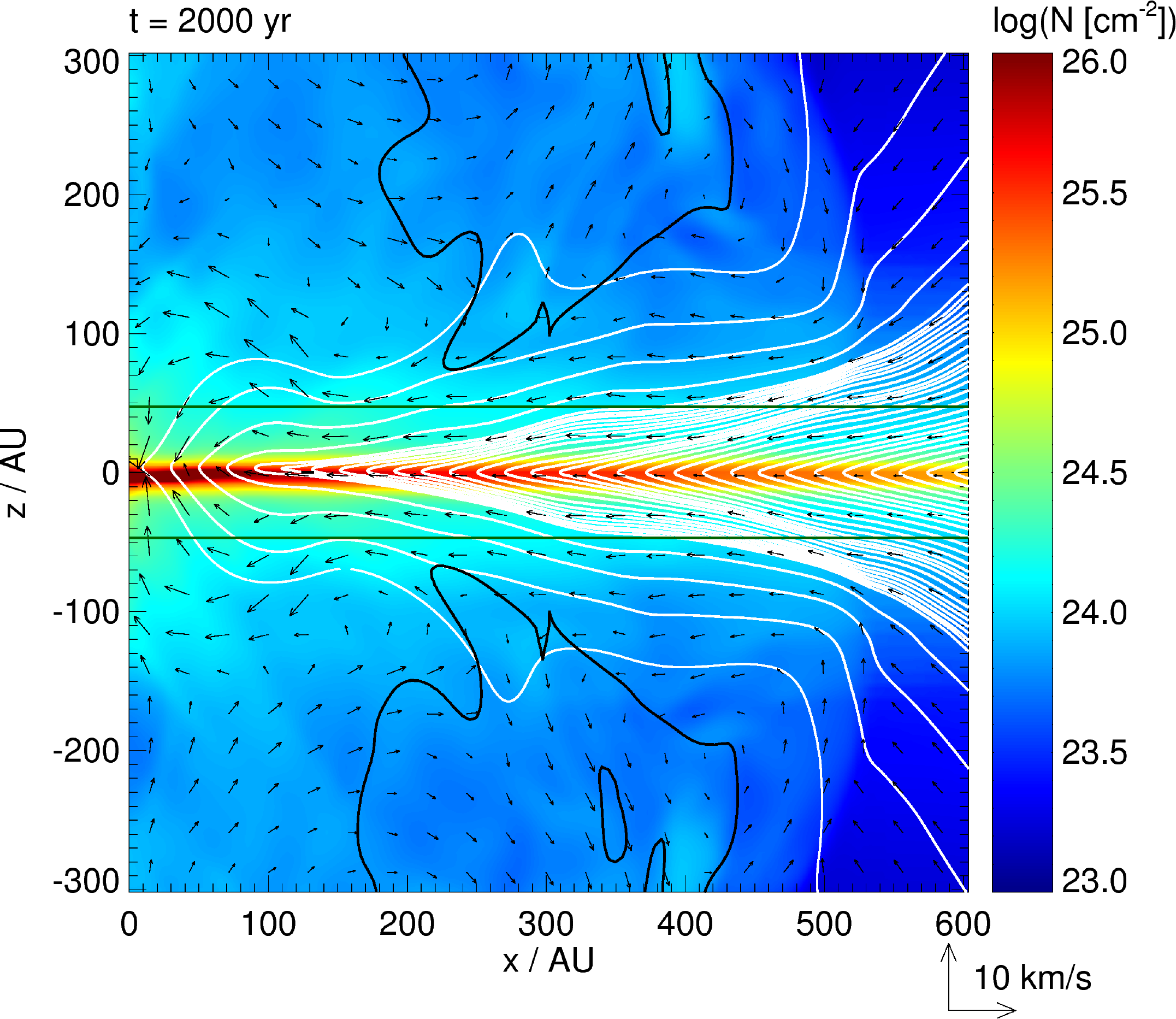} \\
 \includegraphics[width=72mm]{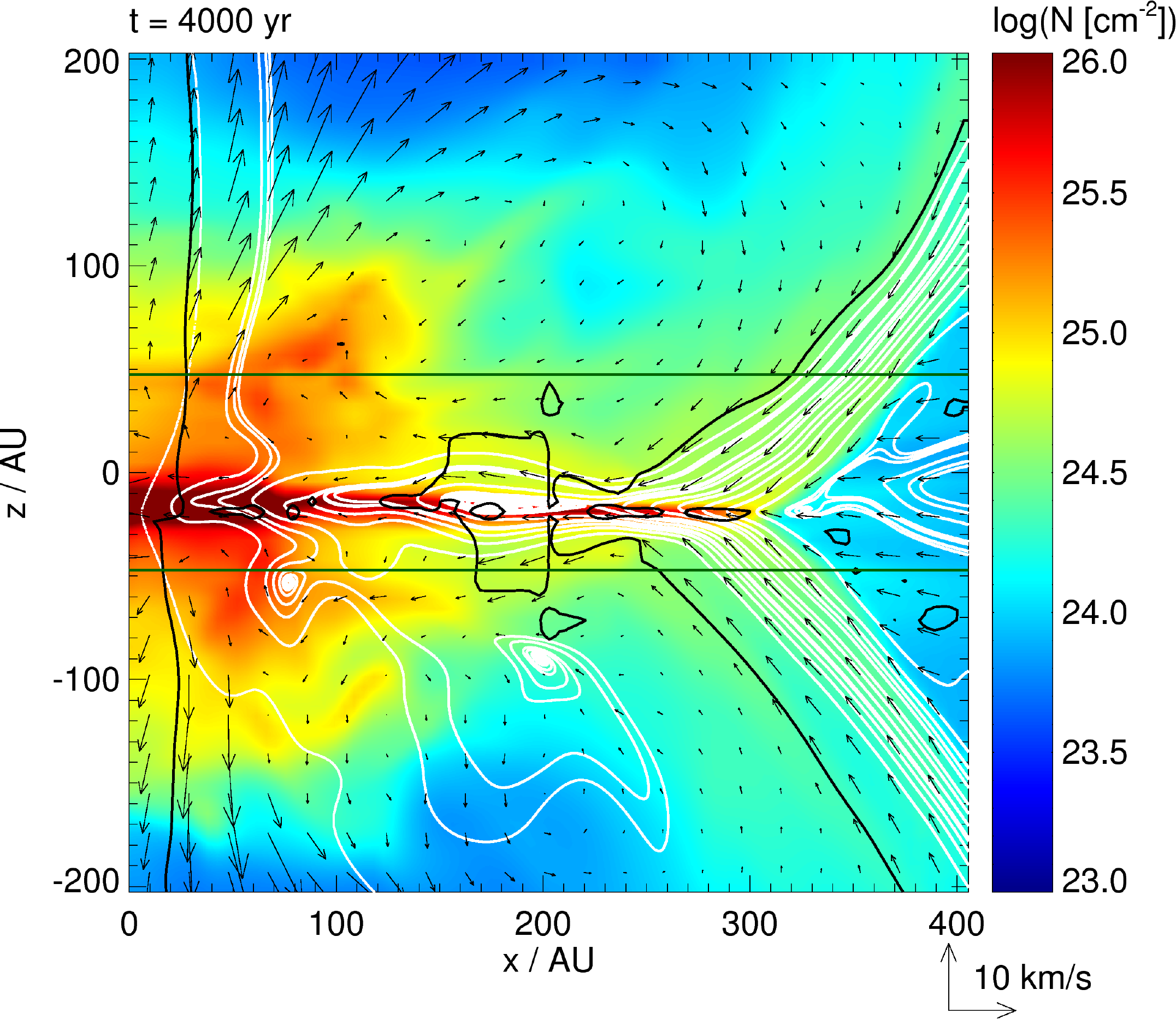}
 \includegraphics[width=72mm]{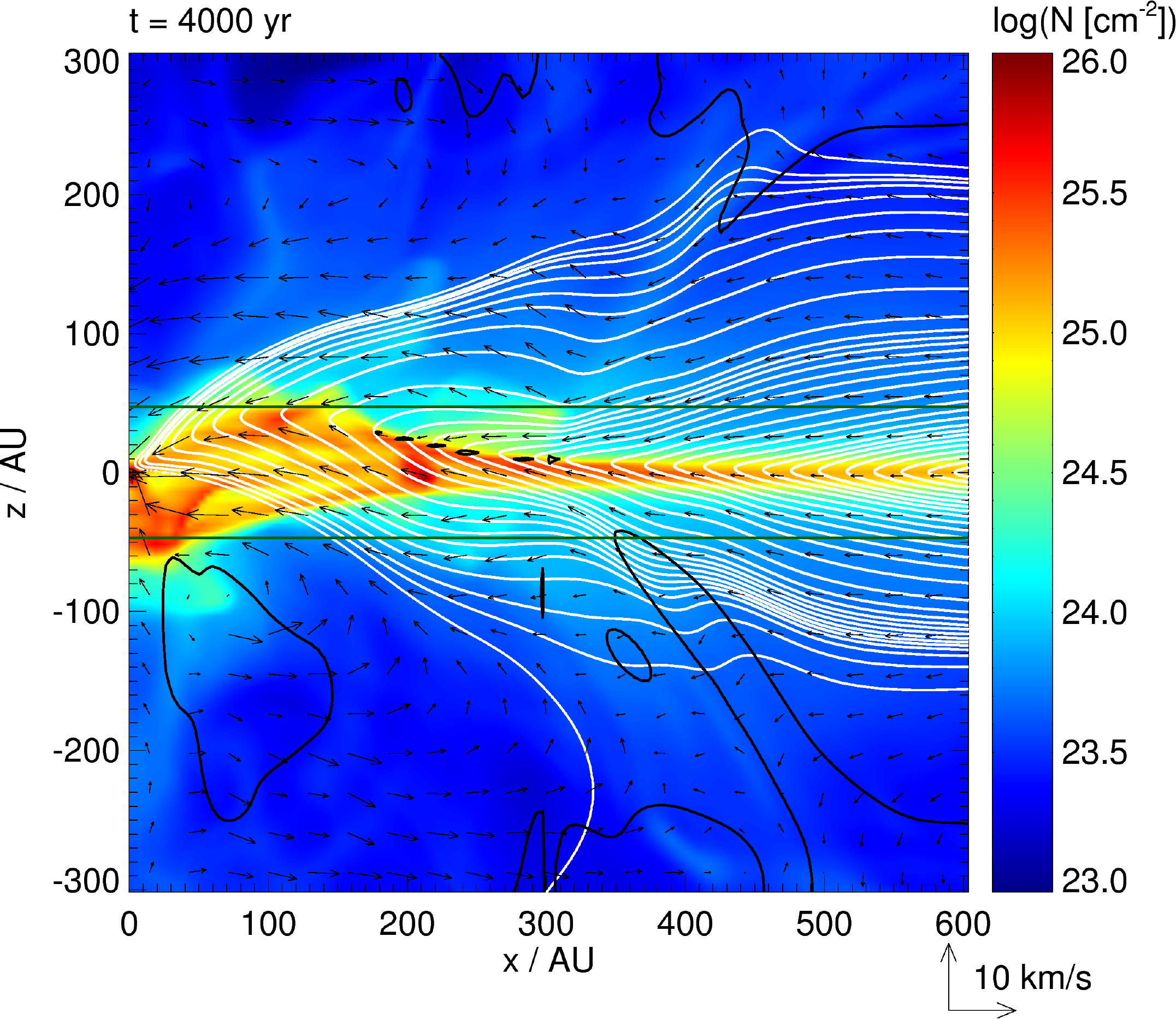}
 \caption{Edge-on view of the central region for run 26-4 (left) and run 5.2-4 (right). Superposed on the column density are the velocity field (black vectors) and the magnetic field lines (white lines). Also shown are the regions where the toroidal magnetic field dominates over the poloidal field (black lines) and the region used for calculating disc properties (dark green line). As can be seen, in run 26-4 the greatest part of the inner region is dominated by the toroidal magnetic field whereas in run 5.2-4 only smaller parts further out are dominated by $B_{\phi}$. Note the different spatial scales in the left and right panel.}
 \label{fig:maggrad}
\end{figure*}

\begin{figure*}
 \includegraphics[width=168mm]{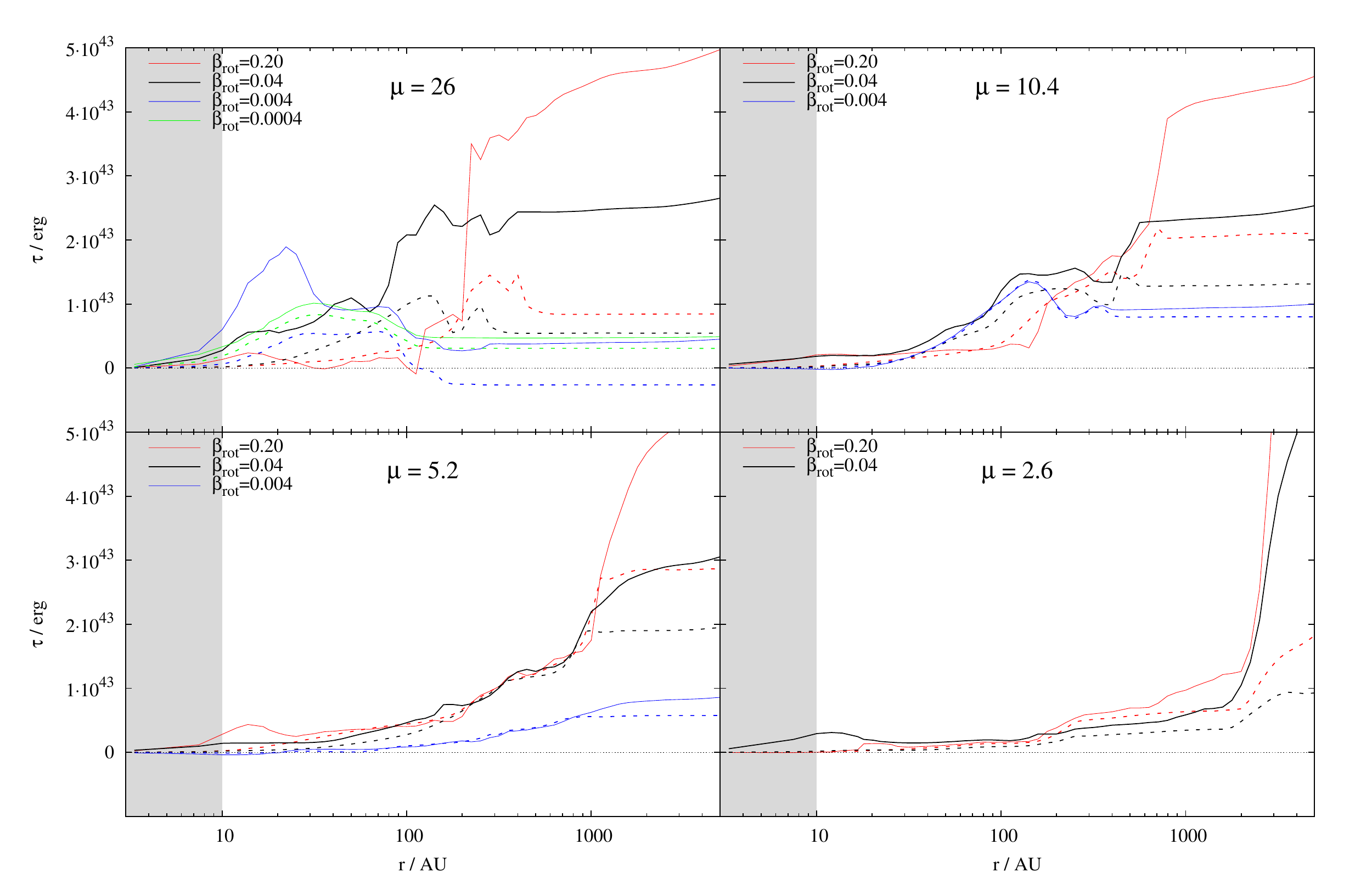}
 \caption{Gas torque (solid lines) and  magnetic field torque (dashed lines) exerted on the disc for runs with $\mu$ = 26 (top left), 10.4 (top right), 5.2 (bottom left) and 2.6 (bottom right) at the end of the simulations.  As $\tau_{\rmn{gas}}$ is negative, we plot $-\tau_{\rmn{gas}}$ for better comparison. For high magnetic field strengths the magnetic torque roughly balances the gas torque resulting in the suppression of Keplerian disc formation.}
 \label{fig:torque}
\end{figure*}

The reason for the effective magnetic braking can be seen in Fig.~\ref{fig:maggrad} showing the edge-on view of the runs 26-4 and 5.2-4 at two different times. As can be seen, the collapse of the gas has dragged in the magnetic field in the midplane producing a long magnetic lever arm~\citep{Allen03} with a large radial component (see also Fig.~\ref{fig:mag_evol}). This lever arm connects the outer, slowly rotating region with the inner, fast rotating region, hence significantly enhancing the magnetic braking efficiency.
As shown before, in the case of strong magnetic fields ($\mu < 10$) magnetic braking is so efficient that large centrifugally supported discs do not form at this early stage. This is known as the magnetic braking catastrophe and was reported previously by several authors for ideal MHD simulations~\citep{Allen03,Mellon08,Hennebelle08,Hennebelle09}. For a better qualitative and quantitative understanding of the magnetic braking effect we calculate the z-component of the two main torques acting on the disc, i.e. the torque exerted by the infalling gas
\begin{equation}
 \tau_{\rmn{gas}} = -\int dV \nabla \cdot \left( \rho \textbfit{v} \cdot \left[ \textbfit{r} \times \textbfit{v} \right]_z \right)
\end{equation}
and the torque exerted by magnetic fields
\begin{equation}
 \tau_{\rmn{mag}} = \frac{1}{4\pi} \int dV \left[ \textbfit{r} \times \left( (\nabla \times \textbfit{B}) \times \textbfit{B}\right) \right]_z.
\end{equation}
For the calculations we integrate over a disc with a height of 47 AU above and below the midplane. We omit the gravitational torque as it is by two to four orders of magnitude smaller than $\tau_{\rmn{gas}}$ and $\tau_{\rmn{mag}}$ due to the nearly axisymmetric gravitational potential. 
\begin{figure*}
 \includegraphics[width=168mm]{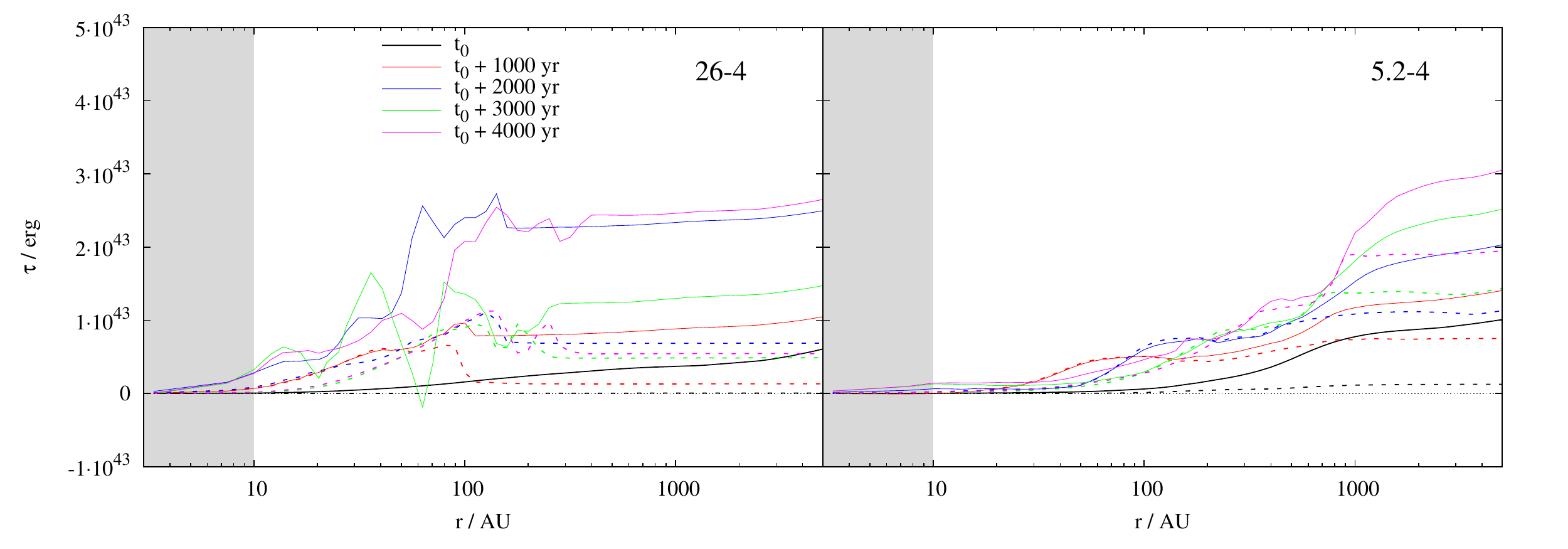}
 \caption{Gas torque (solid lines) and  magnetic field torque (dashed lines) exerted on the disc for run 26-4 (left) and run 5.2-4 (right). The torques are averaged azimuthally and shown at the same times as in Fig.~\ref{fig:vel_evol}. As $\tau_{\rmn{gas}}$ is negative, we plot $-\tau_{\rmn{gas}}$ for better comparison.}
 \label{fig:torque_evol}
\end{figure*}
In Fig.~\ref{fig:torque_evol} the time evolution of $\tau_{\rmn{gas}}$ and $\tau_{\rmn{mag}}$ is shown for the runs 26-4 and 5.2-4. The torques are averaged azimuthally and plotted against the radius. To allow for a better comparison with $\tau_{\rmn{gas}}$ we plot $-\tau_{\rmn{mag}}$. A positive $\tau$ denotes a flux of angular momentum into the disc while for a negative $\tau$ angular momentum is removed from it. Hence, in both runs the magnetic field is removing angular momentum from the disc, i.e. slowing down its rotation ($\tau_{\rmn{mag}} < 0$) whereas the gas exerts a positive torque on the disc. It can be seen that the torques increase steadily with time (except for run 26-4 at 3000 yr). For run 26-4 $\tau_{\rmn{gas}}$ is nearly always larger than the magnetic torque whereas in run 5.2-4 $\tau_{\rmn{gas}}$ is roughly balanced by $-\tau_{\rmn{mag}}$ from the very beginning. These differences show up even more clearly in Fig.~\ref{fig:torque} where the torques at the end of the simulations are shown.
For runs with a weak magnetic field (top left panel of Fig.~\ref{fig:torque}) the torque exerted by the gas is nearly everywhere above the magnetic torque. Thus, there is a net flux of angular momentum into the disc leading to the observed build-up of centrifugally supported discs. Only for run 26-20 $-\tau_{\rmn{mag}}$ equals or even exceeds $\tau_{\rmn{gas}}$ in the inner region although only on a low level. This is due to the large Keplerian disc which has already built up in this run having only very small infall velocities (see top left panel of Fig.~\ref{fig:vel}). The sharp jump of $\tau_{\rmn{gas}}$ around r = 200 AU is caused by the accretion shock at the edge of the disc where $v_{\rmn{rad}}$ drops to zero (see top left panel of Fig.~\ref{fig:vel}).

Analysing the different panels in Fig.~\ref{fig:torque}, it can be seen that at large radii the gas torques are always larger than $-\tau_{\rmn{mag}}$. This implies that the magnetic field has only a small effect on the collapse in the outer parts. In contrast, at smaller radii the effect of magnetic fields gets more and more pronounced. With decreasing $\mu$ the magnetic torque approaches $\tau_{\rmn{gas}}$ and in particular for the runs with $\mu = 5.2$ and 2.6 $-\tau_{\rmn{mag}}$ is very close to $\tau_{\rmn{gas}}$. This is attributed to the fact that an equilibrium between $\tau_{\rmn{gas}}$ and $\tau_{\rmn{mag}}$ is reached where as much angular momentum is removed by magnetic braking as it is added due to the gas infall. Hence, in the case of strong magnetic fields the net angular momentum flux is roughly zero preventing a Keplerian disc from forming.

To summarise, centrifugally supported discs at very early times only form in simulations with $\mu \ga 10$. In these cases magnetic braking is not strong enough to remove angular momentum at high enough rates and hence is not able to prevent Keplerian disc formation (see Fig.~\ref{fig:diskoverview} for an overview).

\subsection{Accretion rates}

\begin{figure}
 \includegraphics[width=84mm]{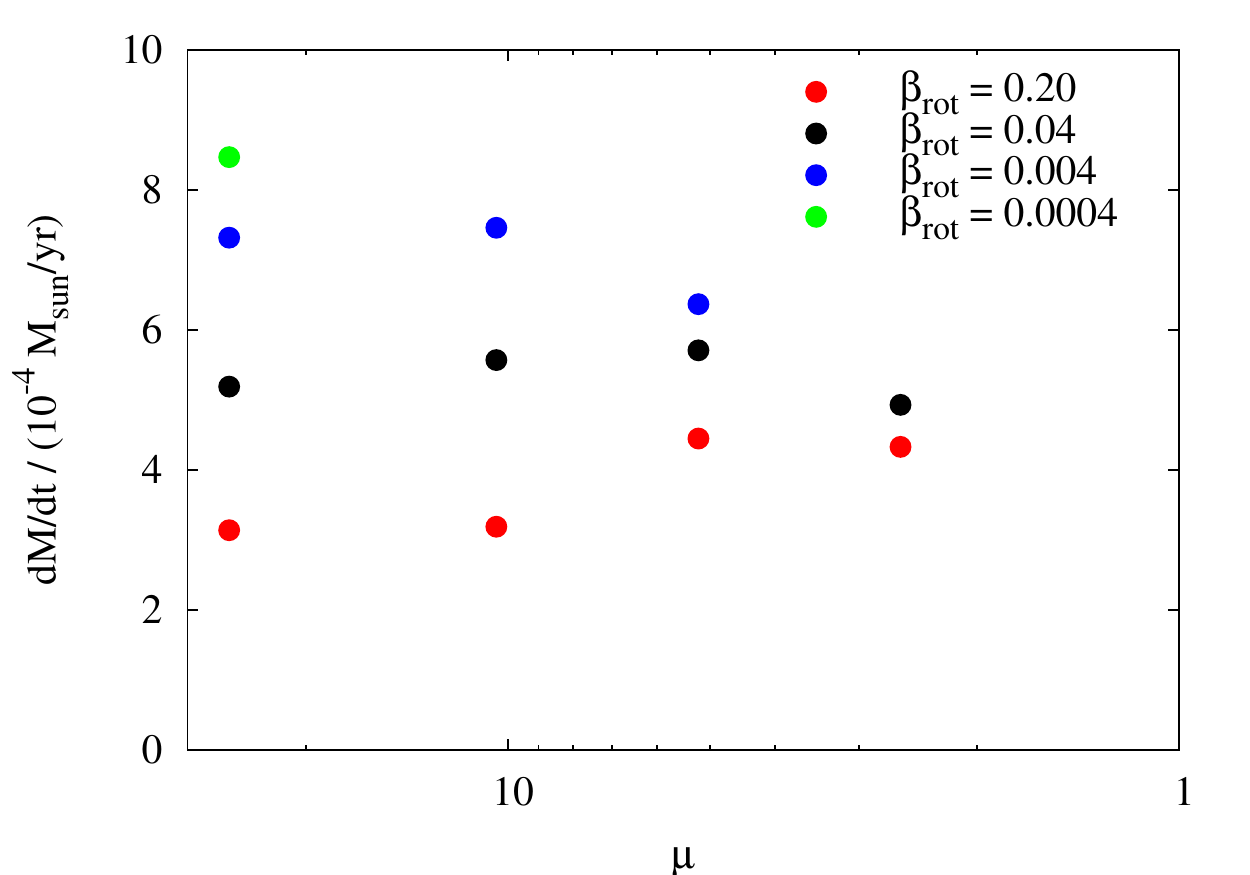}
 \caption{Accretion rates of the different simulations plotted against the initial mass-to-flux-ratio $\mu$. Equal colours denote equal initial rotational energies. The accretion rates seem to converge with decreasing $\mu$.}
 \label{fig:accoverview}
\end{figure}

Closely related to the velocity structure of the matter around the sink particles is the accretion onto the particles themselves. The general behaviour of the accretion rate with varying initial conditions can be inferred from Table~\ref{tab:msinks} where we list the totally accreted masses and the corresponding time averaged accretion rates. Additionally, in Fig.~\ref{fig:accoverview} the accretion rates of all runs are plotted against the initial mass-to-flux ratio.
As listed in Table~\ref{tab:msinks}, in neither of the runs more than 4 M$_{\sun}$ are accreted during the simulation resulting in time averaged accretion rates around a few 10$^{-4}$ M$_{\sun}$ yr$^{-1}$. Interestingly, the accretion rates do not vary by more than a factor of about 3 between the different simulations. This is remarkable, considering the large range in parameter space covered by the initial conditions (Fig.~\ref{fig:models}). For each set of simulations with equal $\mu$ there is also a rough correspondence between the accretion rate and the infall velocity shown in Fig.~\ref{fig:vel}. As expected, higher infall motions result in higher accretion rates.

\begin{figure*}
 \includegraphics[width=84mm]{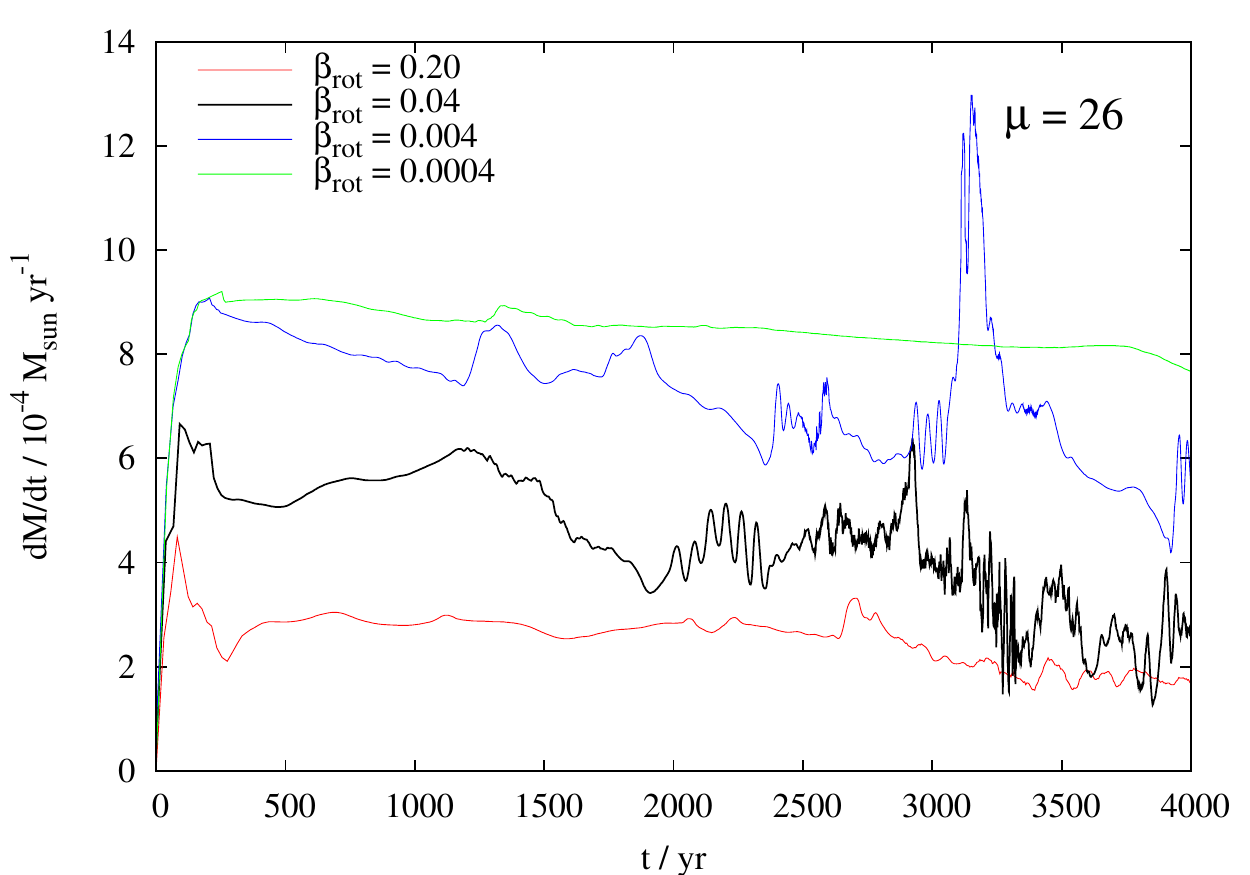}
 \includegraphics[width=84mm]{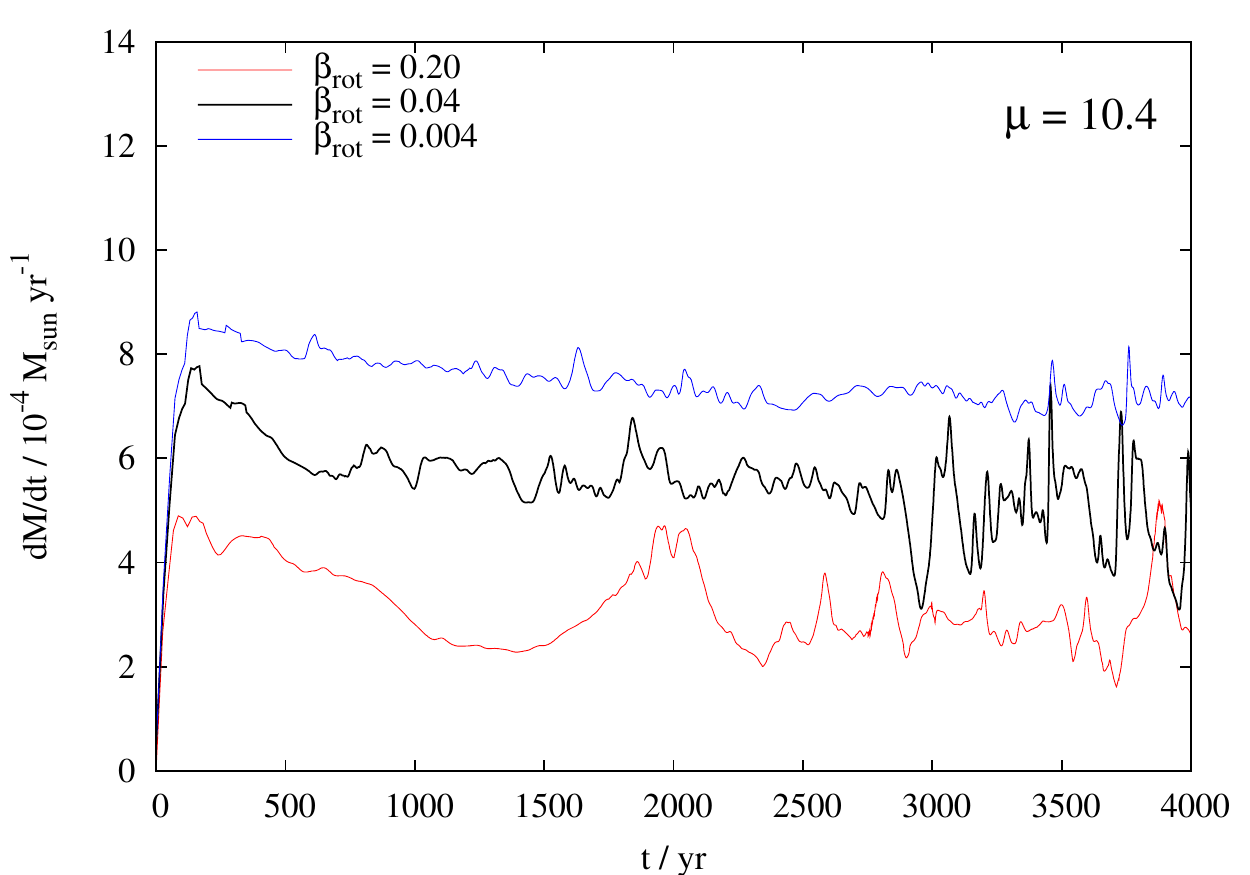} \\
 \includegraphics[width=84mm]{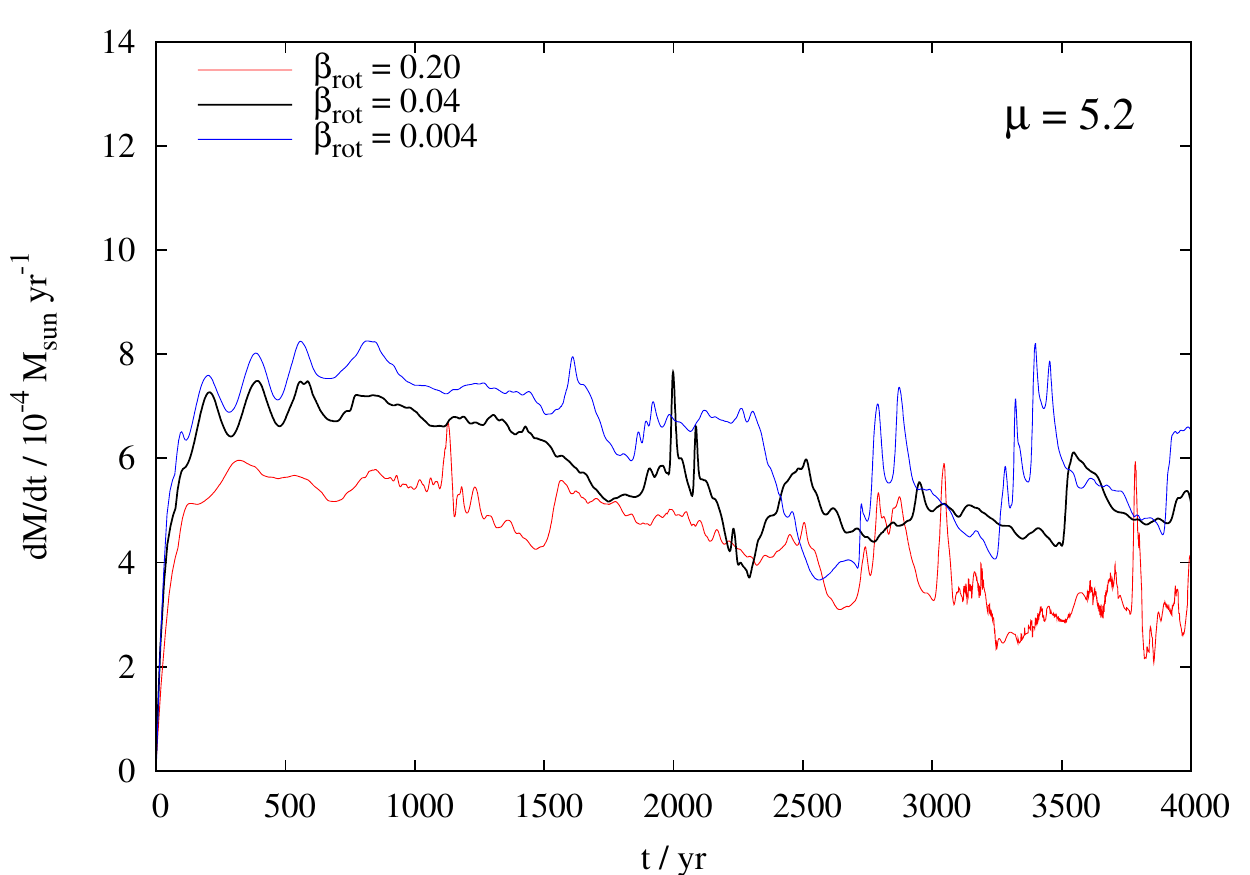}
 \includegraphics[width=84mm]{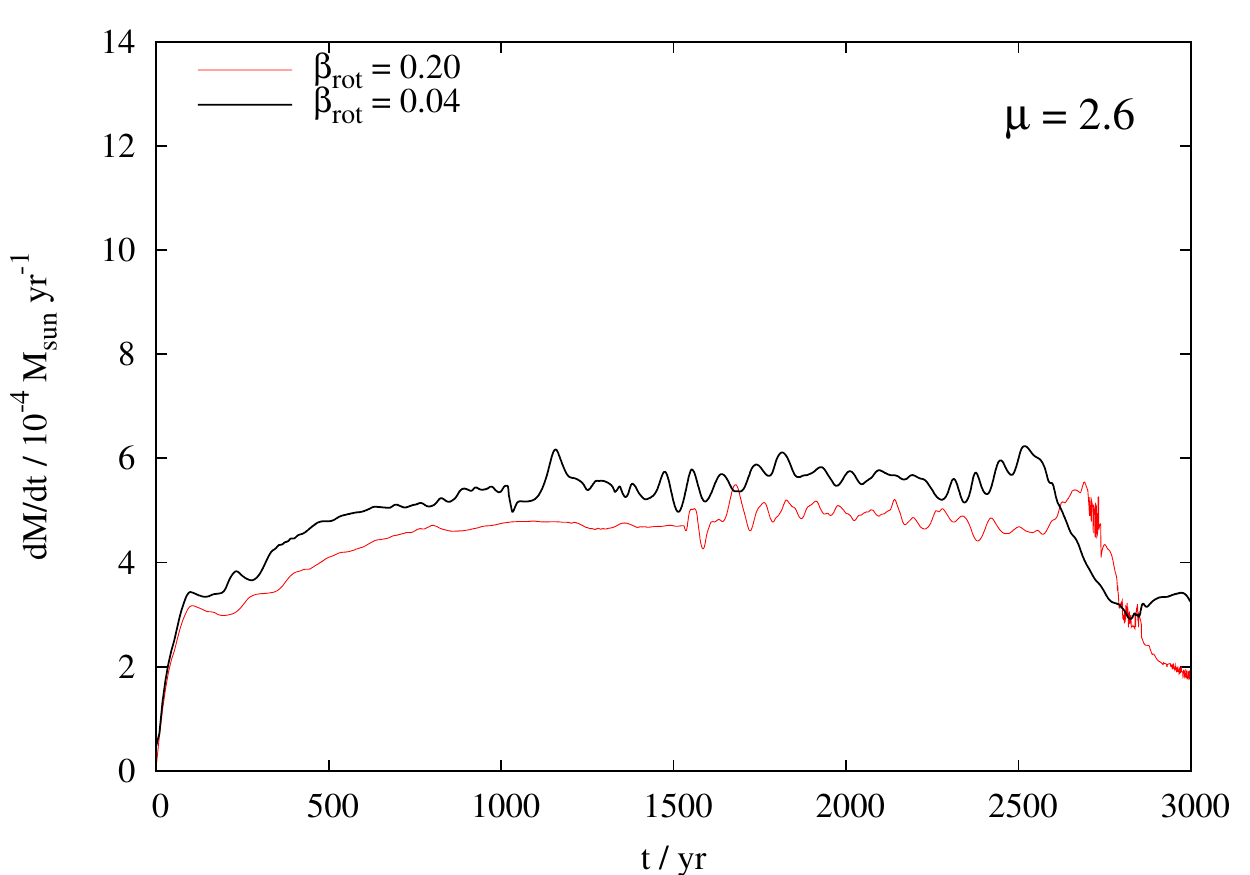}
 \caption{Accretion rates of all runs performed with $\mu = 26$ (top left), 10.4 (top right), 5.2 (bottom left) and 2.6 (bottom right). As expected, the accretion rates decrease with increasing initial rotational energy for a given magnetic field strength. The differences in the accretion rates for runs with fixed $\mu$ but varying $\beta_{\rmn{rot}}$ decrease with increasing magnetic field strength.}
 \label{fig:sink-acc}
\end{figure*}

For a more detailed analysis of the accretion rates, we consider their time evolution in Fig.~\ref{fig:sink-acc}. We mention that for the runs 26-20 and 26-4 where more than one sink particle is formed only the accretion onto the first particle created is shown. A more detailed analysis of the accretion history in the case of fragmentation will be carried out in Section \ref{sec:frag}.
As can be seen in Fig.~\ref{fig:sink-acc}, there seems to be a slight decrease in the accretion rates by a few 10\% over time. Numerical studies~\citep[e.g.][]{Klessen01,Schmeja04} as well as observational results~\citep[see][for an overview]{Andre00} indicate that there is indeed a decline in the accretion rate by orders of magnitude, although timescales for this decline are typically of the order of several 10$^4$ yr and are therefore much longer than in our study. Additionally, there occur fast variations within a factor of about 2 around the mean value. This fast variations are caused by moderate density perturbations developing in the midplane. Each time a perturbation moves through the centre, it causes the accretion to vary around its mean. The variations are generally rather small (with one exception in run 26-0.4) and would probably be smoothed out in time by viscous effects in the inner disc not resolved here. However, we cannot exclude that the varying accretion rates would influence the protostellar evolution as proposed by stellar evolution calculations~\citep[e.g.][]{Wuchterl01,Baraffe10}.

The only exception where the accretion rates decrease significantly over time are the runs with $\mu = 2.6$ showing a sharp drop in the mass accretion after roughly 2500 yr. This is caused by the occurrence of magnetically driven bubbles in the midplane as shown exemplarily in the bottom panel of Fig.~\ref{fig:disk}. However, in the other runs the accretion rates seem to decrease only slightly showing no sign that the magnetically driven outflow, which is launched from the protostellar disc shortly after the formation of the protostar, can significantly reduce mass accretion over time. Similar results in related work on magnetic fields in massive star formation are found by \citet{Peters11} and \citet{Hennebelle11} as well. This behaviour is due to the ongoing accretion through the disc which is nearly unaffected by the shut-off of gas infall from below or above the disc due to the outflow. This can be seen in Fig.~\ref{fig:maggrad} where we plot the velocity structure and magnetic field lines in a slice along the z-axis.

As can be inferred from Table~\ref{tab:msinks} and from Fig.~\ref{fig:accoverview}, there are some systematical trends in the accretion rates with changing initial conditions, although the overall variation is not very large. Increasing the overall rotational support against gravity, i.e. $\beta_{\rmn{rot}}$, for fixed $\mu$ results in lower accretion rates. This is in agreement with the increase in $v_{\rmn{rot}}$/$v_{\rmn{kep}}$ and the decrease in $v_{\rmn{rad}}$ in the surrounding disc as shown in Fig.~\ref{fig:vel}. As already observed for the velocity structure in the midplane, the differences in the accretion rates for runs with different $\beta_{\rmn{rot}}$ but fixed $\mu$ decrease with increasing magnetic field strength as can be seen in Fig.~\ref{fig:accoverview}. This is due to the efficient magnetic braking removing angular momentum at roughly the same rate as it is transported inwards (see Fig.~\ref{fig:torque}). As a consequence, the accretion rates are roughly independent of the initial amount of angular momentum. An even further increase in the field strength would probably result in even lower accretion rates than the 4 - 5 $\cdot 10^{-4}$ M$_{\sun}$ yr$^{-1}$ observed for $\mu$ = 2.6 due to stronger magnetic forces counteracting gravity.

The accretion rates for varying $\mu$ but fixed $\beta_{\rmn{rot}}$ show an interesting and less clear behaviour. For high initial rotational energies, i.e. $\beta_{\rmn{rot}} = 0.20$ and 0.04, respectively, the accretion rates first increase with decreasing $\mu$ and drop again for the case with $\mu = 2.6$. For $\beta_{\rmn{rot}} = 0.004$ the accretion rate increases from $\mu = 26$ to $\mu = 10.4$ before declining again for lower $\mu$. We attribute this behaviour to two competing effects of the magnetic field. On the one hand, magnetic fields act to enhance accretion onto the protostar by magnetic braking reducing the centrifugal support against gravity. Hence, the effect of magnetic braking alone would cause increasing accretion rates with decreasing $\mu$ as it is indeed observed for low field strengths. The second effect influencing the accretion rates is the Lorentz force, i.e. the combination of magnetic pressure and magnetic tension, induced by strongly bent field lines (see Fig.~\ref{fig:maggrad}). Magnetic pressure and magnetic tension counteract gravity by exerting an outward force on the gas resulting in reduced accretion rates. The strength of this effect increases with the field strength thus tending to lower accretion rates with lower $\mu$.

The combination of both effects -- magnetic braking enhancing accretion and the Lorentz force counteracting accretion -- results in the observed behaviour of the accretion rates: By increasing the magnetic field strength at a given $\beta_{\rmn{rot}}$ up to a certain critical value an equilibrium between the torques acting on the disc is reached where the removal of angular momentum by magnetic braking balances its inwards transport (see Fig.~\ref{fig:torque}). Up to this value, an increase in the field strength is associated with increasing accretion rates as observed in our simulations. Further increase in the field strength beyond this point cannot enhance the magnetic braking efficiency any more. In fact, now the increase (decrease, if $\mu$ is considered) leads to declining accretion rates due to the growing strength of the Lorentz force counteracting gravity, in accordance with our findings of declining accretion rates for strong fields with $\mu \la 5$. The exact value of $\mu$, where this turnover occurs, depends on the initial amount of rotational energy and decreases with increasing $\beta_{\rmn{rot}}$.

\subsection{Disc fragmentation} \label{sec:frag}

As mentioned earlier, the runs 26-20 and 26-4 show rapid fragmentation of the protostellar disc after the first protostar has formed (see Fig.~\ref{fig:diskoverview}). Due to a high amount of rotational energy and a weak magnetic field, magnetic braking can only remove a small amount of angular momentum leading to the formation of Keplerian discs with considerable extensions of a few 100 AU (see top left panel of Fig.~\ref{fig:vel} for comparison). As the mass load onto these discs exceeds their capability to transport material inwards by gravitational or viscous torques, the discs become unstable and fragment~\citep[e.g.][]{Kratter10}. At the end of the simulations, i.e. after 4000 yr there are 10 sink particles in run 26-4 and 13 in run 26-20. All other simulations show no fragmentation so far although some of them form a Keplerian disc (compare Fig.~\ref{fig:accoverview}).

\begin{figure*}
 \includegraphics[width=84mm]{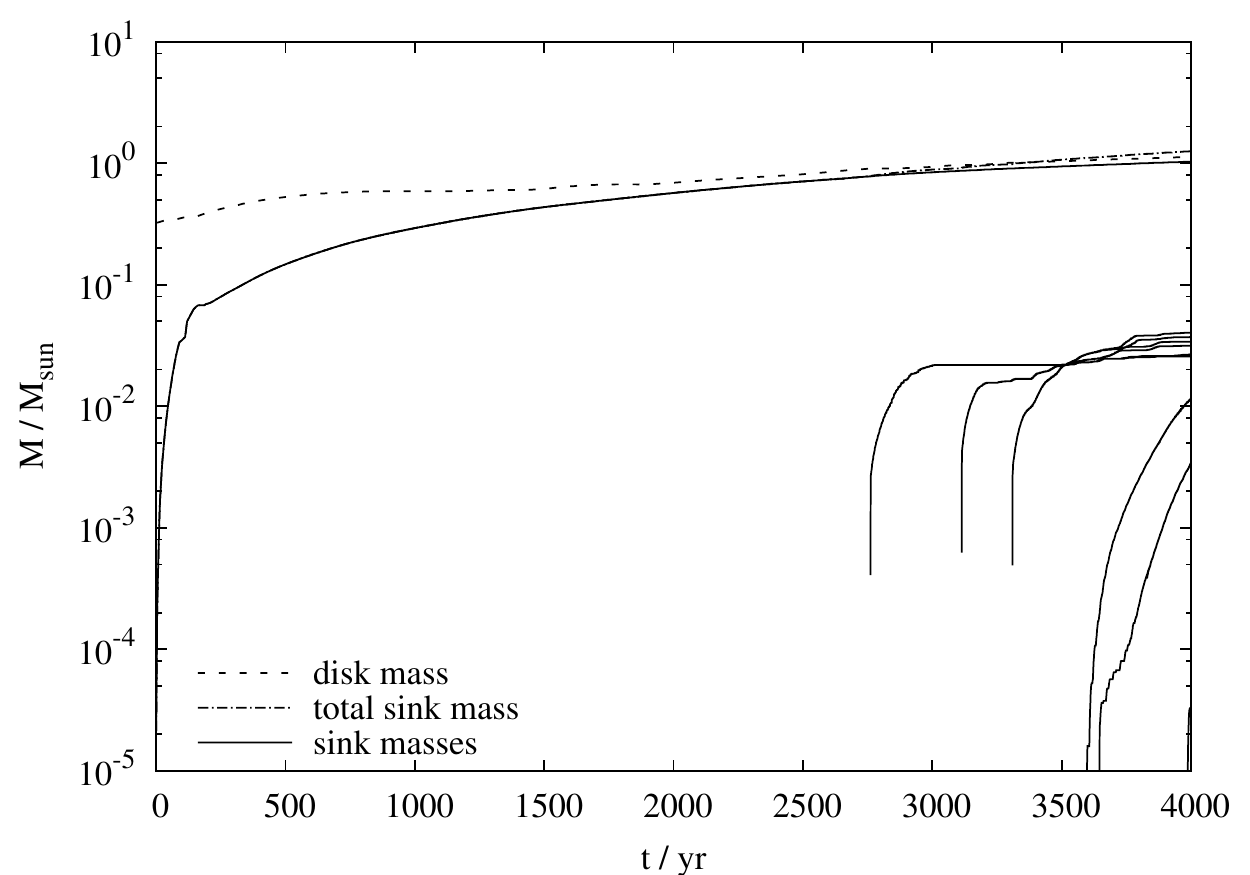}
 \includegraphics[width=84mm]{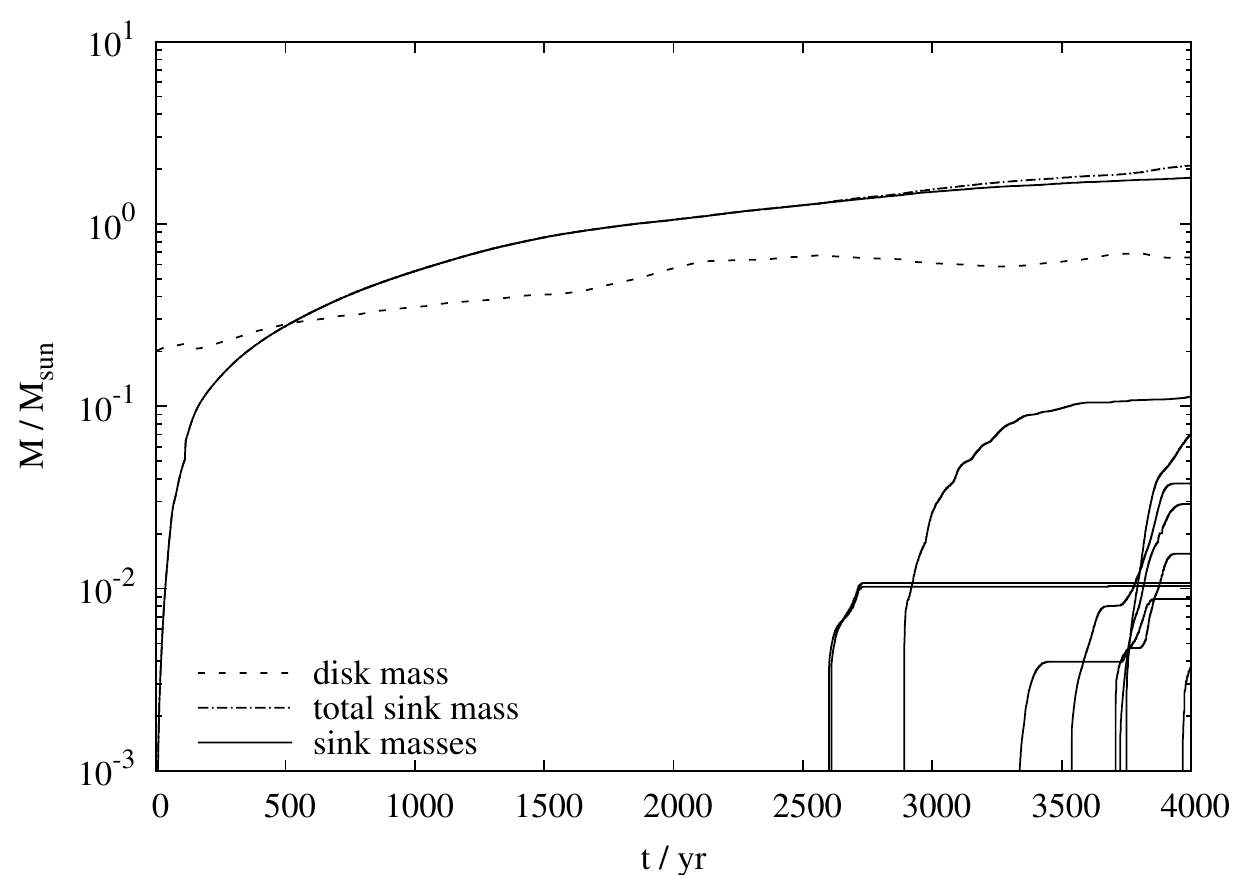}
 \caption{Accretion history of run 26-20 (left) and run 26-4 (right). The dashed-dotted lines show the total mass of all sink particles. The formation of sink particles in run 26-20 occurs pairwise (except the first one) and also the further evolution of each pair is nearly indistinguishable except at the very end so that in the beginning each line represents two particles. Also shown is the evolution of the disc mass with time (dashed lines) which is of the order of or somewhat below the totally accreted mass.}
 \label{fig:frag}
\end{figure*}

The accretion histories for run 26-20 and run 26-4 are shown in Fig.~\ref{fig:frag}. Run 26-20 exhibits a very symmetric disc fragmentation forming pairs of protostars at roughly the same time and opposite positions (as seen from the centre). Even in their further evolution each pair develops very similar as can be seen from the left panel of Fig.~\ref{fig:frag} where the lines of each pair are nearly indistinguishable. For run 26-4 only the evolution of the second and third sink particle is symmetric, while at later times there is no pairwise formation of sink particles any more due to a nonsymmetric evolution of the disc.
We note that in both runs only the first sink particle created has reached a mass above 1 M$_{\sun}$ so far whereas all other particles have masses well below 0.1 M$_{\sun}$. For comparative purposes we also plot the totally accreted mass of all sinks formed in Fig.~\ref{fig:frag} (see also Table~\ref{tab:msinks}). As can be seen, in both runs more than 80\% of the total mass is accreted onto the first sink particle.

After about 2500 yr, i.e. after the creation of further sink particles, there occurs a slight but nevertheless noticeable decrease in the accretion onto the first sink particle (see the red and black lines in the top left panel of Fig.~\ref{fig:sink-acc}). This fragmentation-induced starvation~\citep{Peters10b} is caused by the surrounding sink particles soaking up the infalling material. This behaviour was also observed recently in related work on massive star formation~\citep{Peters10a,Peters11,Girichidis11}.

Beside the sink masses, Fig.~\ref{fig:frag} shows the calculated disc masses as well. For the calculation we define the disc as follows: Firstly, we determine the maximum cylindrical radius R$_{\rmn{max}}$ where the gas falls below a density of $5 \cdot 10^{-15}$ g cm$^{-3}$. The disc mass is then defined as the total mass of gas within a cylinder with a height of 47 AU above and below the midplane and a radius of R$_{\rmn{max}}$ around the centre. As can be seen, the disc masses lie around 1 M$_{\sun}$ and are therefore somewhat below the totally accreted sink masses.

Next, we study the stability of the discs against gravitationally induced perturbations. Disc stability is described by the Toomre parameter  \citep{Toomre64} defined as
\begin{equation}
 Q = \frac{\kappa c_{\rmn{s}}}{\pi \Sigma G}
\end{equation}
with the epicyclic frequency $\kappa$, sound speed $c_{\rmn{s}}$, surface density $\Sigma$, and gravitational constant $G$. Gravitational instability sets in when $Q < 1$. As magnetic fields are present in the discs, one can define the magnetic Toomre parameter~\citep{Kim01}
\begin{equation}
 Q_{\rmn{M}} = \frac{\kappa \left( c_{\rmn{s}}^2 + v_{\rmn{A}}^2 \right)^{1/2}}{\pi \Sigma G},
\end{equation}
where $v_{\rmn{A}}$ is the Alfv\'enic velocity taking into account all components of the magnetic field. In the following, we analyse the stability of the midplane in the runs 26-20, 10-20, 5.2-20 and 26-0.4 as these simulations cover a wide range of initial conditions. In particular, we concentrate on the stabilising effect of the magnetic field by comparing $Q_{\rmn{M}}$ and $Q$. 
\begin{figure*}
 \includegraphics[width=70mm]{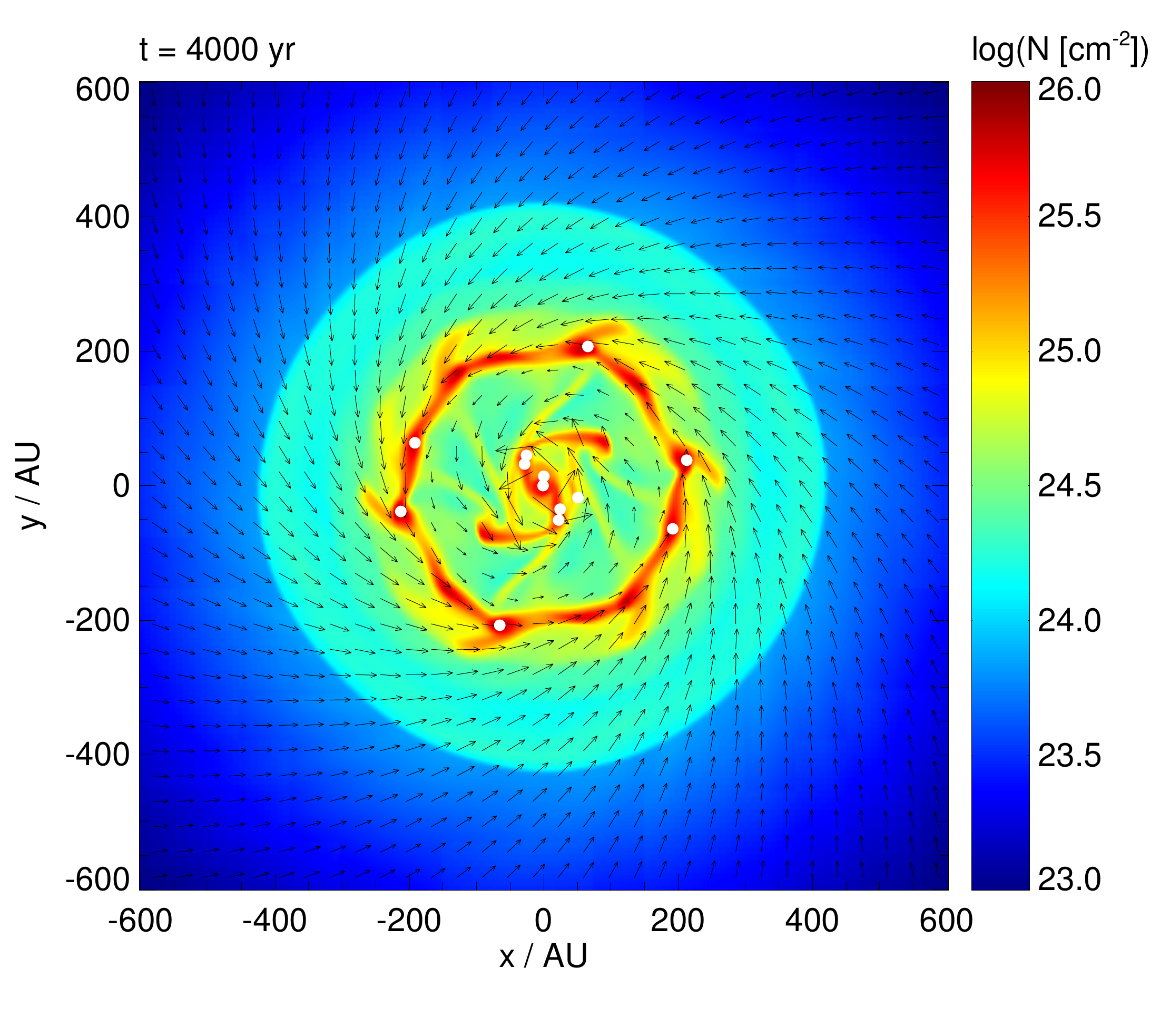} \hspace{10mm}
 \includegraphics[width=70mm]{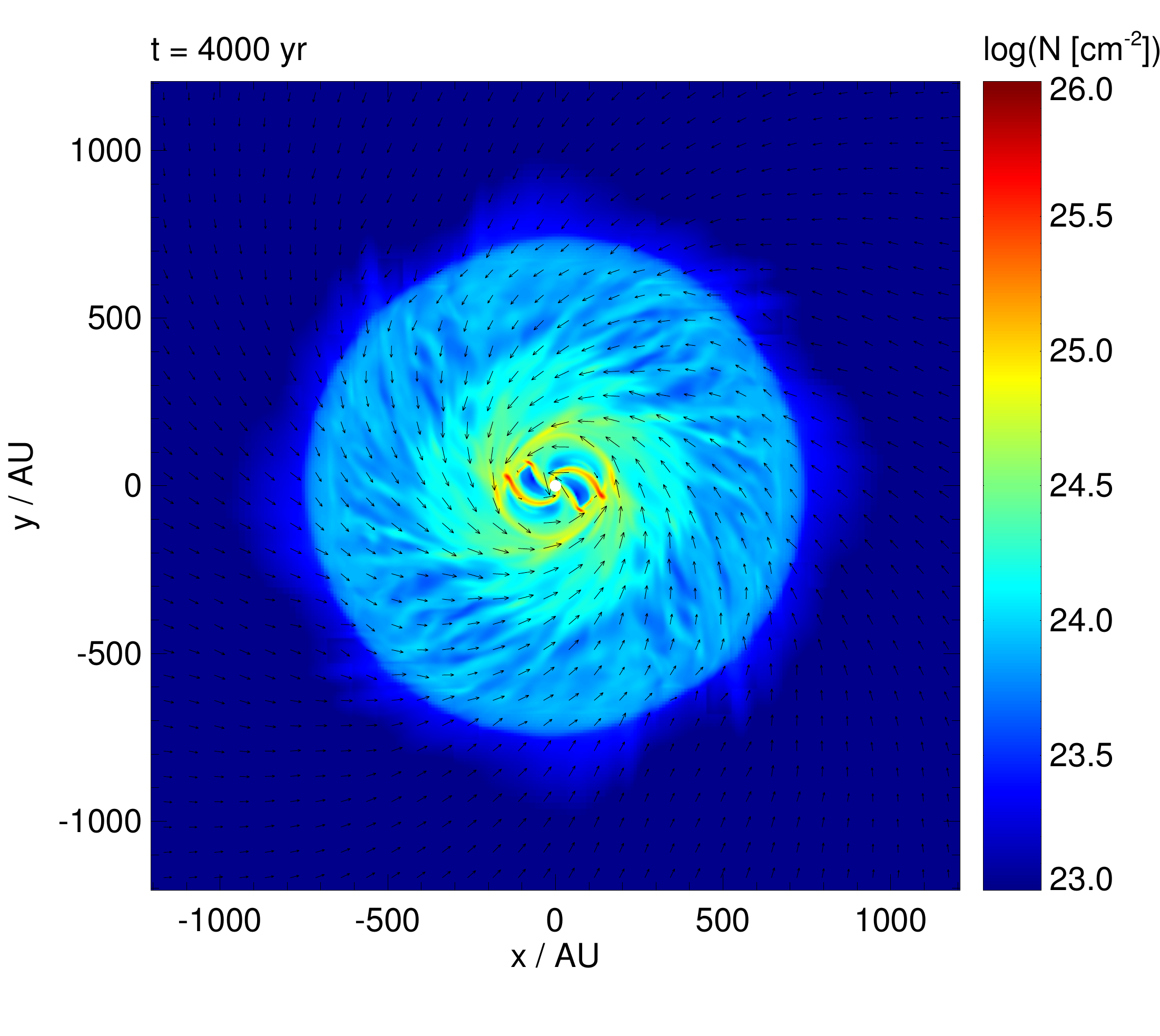} \\
 \vspace{-5mm}
 \includegraphics[width=72mm]{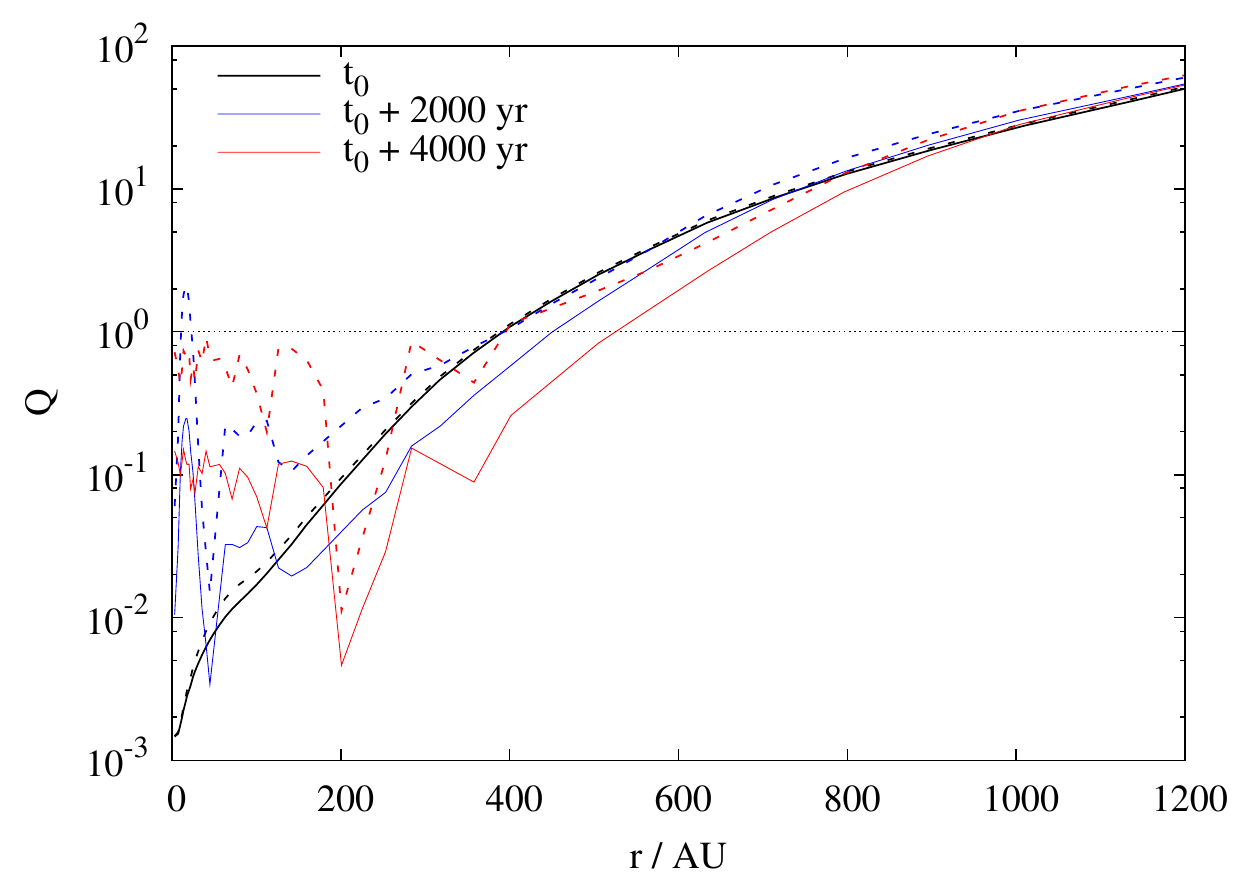} \hspace{10mm}
 \includegraphics[width=72mm]{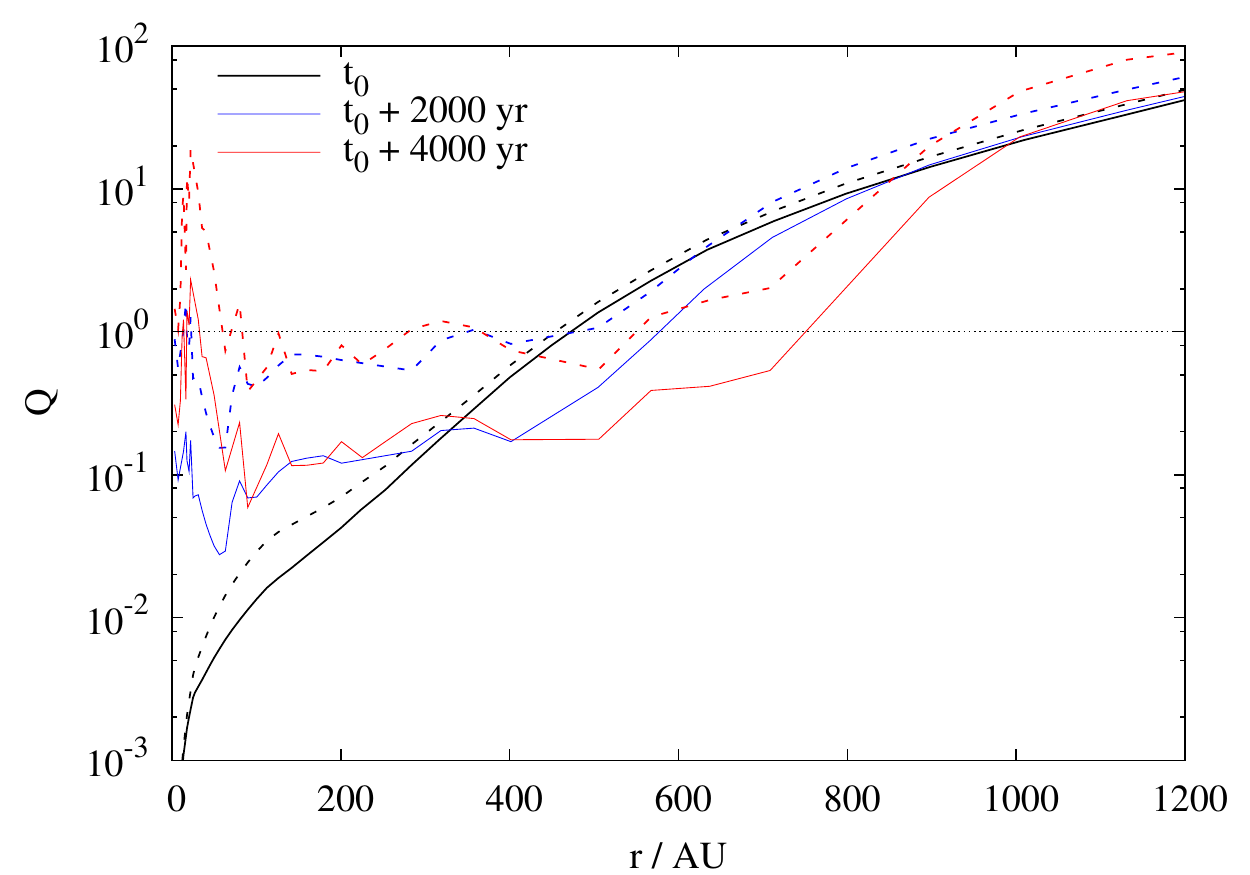} \\
 \includegraphics[width=70mm]{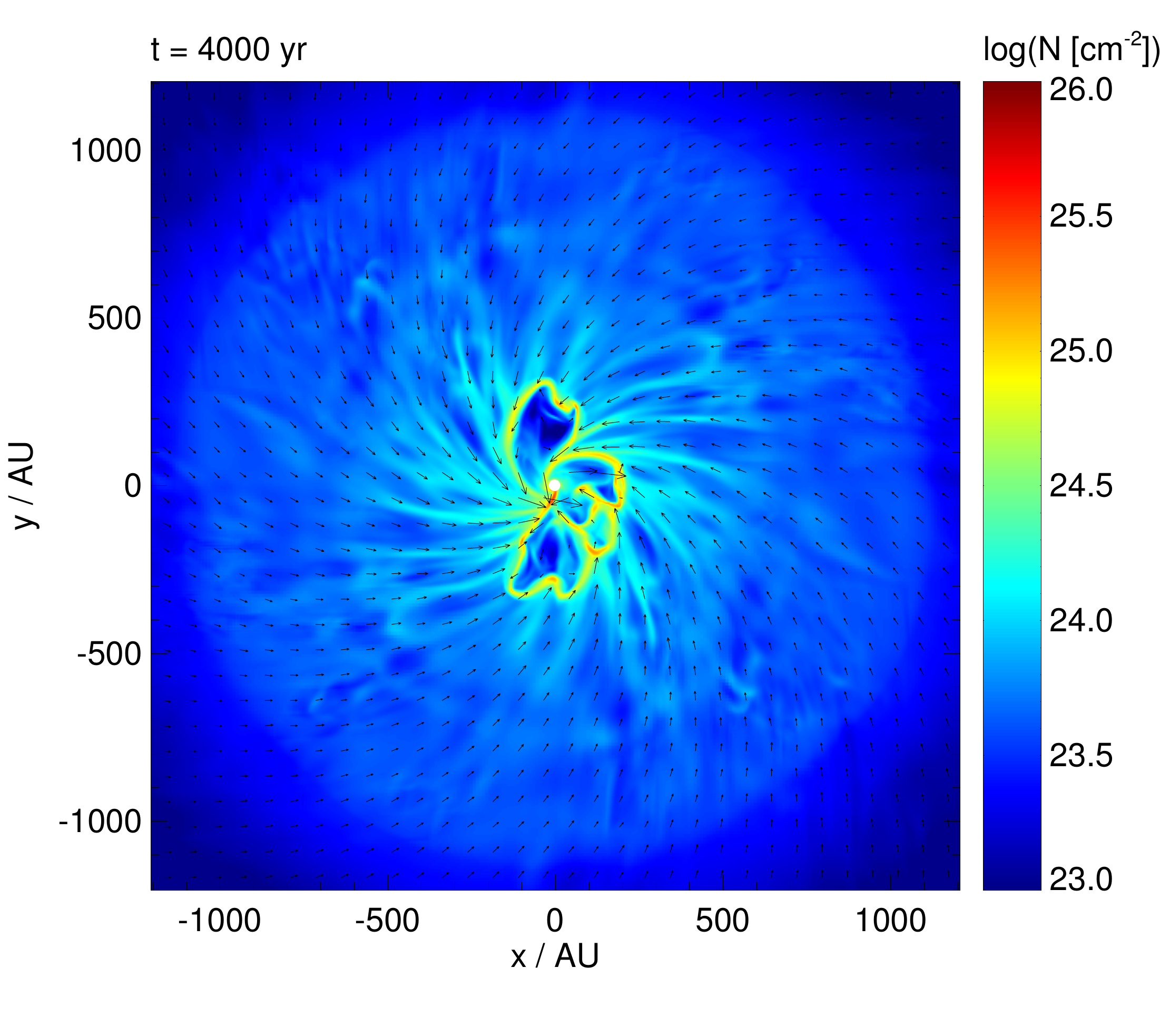} \hspace{10mm}
 \includegraphics[width=70mm]{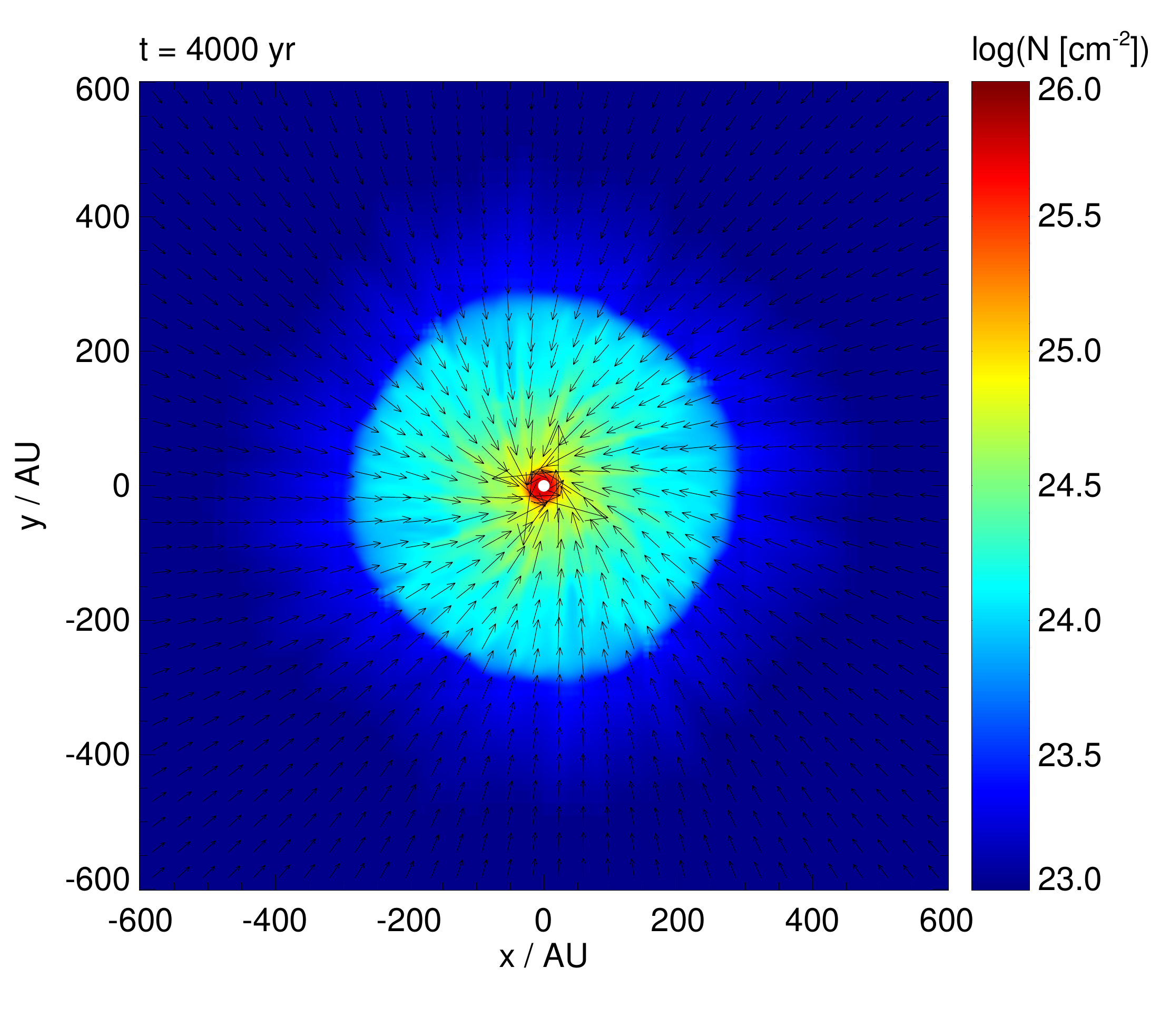} \\
 \vspace{-5mm}
 \includegraphics[width=72mm]{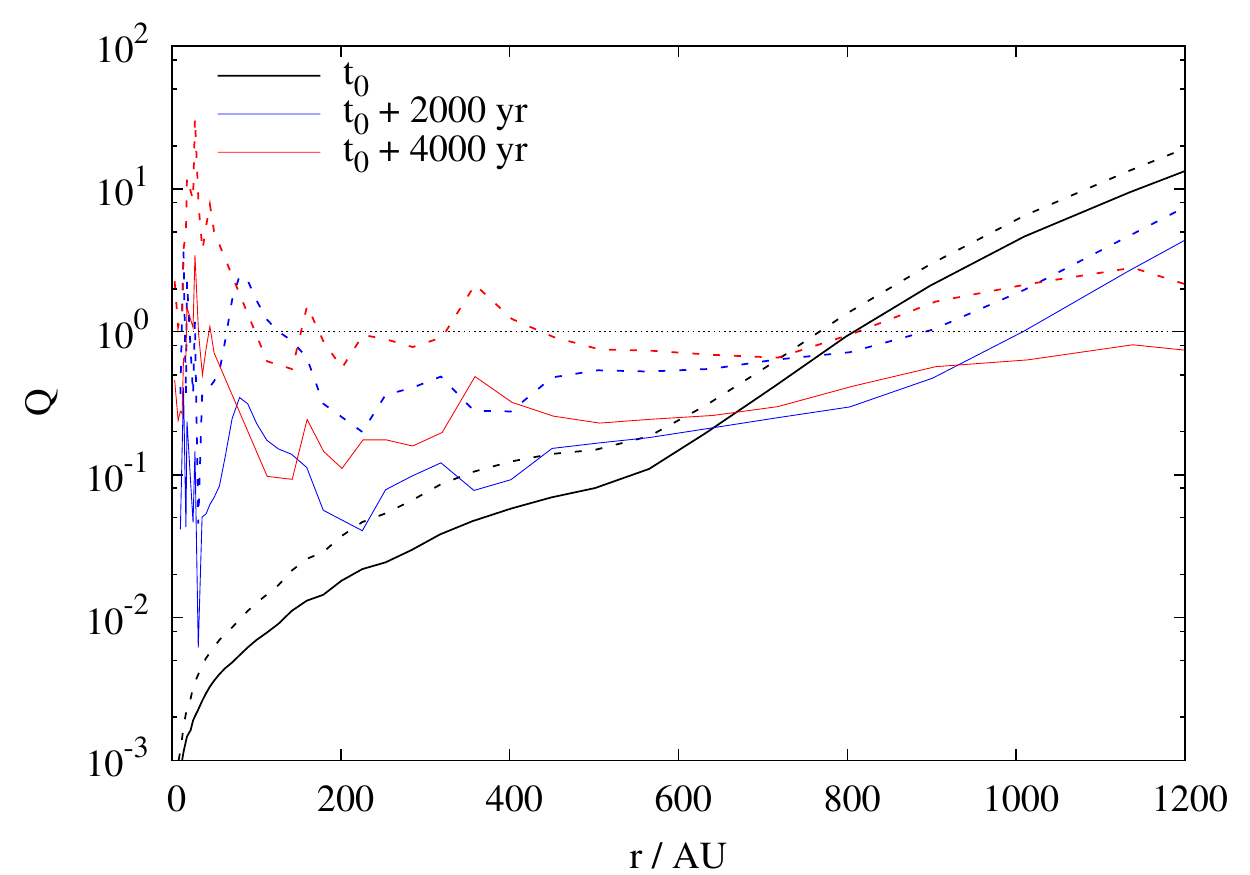} \hspace{10mm}
 \includegraphics[width=72mm]{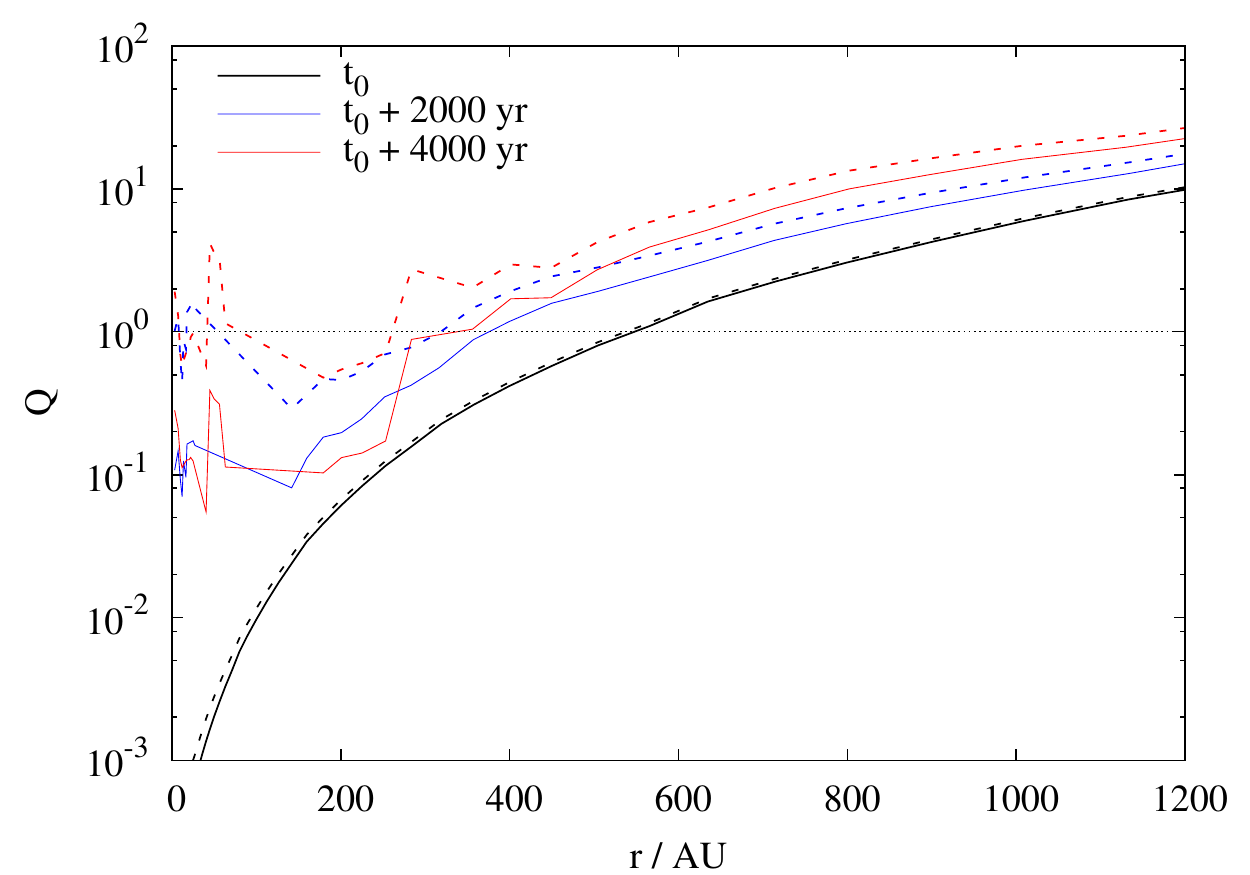}
 \caption{Column density of the disc in the runs 26-20, 10-20, 5.2-20 and 26-0.4 (from top left to bottom right) after 4000 yr. Below each slice the Toomre parameter $Q$ (solid lines) and magnetic Toomre parameter $Q_{\rmn{M}}$ (dashed lines) for t = 0 yr, 2000 yr and 4000 yr plotted against the radius are shown. Within a radius of approximately 500 AU $Q$ is nearly everywhere below 1 for all runs considered thus indicating gravitational unstable discs. Except for run 26-20 fragmentation is suppressed due to the influence of the magnetic field indicated by $Q_{\rmn{M}}$ $\simeq$ 1. In run 26-20 where even $Q_{\rmn{M}}$ is below 1 fragmentation occurs.}
 \label{fig:toomre}
\end{figure*}
In Fig.~\ref{fig:toomre} the face-on view of the discs after 4000 yr as well as the radial dependence of the Toomre parameters are plotted.
$Q$ and $Q_{\rmn{M}}$ are calculated numerically by the azimuthally averaged values of $c_{\rmn{s}}$, $v_{\rmn{A}}$, $\kappa$, and $\Sigma$ and are shown for several times. We neglect the contribution of secondary sink particles to the column density in run 26-0.4 as their only effect would be to lower $Q$ and $Q_{\rmn{M}}$ in regions which are already unstable and fragmenting and therefore they would not qualitatively change the situation. The plot shows that within a radius of r $\simeq$ 500 AU the Toomre parameter $Q$ is almost everywhere below 1 for all runs and all times considered. Hence, in all runs fragmentation would be expected to occur. As this is not the case except for run 26-20, the Q-parameter seems not able to properly describe the stability properties of the discs. Taking into account the effect of magnetic fields, the situation displayed in the column density plots is described much more adequately. The value of $Q_{\rmn{M}}$ is higher than $Q$ by roughly a factor of 10 in all cases thus pointing to much more stable configurations. Nevertheless, for run 26-20 even $Q_{\rmn{M}}$ is below 1 with one spatially confined exception at 2000 yr. Hence, in this case even the combination of thermal and magnetic pressure cannot stabilise the disc against fragmentation as shown in the upper left panel of Fig.~\ref{fig:toomre}. Fragmentation occurs in the innermost part (r $<$ 50 AU) close to the first sink particle and in a filamentary-like ring at r $\simeq$ 200 AU. We note that the situation is somewhat different to run 26-4 where fragmentation seems to occur mainly in spiral like density perturbations closer to the centre (see top panel of Fig.~\ref{fig:disk}). However, the stability analysis of run 26-4 shows that $Q$ and $Q_{\rmn{M}}$ are below 1 as well.

All other runs considered in Fig.~\ref{fig:toomre} show no fragmentation so far. This agrees quite well with the findings of $Q_{\rmn{M}} \simeq 1$. However, there are some situations when $Q_{\rmn{M}}$ is marginally smaller than 1 with the disc still showing no signs of fragmentation. \citet{Kratter10}, analysing the stability of discs under purely hydrodynamical conditions, find that discs can be indeed stable if Q is locally smaller than 1. The authors attribute this to the fact that $Q$ = 1 indicates instability of axisymmetric perturbations in infinitely thin discs~\citep{Toomre64}. For thick discs, as it is the case in our simulations, they argue that the instability criterion is expected to be somewhat relaxed~\citep{Goldreich65}. Hence, we expect our discs to be stable for Toomre values even slightly smaller than 1.

To summarise, the comparison of $Q$ and $Q_{\rmn{M}}$ shows that the stabilising effect of magnetic fields is needed to effectively prevent the discs from fragmenting. For weak magnetic fields ($\mu = 26$) and high amounts of angular momentum, however, even magnetic fields cannot stabilise the discs against fragmentation. Furthermore, we expect that fragmentation would occur for other runs with low magnetic field strengths ($\mu \ge 10$) as well if the evolution of the discs would be followed beyond 4000 yr. This assumption is confirmed by the results that the discs show a continuing growth due to ongoing infall and are only marginally stable ($Q_{\rmn{M}} \simeq$ 1).

\section{Discussion} \label{sec:dis}

\subsection{Numerical caveats} \label{sec:numerics}

The simulations considered so far were run with a maximum resolution of 4.7 AU. To test the resolution dependency of our results we performed two more simulations with the same initial setup as in run 26-4 but with a resolution varied by a factor of 4 in either direction. A detailed comparison of the results is shown in the appendix. The results of the resolution study suggest that the spatial resolution of 4.7 AU used in the simulations presented in this work is sufficient to properly follow the dynamical evolution of the protostellar discs and of the accretion rates.

The bubble like features seen in the lower panel of Fig.~\ref{fig:disk} also occur in other runs with strong initial magnetic fields and are artifacts of our assumption of ideal MHD. In particular, for runs with $\mu = 2.6$ the bubbles show a significant influence on the accretion rates (bottom right panel of Fig.~\ref{fig:sink-acc}) reducing the accretion after 2500 yr. However, for runs with $\mu = 5.2$ the dynamical influence of such bubbles seems to be rather limited as we cannot detect any significant changes in the accretion rates (see bottom left panel of Fig.~\ref{fig:sink-acc}). We therefore conclude that the accretion rates are reliable for runs with $\mu$ down to $\sim$ 5 and up to 2500 yr even for the both runs with $\mu = 2.6$. A possible solution to avoid the formation of bubbles like features could be the inclusion of the effects of non-ideal MHD such as ambipolar diffusion or ohmic dissipation as done in recent work \citep{Duffin09,Mellon09,Dapp10}. However, \citet{Nakano02} argue that ohmic dissipation starts to act efficiently at particle densities above 10$^{-12}$ cm$^{-3}$ which is well above the typical densities found in our discs. Hence, ohmic dissipation is not expected to be capable of significantly reducing the magnetic flux in the centre. Ambipolar diffusion, however, starts to act at lower densities \citep{Duffin09} and thus might be able to reduce the field strength in the centre before magnetically driven bubbles can occur.

As mentioned in Section~\ref{sec:methods}, a minimum density threshold of  $1 \cdot 10^{-15}$ g cm$^{-3}$ is applied after the outflow is launched. The exact amount of mass added artificially during the 4000 yr depends on the individual simulation but never exceeds a few 10$^{-2}$ M$_{\sun}$ corresponding to a mass rate of $10^{-6} - 10^{-5}$ M$_{\sun}$ yr$^{-1}$ which is significantly below the observed accretion rates of a few 10$^{-4}$ M$_{\sun}$ yr$^{-1}$. Hence, we suppose that the dynamical influence of the density threshold should be negligible and its application is acceptable in order to avoid very small timesteps and in consequence significantly higher computational costs.

It is not clear to what extent our results are affected by the chosen density profile $\rho(r) \propto r^{-1.5}$. \citet{Girichidis11} find that their results strongly depend on the initial profile which they attribute to the varying importance of turbulence compared to gravity. As we do not include turbulent motions in our simulations, our results might depend somewhat less on the density profile, in particular with regard to the accretion rates which also vary rather moderately in the work of \citet{Girichidis11}.

\subsection{Disc evolution} \label{sec:diskevol}

As demonstrated in Fig.~\ref{fig:vel}, a transition from early-type, large-scale Keplerian discs to sub-Keplerian discs occurs around a normalised initial mass-to-flux ratios $\mu$ of $\sim$ 10 independent of the initial amount of rotational energy. This is in good accordance with a series of papers studying the evolution of low-mass magnetised disc. While \citet{Hennebelle08} and \citet{Hennebelle09} find a value for $\mu$ between 5 - 10, below which Keplerian disc formation is suppressed, \citet{Allen03} and \citet{Mellon08} find no Keplerian discs for $\mu$ up to at least 10. \citet{Duffin09} studying the possible fragmentation of magnetised discs with an initial $\mu$ = 3.5 find sub-Keplerian rotation profiles as well in agreement with the results mentioned before. Although these simulations apply to low-mass star formation with core masses around 1 M$_{\sun}$, the observed maximum value of $\mu$, for which the formation of large-scale Keplerian discs is suppressed, agrees remarkable well with our finding of $5 < \mu < 10$.

For the sub-Keplerian discs observed in our simulations the infall velocities are generally significantly larger than the rotation speed, i.e. $v_{\rmn{rot}}$/$v_{\rmn{rad}} < 1$, and close to free-fall (see right panel of Fig.~\ref{fig:vel_evol} and Fig.~\ref{fig:vel}). This is attributed to the highly gravitationally unstable configuration of the cores containing about 56 Jeans masses. Hence, they cannot be stabilised against gravitational collapse by thermal pressure alone in contrast to low-mass cores containing only $\sim$ 1 Jeans mass. Consistently, for highly magnetised low-mass cores \citet{Hennebelle08} and \citet{Duffin09} find smaller infall velocities as well as $v_{\rmn{rot}}$/$v_{\rmn{rad}} \ga 1$ contrary to our results.

For $\mu \ga 10$, we find large-scale Keplerian discs in our simulations. This is probably a consequence of not having turbulence in our initial setup. Turbulence would most likely hamper the early formation of large rotating structures and delay it to later stages. However, in recent years there is an increasing number of observations of rotationally supported discs in massive star forming regions~\citep[e.g.][and references therein]{Cesaroni07}. As an example, discs with sizes of a few 100 AU and masses between 0.15 M$_{\sun}$ \citep{Fuller01} and about 10 M$_{\sun}$ \citep{Shepherd01} have been observed. These discs are similar in size to the discs obtained in our runs with weak magnetic fields and also show Keplerian-like rotation profiles. In addition, our disc masses of $\sim$ 1 M$_{\sun}$ calculated in run 26-20 and run 26-4 (see Fig.~\ref{fig:frag}) fit perfectly in the mass interval spanned by these observations. However, the protostars observed have masses around 5 - 10 M$_{\sun}$ and are therefore a factor of a few more massive than our most massive sink particles. We attributed this to the fact that the observed disc/star-systems are in much later evolutionary stages than our systems.

In contrast, for $\mu < 10$ sub-Keplerian discs are observed in our simulations. Again, we mention that our simulations end 4000 yr after sink particle formation, thus in a very early phase. An observer looking at such a system from edge-on will observe a flattened structure similar to a Keplerian disc but without the typical signatures of rotation (see right panel of Fig.~\ref{fig:maggrad}). Indeed, there is a growing number of observations of such flattened structures reported in literature \citep[see][and references therein]{Cesaroni07}, although these observations often refer to more massive structures and more evolved protostellar objects than present in our simulations. Due to the insufficient centrifugal support, these structures are not in equilibrium and show considerable radial infall motions, in accordance with our findings (compare bottom panel of Fig.~\ref{fig:vel}).

The reason for the lack of centrifugal support in such sub-Keplerian discs is the strong magnetic torque acting on the midplane (see Fig.~\ref{fig:torque}). Angular momentum is removed by magnetic braking at roughly the same rate as it is added by the infalling gas. The angular momentum removed from the midplane is partly deposited in the outflow and partly in regions further out which are connected to the inner parts by the magnetic lever arm \citep{Allen03} created through the equatorial pinching of the magnetic field (see Fig.~\ref{fig:maggrad}). A possible way trying to reduce the magnetic braking efficiency would be to include non-ideal MHD effects like ohmic dissipation or ambipolar diffusion in the simulations. However, recent numerical work including ambipolar diffusion \citep{Mellon09,Duffin09}, ohmic dissipation \citep{Dapp10} and also both effects \citep{Li11} show that even in the case of non-ideal MHD it is not possible to form Keplerian discs in such early stages. In fact, these authors find that the formation of rotationally supported discs in the case of strong magnetic fields is suppressed down to scales well below our resolution limit of roughly 10 AU. Hence, at the evolutionary stage considered in this work, we cannot expect to resolve a proper Keplerian disc even by including the effects of ambipolar diffusion or ohmic dissipation in our calculations.

The question now arises how Keplerian discs on 100 AU scale, frequently observed around massive protostars, can form if the cores have mass-to-flux ratios $\mu \la 5$. The situation even intensifies as observed star forming regions usually have mass-to-flux ratios which are only slightly supercritical with a mean $\mu \la 5$~\citep[e.g.][]{Falgarone08,Girart09,Beuther10}. Following the long term evolution of highly magnetised low-mass cores, \citet{Machida10} find large-scale Keplerian discs occurring after roughly 10$^5$ yr. The authors argue that in their case magnetic braking only redistributes the angular momentum within the collapsing cores but does not remove it. Additionally, they find that the outflows are too weak to ultimately remove a significant fraction of mass and angular momentum from the cores. Thus, most of the cores mass and angular momentum finally fall onto the midplane resulting in large-scale Keplerian discs even for high magnetic field strengths. However, we note that the authors might underestimate the total amount of mass and angular momentum transfered into the interstellar medium as only low velocity outflows are modeled in their simulations. Considering high velocity jets and radiation-driven outflows, the amount of material ejected into the interstellar medium could be considerably larger, thus impeding the formation of large Keplerian discs. For massive stars we expect this to be even more severe as here radiation-driven outflows are even more powerful and will be able to eject angular momentum -- deposited in the envelope by magnetic effects -- at even higher rates than low-mass protostellar outflows~\citep[e.g.][]{Arce07}.

At the same time outflows also provide a possible solution of the disc formation problem by diluting the envelope in which the magnetic field lines are anchored~\citep[see also $\S$6.2.2 in][]{Mellon08}. As the magnetic braking time \citep{Mouschovias80}
\begin{equation}
 t_{\rmn{mag}} = \frac{z}{v_{\rmn{A,env}}}\frac{\rho_{\rmn{d}}}{\rho_{\rmn{env}}} \propto \rho_{\rmn{env}}^{-1/2}
\end{equation}
increases with decreasing envelope density, this allows large-scale, centrifugally supported discs to form in particular at later stages not considered in this work when most of the envelope has been blown away by powerful outflows. This would agree with the fact that most of the observed disc/star-systems \citep[see][and references therein]{Cesaroni07} are in a later evolutionary stage than our systems although the observations could also be biased by the fact that massive Class 0 objects are more difficult to detect. The picture of a successive growth of centrifugally supported discs during the evolution into Class I / II is also supported from the theoretical side by \citet{Dapp10}. Therefore, we expect large-scale Keplerian discs to form in our runs as well if the evolution would be followed over a much longer time, thus relaxing the problem of catastrophic magnetic braking.

In summary, our simulations suggest that Keplerian discs do not form around massive protostars in the very early stage except for unusually weak magnetic fields. Nevertheless, we expect that centrifugally supported discs will build up during the subsequent evolution of the collapsing cores.

Another question not addressed so far is the problem of massive binary formation. In neither of our simulations do we see indications for massive binaries so far which of course could be a consequence of the early stage the simulations end. Nevertheless, our results indicate that massive binaries possibly form in later evolutionary stages and that the initial mass ratio should be far from unity with the more massive star sitting in the centre. Observations, however, reveal that a significant fraction of massive binaries consists of components with nearly equal masses~\citep[e.g][]{Mason98,Pinsonneault06}. This apparent contradiction could be resolved by the work of \citet{Artymowicz96} and \citet{Bate97}. These authors simulating circumbinary discs propose that the masses of binary components, even when unequal in the beginning, tend to equal as accretion within the disc occurs preferentially onto the lower-mass companion. Initial density perturbations not consider here could also be a natural way to from massive binaries. However, there are indications that fragmentation induced by initial density perturbations might be suppressed by magnetic fields~\citep{Hennebelle08b}. Thus, different binary formation scenarios might have to be at work.

\subsection{Accretion rates}

As shown in Fig.~\ref{fig:accoverview}, the accretion rates of the different runs vary only by a factor of $\sim$ 3 which is attributed to the two competing effects of the magnetic field namely magnetic braking and the Lorentz force. Adopting accretion rates of a few 10$^{-4}$ M$_{\sun}$ yr$^{-1}$ as observed in our simulations, stars of about 30 M$_{\sun}$ would form within some 10$^4$ up to a few 10$^5$ yr roughly independent of the initial conditions. However, this only holds if the accretion rates stay roughly constant over the whole formation process which might be an oversimplification as indicated by the slight decrease seen in Fig.~\ref{fig:sink-acc} \citep[see also][]{Klessen01,Schmeja04}. A possible way to significantly change accretion rates would be to vary the initial density profile and mass of the molecular cloud core, parameters which are not explored in this work in order to limit the computational costs~\citep[see e.g.][]{Girichidis11}.

However, our observed accretion rates agree well with accretion rates from a number of massive star formation simulations. To begin with, our accretion rates match those necessary to overcome radiation pressure deduced from 1-dimensional calculations~\citep{Kahn74,Wolfire87}. Radiation-hydrodynamical collapse simulations in 2 dimensions~\citep{Yorke02,Kuiper10} and 3 dimensions~\citep{Krumholz07,Krumholz09,Kuiper11} with similar core masses reveal accretion rates of a few 10$^{-4}$ M$_{\sun}$ yr$^{-1}$ up to 10$^{-3}$ M$_{\sun}$ yr$^{-1}$ very similar to ours. \citet{Peters10a,Peters10b,Peters11} simulating the long term evolution of H~\textsc{ii}-regions around massive protostars find similar accretion rates as well. This suggests that in our simulations accretion would continue even if the effect of radiation would be included. \citet{Girichidis11} studying the effect of different initial conditions on massive star formation find accretion rates around 10$^{-3}$ M$_{\sun}$ yr$^{-1}$, somewhat higher than ours possibly caused by the omission of rotation or magnetic fields counteracting gravity and accretion. Similar accretion rates were also found by \citet{Banerjee07} studying the very early evolution of a collapsed cloud core with initial magnetisation and rotation using a significantly higher resolution than in our work. \citet{Hennebelle11} observe accretion rates of the order of 10$^{-5}$ - 10$^{-4}$ M$_{\sun}$ yr$^{-1}$ somewhat smaller than ours which we attribute to the larger core size of $\sim$ 1 pc used in their setup.

Our accretion rates also agree quite well with theoretical estimates and observations. Calculating the accretion rates with the formula given in the theoretical work of \citet{McKee03} adapted to our setup we find a value of about 3 $\cdot$ 10$^{-4}$ M$_{\sun}$ yr$^{-1}$ very similar to the actually observed accretion rates. Observational results for accretion rates in massive star forming regions are hard to obtain and are often calculated only indirectly via observed outflow mass rates. Nevertheless, results from several high-mass star forming regions indicate accretion rates of the order of 10$^{-4}$ - 10$^{-3}$ M$_{\sun}$ yr$^{-1}$ \citep[e.g.][]{Beuther02a,Beuther02c,Beltran06} again in good accordance with our results.

As shown in Fig.~\ref{fig:sink-acc}, the accretion rates do not drop significantly over time (except for runs with $\mu = 2.6$, see Section~\ref{sec:numerics}). Hence, the outflows launched seem to be not capable of significantly reducing mass accretion onto the sinks over time. In order to analyse to what extent the combined effect of magnetic fields and rotation influences mass accretion, we performed the reference calculation inf-0 with no magnetic field and zero rotation. In this run the sink particle accretes 3.55 M$_{\sun}$ within the first 4000 yr corresponding to a time averaged accretion rate of $8.86 \cdot 10^{-4}$ M$_{\sun}$ yr$^{-1}$. Hence, the accretion rates in runs with magnetic fields and rotation are reduced to a level of about 35\% - 95\% of this value (see Table~\ref{tab:msinks}). Our simulations show that feedback due to outflows is not able to stop accretion onto the protostars and hence outflows are not able to determine the low star formation efficiency observed in molecular clouds.

\section{Conclusion} \label{sec:conclusion}

We have studied the collapse of massive molecular cloud cores with varying initial rotational and magnetic energies. The cores are supercritical with mass-to-flux ratios between 2.6 and 26 and have rotational energies well below the gravitational energy. Containing about 56 Jeans masses the cores are highly gravitationally unstable and hence might be sites of protostellar cluster formation. We focussed our discussion on the formation of protostellar discs and on protostellar accretion as measured by sink particles. We find that disc properties are highly sensitive to the initial magnetic field strength whereas protostellar accretion rates are only marginally influenced by varying initial conditions. In the following we summarise our main findings.
\begin{enumerate}
 \renewcommand{\theenumi}{\arabic{enumi}.}
 \item For normalised mass-to-flux ratios $\mu$ below 10 the formation of centrifugally supported discs is completely suppressed. Instead, for $\mu < 10$ sub-Keplerian discs are formed showing nearly no rotational support but radial infall velocities close to free-fall. For weak magnetic fields ($\mu > 10$), however, well defined Keplerian discs with sizes of a few 100 AU build up over time. This agrees well with several studies of collapsing low-mass cores.
 \item Sub-Keplerian rotation for strong magnetic fields ($\mu \la 10$) is caused by magnetic braking. Analysing the torques acting on the midplane we find that angular momentum is removed due to magnetic braking at roughly the same rate as it is added due to the gas infall thereby  preventing Keplerian discs from forming.
 \item Observed accretion rates are of the order of a few 10$^{-4}$ M$_{\sun}$ yr$^{-1}$ varying only within a factor of $\sim$ 3 between the individual runs. This variation is remarkably small considering the large differences in the initial conditions varying over more than two orders of magnitude. We attributed this to two competing effects of the magnetic field. Increasing the magnetic field strength results in an increased accretion rate due to an enhanced magnetic braking efficiency lowering centrifugal support. Above a certain field strength, however, a further increase leads to declining accretion rates due to the effect of magnetic pressure and tension. This results in rather moderate changes in the accretion rates for different initial conditions. Furthermore, accretion rates for different amounts of angular momentum converge with increasing field strength due to the effect of magnetic braking.
 \item For the majority of the simulations disc fragmentation does not occur. Analysing the Toomre parameter $Q$ in these discs reveals that fragmentation is mainly suppressed due to the magnetic pressure. Thermal pressure alone cannot account for the stabilisation as $Q$ is well below 1 whereas the magnetic Toomre parameter $Q_{\rmn{M}}$ settles around 1 indicating stability. In two simulations, which are subject to fragmentation, neither thermal nor magnetic pressure can stabilise the discs. Accordingly, both $Q$ and $Q_{\rmn{M}}$ are well below 1.
 \item In the two runs with disc fragmentation more than 10 sink particles are formed during the first 4000 yr.
Among these sinks only the first one created reaches a mass above 1 M$_{\sun}$ thus containing more than 80\% of the totally accreted mass. All other particles have masses well below 0.1 M$_{\sun}$. The discs formed in both cases reach masses around 1 M$_{\sun}$ somewhat below the totally accreted sink particle masses.
 \item The outflows launched from the protostellar discs are not capable of significantly reducing mass accretion over time. Compared to a run with zero magnetic field and zero rotation, mass accretion is reduced to a level of 35\% - 95\%.
\item Radial profiles of column density and temperature exhibit accretion shock features moving outwards as time evolves. The shocks occur when the infalling gas hits either centrifugal or magnetic barriers thus strongly decreasing the infall speed.

\end{enumerate}

A growing number of observations of discs and bipolar outflows around high-mass protostars \citep[see][for recent reviews]{Beuther05,Cesaroni07} support a high-mass star formation scenario via disc accretion. On the other hand, observations also reveal that prestellar cores with masses ranging from 2 - 2000 M$_{\sun}$ are usually only slightly supercritical with $\mu \la 5$ \citep{Falgarone08, Girart09,Beuther10}. Together with our numerical results this suggest that there should be no Keplerian discs in the very early stages of typical high-mass star forming regions but rather flattened, strongly sub-Keplerian structures. This apparent dichotomy has an important impact on the formation of discs around massive stars. To enable the observed presence of centrifugally supported discs in later stages, effects in the later evolution of the system are required reducing the efficiency of magnetic braking. We discussed possible effects in the context of massive star formation like outflows and non-ideal MHD leading to a successive growth of discs in the later evolution.

So far, the influence of turbulence is completely neglected in our setup. Hence, we plan to include initial velocity perturbations in our setup to obtain more realistic initial conditions as indicated by observations~\citep[e.g.][]{Caselli95}. For subsequent simulation it would be also interesting to study the effect of ambipolar diffusion which is expected to act efficiently in the high density regime of protostellar discs. For this purpose, the existing set of simulations serves as a useful guide to select representative simulations to be repeated with increased resolution and additional physics.

\section*{Acknowledgements}

The authors like to thank the anonymous referee for comments which helped to significantly improve the paper. D.S. likes to thank Philipp Girichidis and Christoph Federrath for many useful discussions and suggestions. The simulations presented here were performed on HLRB2 at the Leibniz Supercomputing Centre in Garching and on JUROPA at the Supercomputing Centre in J\"ulich. The FLASH code was developed partly by the DOE-supported Alliances Center for Astrophysical Thermonuclear Flashes (ASC) at the University of Chicago. D.S. and R.B. acknowledge funding of Emmy-Noether grant BA3706 by the DFG. R.S.K.~acknowledges subsidies from the {\em Baden-W\"urttemberg-Stiftung} (grant P-LS-SPII/18) and from the German {\em Bundesministerium f\"ur Bildung und Forschung} via the ASTRONET project STAR FORMAT (grant 05A09VHA). R.S.K. furthermore gives thanks for subsidies from the Deutsche Forschungsgemeinschaft (DFG) under grants no.\ KL 1358/1, KL 1358/4, KL 1359/5, KL 1358/10, and KL 1358/11, as well as from a Frontier grant of Heidelberg University sponsored by the German Excellence Initiative. D.D. is supported by McMaster University and R.E.P by a Discovery grant from NSERC of Canada.


\section*{Appendix}

Here we compare simulation 26-4 to two more runs with identical initial conditions but a maximum spatial resolution varied by a factor 4 in either direction. In particular, the initial resolution is identical to that in run 26-4, i.e. in the beginning the mesh in the core has a spacing of 302 AU. We list the runs and their corresponding parameters in Table~\ref{tab:res}. The critical value of the density above which sink particles are created is adapted in accordance with the resolution. For all runs performed the refinement criterion applied guaranties that the midplane region is resolved on the highest level used. In particular, we focus in this analysis on accretion properties and radial profiles of different quantities in the midplane. Due to computational cost reasons run 26-4-H is followed for 2000 yr only. Hence, we compare the results at this time.
\begin{table}
 \caption{Performed simulations for the resolution study with maximum spatial resolution, threshold density for sink particle creation, formation time of the first sink particle and total accreted mass after 2000 yr.}
 \label{tab:res}
 \begin{tabular}{@{}lcccc}
  \hline
  Run & d$x$  & $\rho_{\rmn{crit}}$      & t$_0$ & M$_{\rmn{sink}}$ \\
      &  (AU) & (10$^{-12}$ g cm$^{-3}$) & (kyr) & (M$_{\sun}$) \\
  \hline
  26-4-L & 18.9 & 0.0657 & 15.1 & 1.42 \\
  26-4 & 4.7 & 1.78 & 15.2 & 1.05 \\
  26-4-H & 1.2 & 114 & 15.3 & 1.03 \\
  \hline
 \end{tabular}
\end{table}

First, we consider radial profiles of the density, temperature and velocity in the midplane.
\begin{figure*}
 \includegraphics[width=56mm]{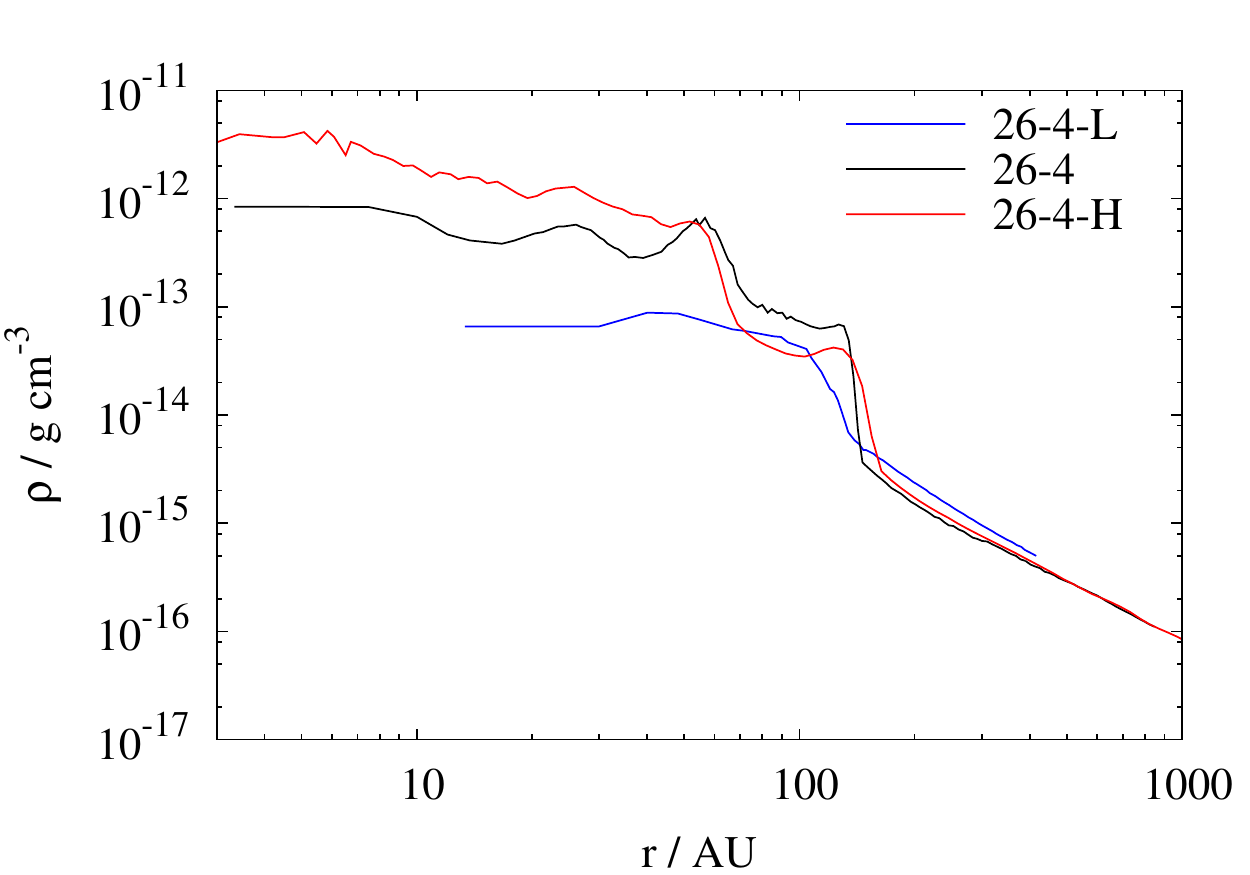}
 \includegraphics[width=56mm]{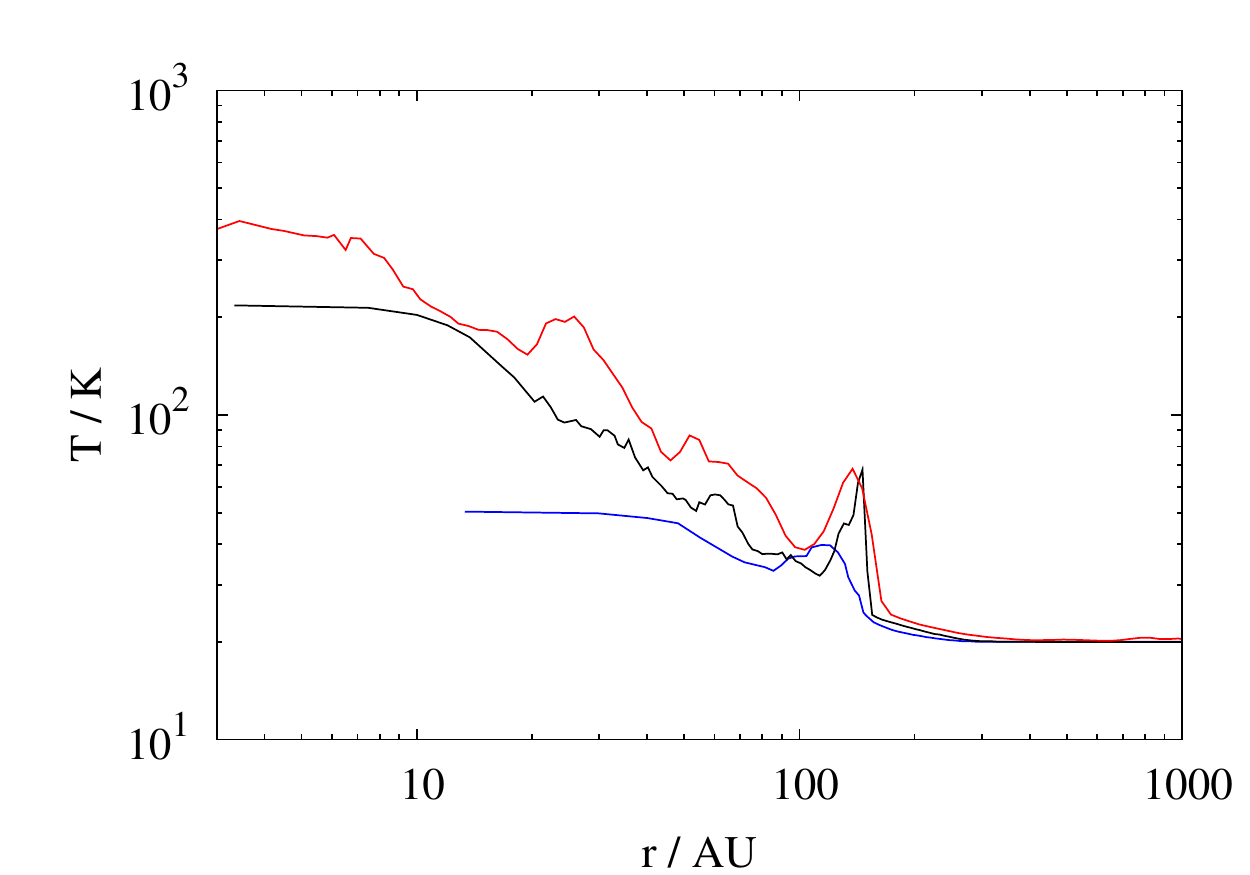}
 \includegraphics[width=56mm]{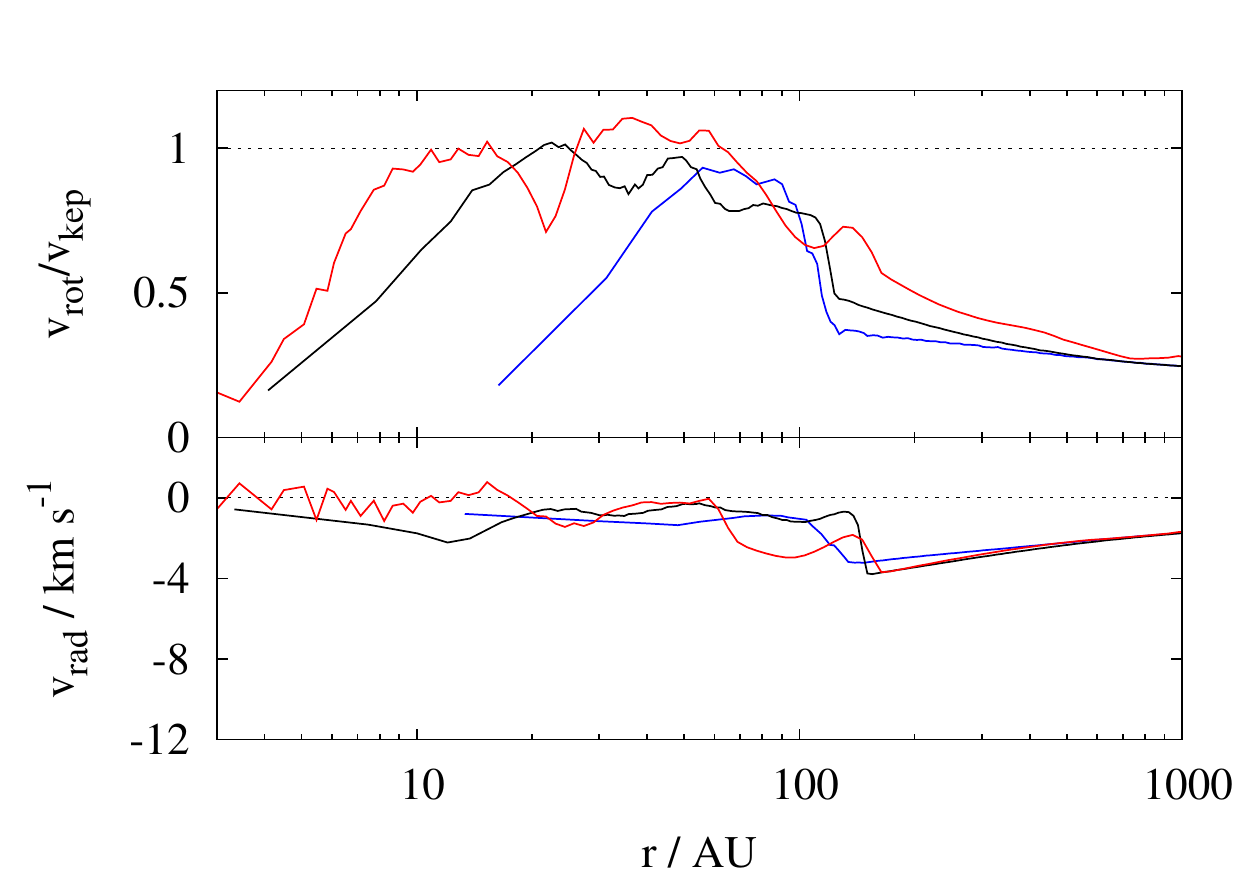}
 \caption{Radial profile of density (left), temperature (middle) and velocity structure (right) after 2000 yr for the runs 26-4-L (blue), 26-4 (black) and 26-4-H (red). 
The accretion shock seen in the left and middle panel is not well resolved for run 26-4-L.}
 \label{fig:res-radial}
\end{figure*}
The quantities in each run are taken in the midplane. In the density profile (left panel of Fig.~\ref{fig:res-radial}) the accretion shock occurring at $\sim$ 150 AU is clearly resolved in the runs 26-4-H and 26-4. In run 26-4-L, however, the shock is somewhat smoothed out due to the limited resolution. Hence, a resolution of 4.7 AU seems required to properly resolve the accretion shock. Within the accretion shock, however, the density increases with resolution. We attribute this to the fact that the vertical structure of the disc is not fully resolved at least in the runs 26-4 and 26-4-L. Here in large parts the vertical disc height is represented by a few grid points only (compare Fig.~\ref{fig:maggrad}). Therefore, to fully resolve the vertical disc structure, a higher resolution, probably even above that in run 26-4-H, would be needed which is not feasible for computational costs reasons.

A similar result holds for the temperature profiles as well (middle panel of Fig.~\ref{fig:res-radial}). The temperature jump at the accretion shock seems be to reasonably well resolved in run 26-4 whereas in run 26-4-L it is smoothed out markedly. Within the accretion shock the temperature reaches as higher values as higher the final resolution is. This is due to the strong coupling of temperature and density above 10$^{-13}$ g cm$^{-3}$ where the gas gets optically thick resulting in higher temperatures at higher gas densities.

Next, we analyse the velocity structure in the midplane (see right panel of Fig.~\ref{fig:res-radial}). In run 26-4-L no Keplerian disc has built up yet. However, we mention that in its further evolution the rotation reaches Keplerian velocities as well. In contrast, in run 26-4 the rotation is already Keplerian up to a radius of $\sim$ 60 AU in good agreement with run 26-4-H. Furthermore, the comparison shows that in run 26-4 the decline in $v_{\rmn{rot}}$/$v_{\rmn{kep}}$ at 20 AU is most likely a resolution effect as in run 26-4-H this decline occurs at a roughly three times smaller radius of $\sim$ 8 AU. This supports the statement made in Section~\ref{sec:time} that the inner 10 AU in runs with a resolution of 4.7 AU are strongly affected by numerical resolution (see also Fig.~\ref{fig:vel}). The radial velocities at radii larger 10 AU show a reasonably well agreement between all runs. Hence, we conclude that regarding the velocity structure in the midplane run 26-4 is reasonably well converged.

\begin{figure}
 \includegraphics[width=84mm]{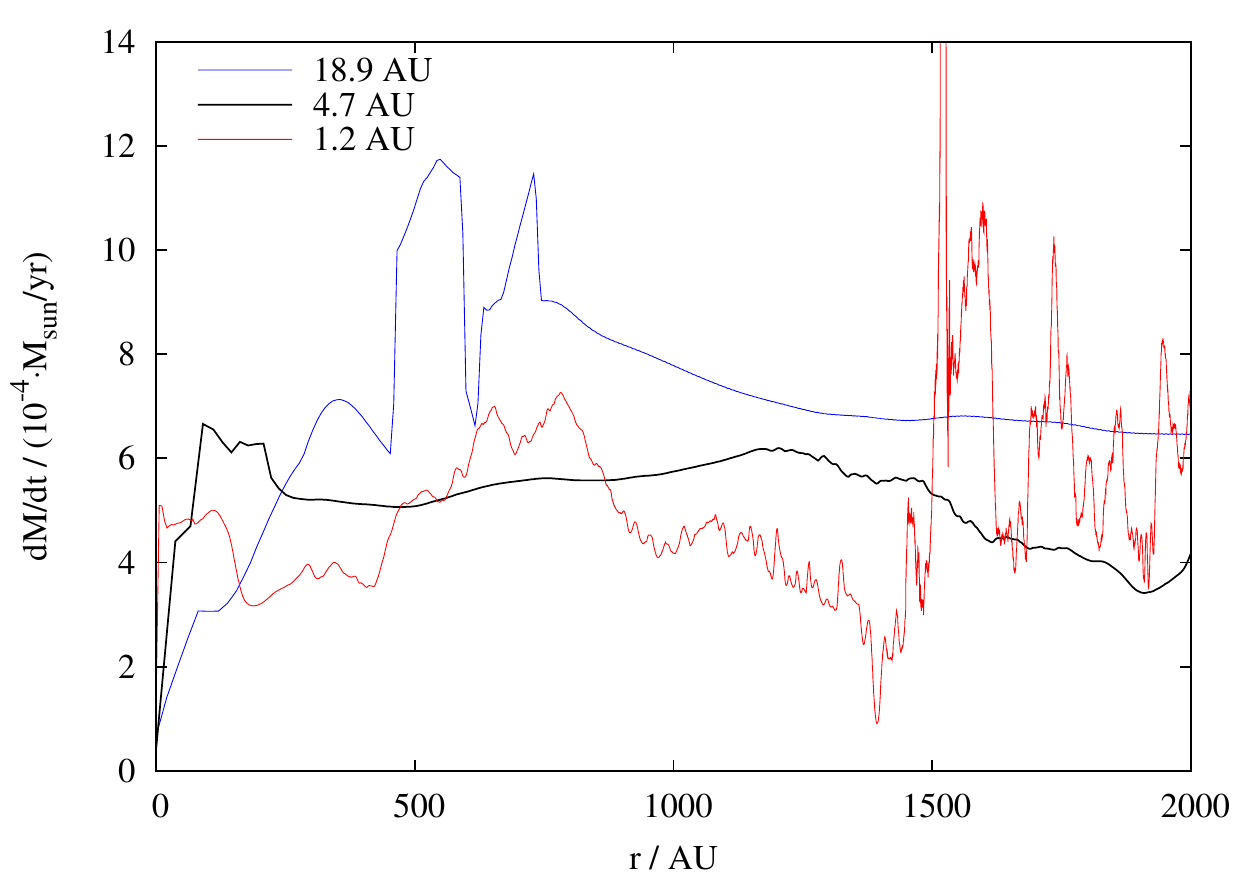}
 \caption{Total accretion rate for the first 2000 yr in the runs 26-4-L (blue line), 26-4 (black line) and 26-4-H (red line) with a maximum spatial resolution of d$x$ = 18.9 AU, 4.7 AU and 1.2 AU, respectively. The mean accretion rates (see also Table~\ref{tab:res}) are as higher as lower the resolution is but seem to converge towards higher resolution.}
 \label{fig:res-acc}
\end{figure}
In Fig.~\ref{fig:res-acc} we show the time evolution of the total accretion rate for the three runs considered. In run 26-4-H two more sink particles are created after roughly 1500 yr which cause the large variations in the accretion rate. In general, however, the accretion rates of the three runs are of the same order of magnitude. From the formation time t$_0$ listed in Table~\ref{tab:res} it can be inferred that t$_0$ increases with spatial resolution. This behaviour is expected as for higher resolution sink particles are created at higher densities and thus later times during the collapse. It can also be inferred from Table~\ref{tab:res} that the mean accretion rate, i.e. the total accreted mass, decreases with increasing spatial resolution. The accretion rate of run 26-4-L is higher than that of run 26-4 by about 35\%. We therefore conclude that regarding accretion properties run 26-4-L is not yet converged. The difference in accreted mass between run 26-4-H and run 26-4 is of the order of 2\% thus significantly lower than the difference between run 26-4-L and run 26-4. Hence, there seems to be a convergence of the accretion rates with increasing resolution and, although run 26-4 seems not fully converged, we conclude that a resolution of 4.7 AU is sufficiently high to properly describe accretion properties of the protostars.

In summary, we can say that a resolution of 4.7 AU seems sufficiently high to properly follow protostellar accretion rates, resolve the accretion shock at the edge of the disc, and display the velocity structure in the disc down to about 10 - 15 AU. In contrast, run 26-4-L with a resolution of 18.9 AU reveals significant differences from run 26-4-H showing that is not yet converged. However, also run 26-4 seems to be not fully converged regarding to the density and temperature structure in the disc. We attribute this partly to the poor resolution of the vertical disc structure. Hence, a higher resolution, probably even above the one used in run 26-4-H, would be necessary to reach convergence regarding this point. However, due to significantly higher computational costs this has not been feasible wherefore a spatial resolution of 4.7 AU is used throughout the paper.
This particular choice is also motivated physically as we want to resolve the first core. Hence, the Jeans length at densities above 10$^{-13}$ g cm$^{-3}$ has to be resolved requiring a resolution of a few AU.

\label{lastpage}

\end{document}